\documentclass{emulateapj}
\usepackage{graphicx}
\usepackage{subfigure}
\usepackage{natbib}
\newcommand\teff{T$_{\rm eff}$}
\newcommand\logg{log\,{\it g}}
\newcommand\kms{km\,s$^{-1}$}
\newcommand\afe{[$\alpha$/Fe] \,}
\newcommand\etal{{\rm et al.\,}}

\slugcomment{In Preparation for the Astrophysical Journal}

\shorttitle{Chemistry of the Carina Dwarf Galaxy} 
\shortauthors{Venn \etal}

\begin{document}


\title{Nucleosynthesis and The Inhomogeneous Chemical Evolution 
   of the Carina Dwarf Galaxy$^*$}

\author{Kim A. Venn$^{1}$, Matthew D. Shetrone$^{2}$, Mike J. Irwin$^{3}$,
   Vanessa Hill$^{4,5}$, Pascale Jablonka$^{5,6}$, 
   Eline Tolstoy$^{7}$, Bertrand Lemasle$^{7}$, Mike Divell$^{1}$, 
   Else Starkenburg$^{7}$, Bruno Letarte$^{8}$, Charles Baldner$^{9}$ 
   Giuseppina Battaglia$^{10}$, 
   Amina Helmi$^{7}$, Andreas Kaufer$^{11}$, Francesca Primas$^{10}$.
}

\affil{$^{1}$Department of Physics \& Astronomy, University of Victoria, 3800 Finnerty Road, 
             Victoria, BC, V8P 1A1, Canada 
  $^{2}$McDonald Observatory, University of Texas at Austin, HC75 Box 1337-McD, 
        Fort Davis, TX, 79734, USA
  $^{3}$Institute of Astronomy, University of Cambridge, Madingley Road, Cambridge, CB03 0HA, UK
  $^{4}$Laboratoire Cassiop\'ee UMR 6202, Universit\'e de Nice Sophia-Antipolis, CNRS,
        Observatoire de la C\^ote d'Azur, France
  $^{5}$GEPI, Observatoire de Paris, CNRS UMR 8111, Universit\'e Paris Diderot, F-92125,
        Meudon, Cedex, France
  $^{6}$Laboratoire d'Astrophysique, \'Ecole Polytechnique
        F\'ed\'erale de Lausanne (EPFL), 1290 Sauverny, Switzerland,
  $^{7}$Kapteyn Astronomical Institute, University of Groningen, PO Box 800, 9700 AV Groningen,
        The Netherlands
  $^{8}$South Afrian Astronomical Observatory, Observatory Road, 7935 Observatory, South Africa
  $^{9}$Dept of Astronomy, Yale University, P.O. Box 208101, New Haven, CT 06520-8101, USA
  $^{10}$European Southern Observatory, Karl-Schwarzschild-Strasse 2, 85748 Garching, Germany
  $^{11}$European Southern Observatory, Alonso de Cordova 3107, Santiago, Chile 
}

\email{email: kvenn@uvic.ca}

\thanks{$^*$ This project is partially based on VLT FLAMES 
spectroscopic observations obtained at the European Southern
Observatory, proposals 074.B-0415 and 076.B-0146, and partially based on MIKE spectra
gathered at the 6.5-m Magellan Telescopes, in Chile.} 
 
\begin{abstract}

The detailed abundances of 23 chemical elements in nine bright RGB stars in the 
Carina dwarf spheroidal galaxy are presented based on high resolution 
spectra gathered at the VLT and Magellan telescopes.     A spherical 
model atmospheres analysis is applied using standard methods (LTE and
plane parallel radiative transfer) to spectra ranging from 380 to 680 nm.   
Stellar parameters are found to be consistent between photometric and
spectroscopic analyses, both at moderate and high resolution.
The stars in this analysis range in metallicity from $-2.9 < $ [Fe/H] $< -1.3$, 
and adopting the ages determined by Lemasle \etal (2012), we are able to 
examine the chemical evolution of Carina's old and intermediate-aged
populations.   One of the main results from this work is the evidence
for inhomogeneous mixing in Carina;  a large dispersion in 
[Mg/Fe] indicates poor mixing in the old population, an offset 
in the [$\alpha$/Fe] ratios between the old and intermediate-aged 
populations (when examined with previously published results) 
suggests that the second star formation event occurred in $\alpha$-enriched 
gas, and one star, Car-612, seems to have formed in a pocket enhanced in 
SN Ia/II products.   This latter star provides the first direct link 
between the formation of stars with enhanced SN Ia/II ratios in dwarf 
galaxies to those found in the outer Galactic halo (Ivans \etal 2003).  
Another important result is the potential evidence for
SN II driven winds.   We show that the very metal-poor stars in Carina
have not been enhanced in AGB or SN Ia products, and therefore their 
very low ratios of [Sr/Ba] 
suggests the loss of contributions from the early SNe II. 
Low ratios of [Na/Fe], [Mn/Fe], and [Cr/Fe] in two of these stars
support this scenario, with additional evidence from the low [Zn/Fe]
upper limit for one star.  
%
It is interesting that the chemistry of the metal-poor stars in Carina 
is not similar to those in the Galaxy, most of the other dSphs, or the UFDs, 
and suggests that Carina may be at the critical mass where {\it some}
chemical enrichments are lost through SN II driven winds.

\end{abstract}


\section{Introduction} 

Chemical analyses of stars in nearby classical dwarf galaxies have
swelled in the past decade, with the advent of large aperture 
telescopes, high efficiency spectrographs, and multiplexing 
capabilities.    
%
%
Detailed abundances are now available for dozens to hundreds of 
stars in classical dwarf spheroidal galaxies (dSph), 
ultra faint dwarf (UFD) systems, and even a few massive
stars in some low mass dwarf irregular galaxies (dIrr);
see the review by Tolstoy \etal (2009).
These chemical studies have shown that 
low mass dwarf galaxies have had a slower chemical 
evolution than the stellar populations in the Milky Way.
For example, the majority of stars in dSphs have mean 
metallicities and [$\alpha$/Fe]\footnote{ 
[X/Fe] = log(X/Fe)$_*$ - log(X/Fe)$_\odot$.}
ratios that are lower than those of the Sun. 
Stars in dwarf galaxies also tend to have different 
heavy element ratios (e.g., higher [Ba/Y] or [La/Eu])
than similar metallicity stars in the Galaxy 
(e.g., Venn \etal 2004, \citealt{Aoki09}, Letarte \etal  2010).  
On the other hand, the most metal-poor stars in dwarf galaxies, 
with [Fe/H] $\le -2.5$, tend to have similar abundance ratios 
to Galactic halo stars, even for the $\alpha$ and heavy element 
ratios
(e.g., Sculptor, Fornax, and Sextans, Tafelmeyer \etal  2010).   
This is also true of the very metal-poor stars in the UFDs 
(e.g., Com Ber, Bo\"otes I, and Leo IV, Norris \etal 2008, 
Feltzing \etal 2009,  Frebel \etal 2010a, Simon \etal 2010). 
One exception are the $\alpha$ and heavy element abundances
for metal-poor stars in Sextans, which show both offsets and 
larger dispersions than the Galaxy and other dwarf galaxies 
\citep{Aoki09}. 


The Carina dwarf galaxy provides a new opportunity for chemical 
evolution and nucleosynthetic studies.  It has a low mass that is 
similar to that of Sextans (Walker \etal 2009), and it has had 
an unusual, episodic star formation history.  Its colour-magnitude 
diagram (CMD) shows at least three main sequence turn-offs 
(Monelli \etal  2003, Bono \etal  2010) and is best described 
by a star formation history with a well-defined old population 
($\sim$10-12 Gyr), a dominant intermediate-aged population 
($\sim$5-7 Gyr), and a trace young population ($\sim$2 Gyr;
although the specific timescales are uncertain).   
These star forming episodes are separated by 
long quiescent periods, seen as gaps between the main sequence turn-off 
points.  It is estimated that 70 to 80\% of the stars in Carina have 
intermediate-ages, while most of the remaining stars are old and 
associated with the first star forming episode (also see Dolphin 2002, 
Hernandez \etal  2000, Hurley-Keller \etal  1998, and Mighell 1997).  
One wonders how this unique star formation history may have 
affected the chemical evolution of Carina. 
 
In spite of its punctuated star formation history, Carina has 
an extremely narrow red giant branch.   One possibly is that the
RGB stars in Carina have a fortuitous alignment in the age-metallicity 
degeneracy, such that the older metal-poor stars overlay the 
metal enhanced intermediate-aged stars. 
Alternatively, the narrow red giant branch could be dominated by
intermediate-aged stars only, if that population formed on a 
relatively short timescale and with only a modest metallicity spread
(Rizzi \etal 2003, Bono \etal 2010).
Low resolution spectra of 437 red giant branch (RGB) stars in Carina 
analysed by Koch \etal (2006) showed the mean metallicity in Carina is
[Fe/H] $\sim -1.7 \pm0.9$, in agreement with similar analyses
by Helmi \etal (2006) and Starkenburg \etal (2010).  This metallicity
spread is larger than predicted by \citet{Bon10} from their CMD
analysis ($\Delta$[Fe/H] $\le \pm0.5$ dex). 
Koch \etal (2006) also found no significant gradients in metallicity or
stellar population with position in Carina, unlike the results for
Sculptor (Tolstoy \etal 2004) and Fornax (Battaglia \etal 2006).

High resolution analyses of five RGB stars (Shetrone \etal  2003) 
and ten RGB stars (Koch \etal  2008) supported the larger range in
metallicity found by the low resolution spectral analyses.   These
analyses also showed that Carina is like the other dSph galaxies in 
that the [$\alpha$/Fe] ratios are lower than in Galactic halo stars
with similar metallicity.   Lanfranchi \etal (2003, 2004, 2006) used 
these datasets to develop chemical evolution models for Carina, tuning 
the high wind efficiency and low star formation efficiency to reproduce 
Carina's metallicity distribution function, [$\alpha$/Fe] ratios, 
and low gas content.   

In this paper, we present a new detailed abundance analysis of up to
23 elements in nine stars in the Carina dwarf galaxy.    This work
increases the number of elements, the number of stars, and the 
metallicity range previously explored.   We also report the abundances
for two newly discovered very metal-poor stars (with [Fe/H]$\sim -2.85$),
which allow us to examine the earliest epoch of star formation.
These results, with previously published abundances, are used to
explore the unique star formation history and chemical evolution in Carina.

\begin{figure}[t]
\includegraphics[width=90mm, height=90mm]{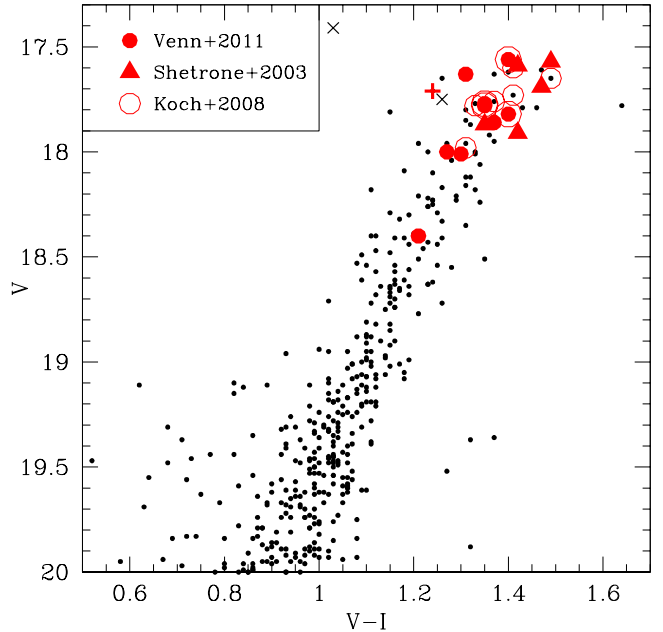}
\caption{Colour-magnitude diagram, V vs (V-I), for Carina
from our ESO WFI data.   Only stars with radial velocities 
$>$ 200 \kms\ (from Koch \etal 2006) are shown.
Stars with high resolution spectral analyses are identified
by red symbols, including nine stars from this paper, five
from Shetrone \etal (2003), and ten from Koch \etal (2008; 
four are in common with this analysis). 
One carbon-star is noted by the red cross,
and two foreground objects are shown as black crosses.
The faintest star in our sample is Car-7002. \\
}
\label{cmd}
\end{figure}


\section{Observations and Data Reductions} 

\begin{figure}[t]
\begin{center}
\includegraphics[width=80mm, height=40mm]{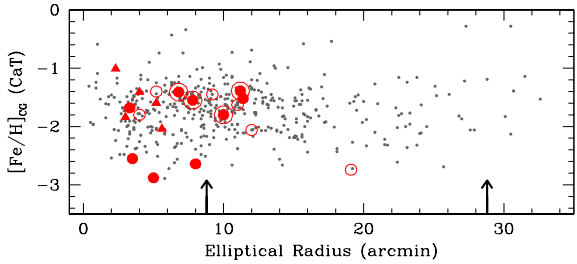}
\caption{Spatial location and metallicities of the nine Carina stars 
in this paper, as well as five from Shetrone \etal (2003), and ten 
stars from Koch \etal (2008; four are in common with this analysis). 
Symbols are the same as in Figure~\ref{cmd}.  
These are compared to the distribution of 437 RGB 
stars from Koch \etal (2006) which have a high (3~$\sigma$) 
probability of Carina membership.   
The metallicities
on this plot are those determined by Koch \etal (2006).
Elliptical radii are calculated using the structural 
parameters for Carina from Irwin \& Hatzidimitrou (1995; 
central coordinates 6$^h 40.6'$ and $-50^o 56'$ for epoch 
B1950, position angle 65$^o$, ellipticity $\epsilon$=0.33),
and listed in Table~\ref{obs}, 
and core and tidal radii are indicated.
}
\label{fegrad}
\end{center}
\end{figure}

The data presented in this paper were acquired at two observatories 
during separate time allocations in January 2005. 
The FLAMES (Fibre Large Array Multi Element Spectrograph)  
multiobject spectrograph at the 8.2 meter UT2 
(Kueyen) at the Very Large Telescope (VLT) 
of the European Southern Observatory (ESO) was used
to collect high resolution spectra using both the 
UVES and
GIRAFFE fiber modes (Pasquini \etal  2002).
The analysis of the FLAMES/GIRAFFE spectra is presented 
by Lemasle \etal (2012). 
Also, the the MIKE (Magellan Inamori Kyocera Echelle) spectrograph 
at the Magellan Landon 6.5m Clay Telescope at the Las Campanas Observatory
was used to collect high resolution spectra of individual stars outside
of the central field of Carina.  
Our targets and those previously analysed with high resolution spectral
analyses by Koch \etal (2008) and Shetrone \etal (2003) are shown in 
Fig.~\ref{cmd} on the $V$ vs $(V-I)$ CMD.
These stars are also shown in Fig.~\ref{fegrad} as metallicity 
(from Koch \etal 2006) versus their location within Carina.

\subsection{Magellan MIKE Spectra }

Target selection for the Magellan run was from low resolution 
spectra of the \ion{Ca}{2} 8600 \AA\ feature (=CaT) for hundreds of stars 
in the Carina dSph in the ESO archive (Koch \etal 2006, Helmi \etal 2006).    
The feature was used to ascertain membership from radial velocities, 
as well as estimate an initial metallicity for each target based 
on the CaT-metallicity calibration available at that time 
\citep{Bat08}.
Exposure times, 
radial velocities (\kms), 
the signal-to-noise ratio (SNR) at three wavelengths, 
the elliptical radius from the center of Carina (in arcminutes),
and alternative names for the targets are summarized in Table~\ref{obs}. 
Spectra for
three objects in the metal-poor globular cluster M68 were also taken
as standard stars.   M68 was chosen as a standard since it has
low metallicity (Harris 1996) and low reddening.   BVI magnitudes
are from Stetson (2000, with corrections onto the Johnson Kron-Cousins
scale of the Landolt standard stars as in Stetson 2005), which places
these stars on the upper RGB of M68 and suggests they have 
intrinsic luminosities similar to our program stars.

On the red side, the double echelle design covers 4850 - 9400 \AA, 
and on the blue side, 3800 -5050 \AA. The quality of the blue spectra
is significantly higher in the 
overlapping wavelength region ($\sim$ 4850 to 5050 \AA).
%
Using a 1.0'' slit, a spectral resolution R = 28,000
(blue) to R = 22,000 (red) was obtained.
The red chip had a gain of 1.0 electrons ADU$^{-1}$ and 
read noise of 4 electrons; the blue chip had a gain of 
0.47 electrons ADU$^{-1}$ and a read noise of 2 electrons.  
We binned on-chip with 2x2 pixels.   Table~\ref{obs} lists
the SNR per pixel achieved in the final combined spectra near 
4200, 5200, and 6200~\AA.

The data were reduced using standard IRAF\footnote{IRAF is 
distributed by the National Optical Astronomy Observatory, 
which is operated by the Association of Universities for Research in 
Astronomy, Inc., under cooperative agreement with the National Science 
Foundation.} routines.   
Sky subtraction was done with a smooth fit perpendicular to the dispersion 
axis.    Heliocentric corrections were applied to each spectrum before
determining their radial velocities.
Multiple spectra were median combined, which helped to remove
cosmic ray strikes.
Spectra taken of the Th-Ar lamp provided the wavelength
calibration.   Both quartz flats and screen flats were taken at varying
exposure levels; no significant offsets were found between the 
well-illuminated quartz flats and the science exposures, thus 
the quartz flats were adopted for the data reductions.   
A small dark current was noticed
on one corner of the blue chip in a region that did not receive 
starlight and therefore did not affect this analysis.
The final spectra were normalized using k-sigma clipping with a non-linear
filter (a combination of a median and a boxcar).   The effective scale
length of the filter was set from 8 to 15 \AA, dependent on the crowding 
of the spectral lines, and we found that this was sufficient to follow the 
continuum without affecting the presence of the lines when used in 
conjunction with iterative clipping.
This method was also used by \citet{Bat08} to normalize CaT
spectra.   

\begin{center}
\begin{deluxetable*}{llrrcrrrccc}
\footnotesize
\tablecaption{Observing Information \label{obs}}  
\tablewidth{0pt}
\tablehead{
\colhead{Star} & \colhead{Other} & \colhead{RA} & \colhead{DEC} & \colhead{R$_{\rm ell}$} & \colhead{Exp} & \colhead{RV} & \colhead{RV} & \colhead{SNR} & \colhead{SNR} & \colhead{SNR} \\
\colhead{} & \colhead{Name} & \colhead{(2000)} & \colhead{(2000)} & \colhead{(arcmin)} & \colhead{Time} & \colhead{CaT} & \colhead{HR} & \colhead{4200} & \colhead{5200} & \colhead{6200} \\
\colhead{} & \colhead {} & \colhead{} & \colhead{} & \colhead{} & \colhead{(s)} & \colhead {(\kms)} & \colhead{(\kms)} & \colhead{} & \colhead{} & \colhead{}
} 
\startdata
M68-6022 & S195  &  12 39 17.0 & -26 45 39 & \nodata & 2x 600 & \nodata & -95.1 $\pm$ 1.5 & 60 & 73  & 75 \\
M68-6023 & S239  &  12 39 37.6 & -26 45 15 & \nodata & 1x 600 & \nodata & -94.8 $\pm$ 1.3 & 55 & 66  & 65 \\
M68-6024 & S225  &  12 39 30.1 & -26 42 47 & \nodata & 1x 600 & \nodata & -96.2 $\pm$ 1.6 & 30 & 44  & 46 \\
\\
Car-1087    & S12924     &  6 41 15.4 & -51 01 16 & 5.0 & 4x 3600 & 229.3 & 220.1 $\pm$ 4.1 & 20 & 30 & 32 \\
            & LGO4c-006621 & \\ 
Car-5070    & S24846     &  6 41 53.8 & -50 58 11 & 3.5 & 3x 3600 & 213.5 & 211.5 $\pm$ 3.5 & 20 & 30 & 32 \\
            & car1-t213  & \\
Car-7002    & S06496     &  6 40 49.1 & -51 00 33 & 8.0 & 3x 3600 & 226.6 & 224.2 $\pm$ 3.7 & 18 & 30 & 27 \\
            & LGO4c-000826 \\
            & car0619-a2 \\
Car-484     & car1-t057 (K)  &  6 41 39.6 & -50 49 59 & 11.2 & 11x 3600 & 232.6 & 229.4 $\pm$ 0.7 & \nodata &  20 & 35 \\
            & LGO4a-002181 & \\
Car-524     & car1-t083  &  6 41 14.6 & -50 51 10 & 11.4 & 11x 3600 & 219.4 & 218.6 $\pm$ 0.7 & \nodata &  20 & 40 \\
            & LGO4a-002065 & \\
Car-612     & car1-t076 (K)  &  6 40 58.6 & -50 53 35 & 10.0 & 11x 3600 & 223.1 & 222.9 $\pm$ 0.7 & \nodata &  19 & 29 \\
            & LGO4a-001826 & \\
Car-705     & car1-t048 (K)  &  6 42 17.3 & -50 55 55 & 6.8 & 11x 3600 & 221.3 & 220.9 $\pm$ 0.9 & \nodata &  18 & 28 \\
            & LGO4a-001556 &  \\
Car-769     & car1-t069  &  6 41 19.7 & -50 57 26 & 3.3 & 11x 3600 & 219.4 & 218.9 $\pm$ 0.6 & \nodata &  17 & 27 \\
            & LGO4a-001364 & \\
Car-1013     & car1-t152 (K)  &  6 41 22.0 & -51 03 43 & 7.8 & 11x 3600 & 218.5 & 218.4 $\pm$ 1.5 & \nodata &  11 & 15 \\
            & LGO4c-006477 & \\
Car-837     & car1-t191 &  6 41 46.3 & -50 58 56 & \nodata & 11x 3600 & 232.6 & 222.4 $\pm$ 0.7 & \nodata & 20  & 30 \\
            & (C-star) \\
Car-489     & (non-member) &  6 41 37.0 & -50 50 07 & \nodata & 11x 3600 & 236.5 & \nodata   &  \nodata &  16 & 25 \\ 
Car-X       & (non-member) &  6 41 26.9 & -51 00 34 & \nodata & 11x 3600 & \nodata & 4.9 $\pm$ 0.8 & \nodata & 15  & 25 
\enddata
\tablecomments{
Data for M68 and the first three Carina stars are from
the Magellan MIKE spectrograph,
with the remaining nine Carina stars observed 
with the FLAMES spectrograph at the VLT.   Stars partially
analysed by Koch \etal (2008) are noted with ``(K)'' beside
their alternative names.
%
%
The S\# target names are from P.B. Stetson's online database
of homogeneous photometry (at http://cadcwww.hia.nrc.ca/stetson).   
Car-X is a bright target right next to Car-909, which was the actual target.
SNR are the signal-to-noise ratios {\it per pixel}. 
}
\end{deluxetable*}
\end{center}

\subsection{VLT FLAMES/UVES Data}

Observations with the multi-object FLAMES spectrograph at the
UT2 Kueyen ESO-VLT (Pasquini \etal  2002) were carried out for nine
targets in the central region of Carina.  In two separate configurations,
the FLAMES/UVES fibers were placed on bright RGB stars resulting in 
nine stellar spectra and two sky observations. 
Two of the nine targets proved to be foreground RGB stars, 
while one is a carbon star that is unsuitable for our analysis.    
Target coordinates, exposure times, 
radial velocities (\kms), 
elliptical radius from the center of Carina (in arcmins), and the SNR
at two wavelengths for the final combined spectra 
are listed in Table~\ref{obs}.     
Other names for each target are also listed.
FLAMES/GIRAFFE spectra were simultaneously obtained at  
high resolution (R=20,000) for 36 more stars in Carina, 
but over much shorter wavelength regions (3 wavelength 
settings that yielded $\sim$250 \AA\ each) - 
these are presented in a separate paper by Lemasle et al. (2012).
Due to variable weather conditions, only eleven of the one-hour 
exposures had sufficient signal to be used further in this analysis.   

On the red side, the double echelle design covers 5840$-$6815 \AA, 
and on the blue, 4800$-$5760\AA.  A 1.0'' slit yields 
a spectral resolution R = 47,000.   Table~\ref{obs} lists
the exposure times and SNR per pixel in the final 
combined spectra near 5200 and 6200~\AA.   Unfortunately, 
six of the one-hour exposures did not have enough signal in
the FLAMES/UVES fibers to be useful in this analysis.
The data were reduced using the ESO FLAMES pipeline.\footnote{The
GIRAFFE Base-Line Data Reduction, girBLDRS, was written at the
Observatory of Geneva by A. Blecha and G. Simond, and is available 
through SourceForge at http://girbldrs.sourceforge.net/.}
The sky spectrum was fit with a smoothly varying function, and
this was subtracted from the stellar exposures per wavelength 
set up (to reduce adding more noise in already low SNR spectra).   
Each spectrum was then heliocentric velocity corrected, radial
velocities were determined, and the spectra were coadded 
(weighted by the SNR).   The final spectra were normalized using 
an iterative asymmetric k-sigma clipping routine with a
non-linear filter (like for the Magellan/MIKE spectra).

\section{Stellar Analyses}

\subsection{Line List}

A range of elements are detectable in our spectra, which enables 
a comprehensive abundance analysis.
Atomic lines for this analysis were selected from the literature,
including the line lists from Shetrone \etal (2003), Cayrel et al. (2004),
\citet{Aoki07}, Cohen \etal (2008), Letarte \etal (2009), 
Tafelmeyer \etal (2010), and Frebel \etal (2010a); see Table~\ref{ukvlines}.
Atomic data for the \ion{Fe}{1} lines were updated from O'Brian \etal (1991)
when available, or the atomic data was updated to the latest values 
in the National Institute of Standards and Technology 
(NIST) database.\footnote{http://physics.nist.gov/PhysRefData/ASD/index.html}

\begin{center}
\begin{deluxetable*}{lrrrrrrrr}
\footnotesize
\tablecolumns{10}
\tablewidth{0pt}
\tablecaption{Line List \label{ukvlines}}
\tablehead{
\colhead{Wave} & \colhead{$\chi$} & \colhead{log~gf} & \colhead{CAR} & \colhead{CAR} & \colhead{CAR} & \colhead{CAR} & \colhead{CAR} & \colhead{CAR} \\
\colhead{(\AA)} & \colhead{(eV)} & \colhead{} & \colhead{484} & \colhead{524} & \colhead{612} & \colhead{705} & \colhead{769} & \colhead{1013} 
}
\startdata
Fe I \\
 5001.864 & 3.88 & 0.010 & 124.9 & 116.7 & 124.0 & 134.0 & 104.3 & 165.1 \\
 5006.120 & 2.83 & -0.615 & 194.7 & 166.1 & 185.2 & 193.8 & 158.2 & \nodata \\
 5012.070 & 0.86 & -2.642 & \nodata & 195.0 & \nodata & \nodata & 179.3 & \nodata \\
 5014.940 & 3.94 & -0.300 & 113.4 & 103.4 & 117.0 & 114.3 & 65.4 & 97.2 
\enddata
\tablecomments{Equivalent widths are in m\AA, and upper limits are noted.   
When spectrum syntheses have been used in the abundance analysis, 
then an ``S'' is noted.  A second table includes the stars observed with the
MIKE spectrograph at Magellan. Both tables are 
published in their entirety in the electronic edition.   
A portion is shown here for guidance regarding its form and content.}
\end{deluxetable*}
\end{center}

\subsection{Equivalent Width Measurements \label{eqws}}

Most of the elemental abundances in this analysis are determined
from equivalent width (EW) measurements.   Spectrum syntheses were 
used only for lines affected by hyperfine splitting or for elements
with line measurements from low SNR spectra.  

\begin{figure}[hb]
\includegraphics[width=90mm, height=100mm]{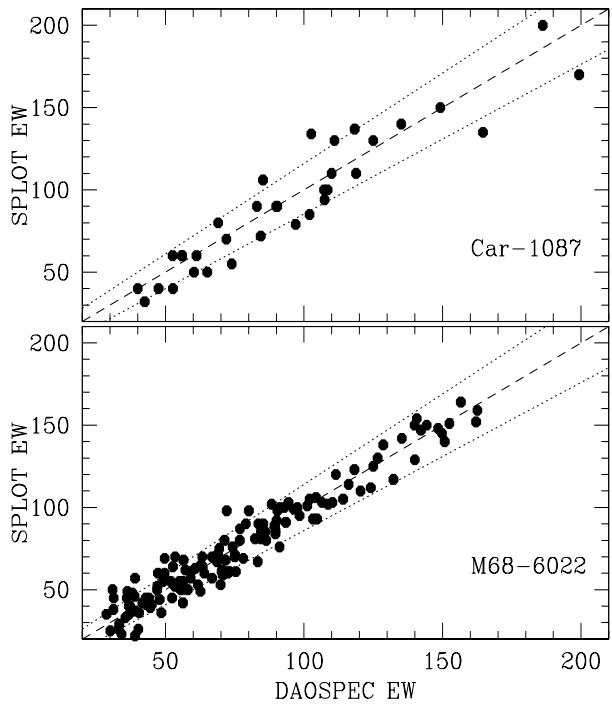}
\caption{Equivalent width measurement comparisons 
for M68-6022 and Car-1087, both observed with the 
MIKE spectrograph.   The three lines represent equal 
EW values and EW $\pm$ $\Delta$EW, where
$\Delta$EW = EW$_{\rm rms}$ + 10\% EW, and 
EW$_{\rm rms}$ = 3 m\AA\ and 6 m\AA, respectively.
}
\label{ew}
\end{figure}

Equivalent widths were measured using the Gaussian fitting routine
DAOSPEC (Stetson \& Pancino 2008).    
The placement of the continuum is a critical step, as 
it influences the measurements of the equivalent widths; thus 
some care was taken in assigning and testing the DAOSPEC fitting 
parameters. For example, a low order polynomial (order 13) was adopted 
to allow DAOSPEC to measure the effective continuum, and tests of 
other low order values (order =5 and =1) resulted in nearly 
identical equivalent widths, with  $\Delta$EW $<$ 1 \%.
As a final exercise, the continuum was set to one so that DAOSPEC 
would not relocate the global continuum for the measurement 
(order = -1).  Results from this test showed a slight offset 
$\le +12\%$ in the equivalent width measurements, as expected.
We note this offset is similar to or less than our adopted equivalent 
width measurement errors from the Cayrel formula (described below).    
For testing purposes, equivalent widths were also measured with 
{\it splot} in IRAF to determine the area under the continuum.
A comparison of these measurements with those from DAOSPEC for two 
of the stars observed with the MIKE spectrograph are shown in 
Fig.~\ref{ew}.

The expected {\it minimum} measurement uncertainties on the equivalent 
widths (EW$_{\rm rms}$) have been estimated using a revised Cayrel formula,
i.e., a new derivation of the measurement errors on the equivalent 
widths was presented by \citet{Bat08}, where it was 
found that the factor of 1.5 in the Cayrel (1988) formula is actually 
within the square root; thus,

\vspace{0.1in}
EW$_{\rm rms}$ = SNR$^{-1}$ x \ $\sqrt[]{1.5 \times FWHM \times \delta{\rm x}}$
\vspace{0.1in}

\noindent
where SNR is the signal-to-noise ratio per pixel, FWHM is the line full
width at half maximum, and $\delta{\rm x}$ is the pixel size.
An extra 10\% of the EW was added to this for a more conservative 
measurement error, such that 

\vspace{0.1in}
$\Delta$EW = EW$_{\rm rms}$ + 0.10 $\times$ EW.
\vspace{0.1in}

\noindent

For the M68 stellar spectra taken at Magellan the revised 
Cayrel formula gives EW$_{\rm rms}$ = 3 m\AA, but EW$_{\rm rms}$ = 6 m\AA\
for the lower SNR spectra of the Carina stars observed with MIKE. 
For the spectra taken at the VLT with FLAMES/UVES, 
the higher resolution yields a minimum error of
EW$_{\rm rms}$ = 4 m\AA, with an exception for Car-1013 where
EW$_{\rm rms}$ = 7 m\AA\ due to its lower SNR spectrum.

The reported errors from the DAOSPEC program were lower than 
those from the revised Cayrel formula, by 1/3 to 1/2.  This
is due to correlations in the noise estimates in DAOSPEC 
when the pixel data are rebinned/interpolated during the
spectral extractions and wavelength calibrations
(P. Stetson, priv. communications). 
When individual line measurements were significantly different 
between the DAOSPEC and {\it splot} measurements, those lines 
were checked by eye.   These lines tended to be unresolved
blends and/or suffered from difficult continuum placement.
Since DAOSPEC uses a fixed FWHM and a consistent prescription 
for continuum placement, those EW measurements were adopted for 
this analysis, although the $\Delta$EW errors are from the revised Cayrel formula.

\begin{center}
\begin{deluxetable*}{lrrr|rr|rr|rrrrr}
\footnotesize
\tablecaption{Target Information \label{photom}}  
\tablewidth{0pt}
\tablehead{
\colhead{Star} & 
\colhead{B} & \colhead{V} & \colhead{I} & \colhead{V} & \colhead{I} & 
\colhead{J} & \colhead{K$_{s}$} & 
\colhead{B} & \colhead{V} & \colhead {I} & \colhead{J} &  \colhead{K}  \\
\colhead{} & 
\colhead{PBS} & \colhead{PBS} & \colhead{PBS} & \colhead{WFI} & \colhead{WFI} & 
\colhead{2MASS} & \colhead{2MASS} & 
\colhead{Gul} & \colhead{Gul} & \colhead{Gul} & \colhead{Gul} & \colhead{Gul}  \\
} 
\startdata
M68-6022 &   15.127 & 14.205 & 13.106 & $-$ & $-$ & 
                 12.324 & 11.659 & $-$ & $-$ & $-$ & $-$ & $-$ \\
M68-6023 &   15.192 & 14.284 & 13.191 & $-$ & $-$ &
                 12.363 & 11.712 & $-$ & $-$ & $-$ & $-$ & $-$ \\
M68-6024 &   15.230 & 14.354 & 13.289 & $-$ & $-$ &
                 12.479 & 11.858 & $-$ & $-$ & $-$ & $-$ & $-$ \\
Car-1087 &       19.146 & 18.031 & 16.735 & 18.00 & 16.73 & 
                 15.802 & 15.301 & $-$ & $-$ & $-$ & $-$ & $-$  \\
Car-5070 &       19.125 & 17.904 & 16.520 & 17.86 & 16.49 & 
                 15.641 & 14.758 & $-$ & $-$ & $-$ & $-$ & $-$  \\
Car-7002 &       19.284 & 18.344 & 17.146 & 18.40 & 17.19 & 
                 16.244 & 15.562 & $-$ & $-$ & $-$ & $-$ & $-$  \\
Car-484  &       19.009 & 17.603 & 16.179   & 17.56 & 16.16 & 
                 15.322 & 14.627 &
                 19.055 & 17.607 & 16.178 & 15.290 &  14.406 \\
Car-524  &       18.934 & 17.645 & 16.308   & 17.63 & 16.32 & 
                 15.388 & 14.473 &
                 18.921 & 17.598 & 16.278 & 15.398 & 14.574 \\
Car-612  &       19.077 & 17.811 & 16.416   & 17.82 & 16.42 & 
                 15.456 & 14.763 &
                 19.103 & 17.779 & 16.403 & 15.609 &  14.828 \\
Car-705  &       19.101 & 17.828 & 16.430   & 17.78 & 16.43 & 
                 15.687 & 14.597 &
                 19.104 & 17.788 & 16.424 & 15.509 & 14.663 \\
Car-769  &       19.202 & 18.002 & 16.693   & 18.01 & 16.71 & 
                 15.698 & 14.921 &
                 19.224 & 18.025 & 16.725 & 15.758 &  15.040 \\
Car-1013  &      19.048 & 17.783 & 16.424   & 17.77 & 16.42 & 
                 15.515 & 14.888 &
                 19.056 & 17.767 & 16.421 & 15.528 &  14.726 \\
Car-489  &       $-$ & $-$ & $-$   & 17.85 & 16.53 & 
                 15.544 & 14.933 &  $-$ & $-$ & $-$  & $-$ & $-$ \\
Car-837  &       $-$ & $-$ & $-$   & 17.71 & 16.47 & 
                 15.456 & 14.509 &
                 19.111 & 17.705 & 16.509 & 15.427 &  14.621 \\
Car-X    &       $-$ & $-$ & $-$   & 17.51 & 16.42 & 
                 15.748 & 15.223 & $-$ & $-$ & $-$ & $-$ & $-$ 
\enddata
\tablecomments{
BVI from P.B. Stetson's website of homogeneous photometry,  
VI from our WFI data, JK$_s$ from the 2MASS database, and
BVIJK from M. Gulleuszik (private communications).
}
\end{deluxetable*}
\end{center}

A maximum EW of 200 m\AA\ was adopted for all stars since deviations 
from the Gaussian profiles used in DAOSPEC become more significant 
for stronger lines.   In fact, for lines $\ge$200 m\AA, we found 
that DAOSPEC would occassionally divide these lines into 2-3 lines, 
each with lower EWs.   The final line lists and measurements were
carefully reviewed during this analysis.

\subsection{Radial Velocities}

Radial velocities were initially measured from 5-7 strong 
lines and refined using DAOSPEC (which cross correlates 
the measured positions of the detected spectral lines
with reference wavelengths from an input linelist).
Heliocentric corrections were then applied.   The radial
velocities for the Carina stars are listed in Table~\ref{obs};
the results for the Magellan/MIKE blue and red spectra were 
averaged together since each arm was reduced independently.
The three M68 stars have a mean radial velocity of 
$-95.4 \pm 1.0$ \kms, with an average measurement error
of 1.5 \kms.     This is in agreement with results in
the literature, e.g., Lane \etal (2009) report a radial velocity
of  $-94.93 \pm 0.26$ \kms\ from 123 RGB stars in M68 taken 
at the 3.9m Anglo Australian Telescope with the AAOmega
spectrograph.
The mean radial velocity of our nine Carina stars is 
$220.5 \pm 4.9$ \kms; the average measurement error of
the FLAMES/UVES data are 0.9 \kms, whereas it is 
3.8 \kms\ for the metal-poor stars with lower 
quality Magellan/MIKE spectra. 
These results are in agreement with the mean radial
velocity of 223.9 \kms\ found by Koch \etal (2006) for 
437 RGB members of Carina, with a dispersion of 7.5 \kms.
The radial velocities from both the high resolution
(this paper) and lower resolution spectra (from 
Koch \etal) are shown in Table~\ref{obs}; again, these are 
in good agreement, with a mean difference in 
$\langle$ RV$_{\rm CaT}$ $-$ RV$_{\rm HR}$ $\rangle$ = 2.1 $\pm$ 3.4 \kms.

\subsection {Photometric Parameters (physical gravity) \label{photometry}}

BVIJK$_{s}$ photometric values for the Carina targets and 
M68 standard stars are listed in Table~\ref{photom}.  
%
BVI magnitudes for the M68 stars are from Stetson 
(2000, on the Johnson Kron-Cousins scale,
Stetson 2005).  J and K$_{s}$ magnitudes are from 
the 2MASS database\footnote{This publication makes use
of data products from the Two Micron All Sky Survey which is a
joint project of the University of Massachusetts and the Infrared
Processing and Analysis Center/California Institute of Technology,
funded by the National Aeronautics and Space Administration and
the National Science Foundation.   The database can be found at
http://www.ipac.caltech.edu/2mass/releases/allsky.}.
%
For the Carina members, VI magnitudes are from the ESO 2.2m WFI
photometry; initial calibrations were on the default ESO zero-point
values, but have been updated to the Johnson Kron-Cousins scale with
Stetson's database and Gullieuszik's photometry (private communications).   

Initial metallicities for the Carina stars are from the CaT 
measurements (Koch \etal 2006). For the three M68 stars, the
average metallicity from Lee \etal (2005) is adopted,
[Fe/H]=$-2.16$. 
\footnote{
Lee \etal (2005) found a mean [\ion{Fe}{2}/H]=$-2.16$ from a high 
resolution spectral analysis of eight stars in M68, using photometric 
gravities.   They also find a mean [\ion{Fe}{1}/H]=$-2.56$ using 
spectroscopic gravities.}.
 
Color temperatures were found using the Ramirez \& Melendez (2005)
calibration, and adopted as the initial effective temperatures;
see Table~\ref{photomparams}. 
%
%
Deviations between the different color temperatures are 
quite small for the three M68 globular cluster stars, 
i.e.,  $\sigma_T \le$ 20~K from T$(B-V)$, T$(V-I)$, 
T$(V-J)$, and T$(V-K)$.
Deviations are larger for the Carina stars, 
with $\sigma_T \le$ 134~K, and $\langle \sigma_T \rangle$ = 74~K.
Physical (photometric) gravities are determined using the 
standard relation, 

\vspace{0.1in}
\logg \, = \logg$_\sun$  + log $\left({\rm M}_*\over{{\rm M}_\sun}\right)$ + 4$\times$log $\left({T_{\rm eff*}}\over{T_{\rm eff}\sun}\right)$

\vspace{0.05in}
\hspace{1.0in} + 0.4$\times$(M$_{\rm Bol*}$ - M$_{\rm Bol\sun}$)

\vspace{0.1in}
\noindent
with \logg$_\sun$ = 4.44, T$_{\rm eff\sun}$ = 5777 K, M$_{\rm Bol_\sun}$ = 4.75,
and adopting a reddening law of A(V)/E(B-V) = 3.24.    The values of
M$_{\rm Bol*}$ are based on the bolometric correction to the V magnitude 
from \citet{Alo99}, with an assumed distance and stellar mass.

\begin{center}
\begin{deluxetable*}{lrrrrrrrrr} 
\footnotesize
\tablecaption{Photometric Parameters \label{photomparams}}
\tablewidth{0pt}
\tablehead{
\colhead{Target} & 
\colhead{Tbv} & \colhead{Tvi} & \colhead{Tvj} & \colhead{Tvk} &
\colhead{$\langle$T$\rangle$} & \colhead{$\sigma_T$} &
\colhead{\logg} & \colhead{Mv} & \colhead{Mbol}   \\
\colhead{} &
\colhead{(K)} & \colhead{(K)} & \colhead{(K)} & \colhead{(K)} &
\colhead{(K)} & \colhead{(K)} & \colhead{} &
\colhead{} & \colhead{} 
} 
\startdata
M68-6022   & 4663 & 4671 & 4677 & 4694 & 4676 &  13 & 1.53 & -0.98 & -1.40  \\ 
M68-6023   & 4686 & 4682 & 4654 & 4647 & 4667 &  20 & 1.55 & -0.91 & -1.33  \\
M68-6024   & 4741 & 4732 & 4722 & 4701 & 4724 &  17 & 1.61 & -0.84 & -1.23  \\
Car-1087   & 4391 & 4457 & 4572 & 4416 & 4459 &  80 & 0.95 & -2.05 & -2.59  \\ 
Car-5070   & 4303 & 4314 & 4367 & 4364 & 4337 &  33 & 0.83 & -2.19 & -2.79  \\
Car-7002   & 4677 & 4552 & 4492 & 4473 & 4548 &  92 & 1.16 & -1.65 & -2.15  \\
Car-484    & 4073 & 4245 & 4383 & 4320 & 4255 & 134 & 0.65 & -2.49 & -3.14  \\
Car-524    & 4228 & 4346 & 4235 & 4311 & 4280 &  58 & 0.70 & -2.42 & -3.05  \\
Car-612    & 4229 & 4253 & 4301 & 4224 & 4252 &  35 & 0.75 & -2.23 & -2.88  \\
Car-705    & 4194 & 4305 & 4221 & 4460 & 4295 & 120 & 0.77 & -2.27 & -2.89  \\
Car-769    & 4308 & 4361 & 4278 & 4254 & 4300 &  46 & 0.86 & -2.04 & -2.65  \\
Car-1013   & 4208 & 4280 & 4372 & 4250 & 4278 &  70 & 0.75 & -2.28 & -2.91  
\enddata
\tablecomments{
Photometric temperatures are from the Ramirez \& Melendez (2005) calibration.  
Each temperature is determined from a photometric colour, e.g., Tbv represents
the temperature determined from the (B$-$V) colour.   The average of the four
colour temperatures ($\langle$T$\rangle$) and standard deviation ($\sigma_T$) are listed
along with other parameters derived from the magnitudes (see Section~\ref{photometry}).  
}
\end{deluxetable*}
\end{center}

For M68, we adopt the distance and reddening in the Harris catalogue (1996), 
(m-M)v = 15.19 $\pm$0.10 and E(B-V)=0.05 
from isochrone fitting by McClure \etal (1987).  
This value is in good agreement with \citet{Broc97}, but the 
proper motion distance modulus, (m-M)v = 14.91 (Dinescu \etal 1999) 
is quite short if this reddening is adopted.
A turn off mass of 0.83 $\pm$0.03 M$_\odot$ is found for M68
using isochrones by Vandenberg \& Bell (1985). 

For Carina, Mighell (1997) determined the distance modulus and reddening
from WFPC2 V and I band imaging, but also compares his results in an
appendix to the many estimates available in the literature.   We adopt 
Mighell's determination of (m-M)v=20.05 $\pm$0.11 and E(B-V)=0.06 $\pm$0.02.   
This reddening value is the same as from the Schlegel \etal (1998) maps.
One of the highest values for the distance modulus was found by 
Monelli \etal (2003; (m-M)v = 20.24, with E(B-V)=0.03), but as discussed below
(also see Section 5); these differences have negligible effects 
on our spectral analysis. 
An estimate of the turn off mass is more difficult in Carina
due to its separate episodes of star formation, each of which has 
a different turn off mass; these estimates can have a larger effect 
on the physical gravity than the distance modulus.  
Since Carina is dominated by an intermediate aged population 
(5-7 Gyr old), we examined isochrones (Fagotto \etal 1994, 
with Y=0.23 and Z=0.0004) to find that the age range of 
5 to 12 Gyr corresponds to turn off masses of 1.0 to 0.8 M$_\odot$,
respectively.   
We adopted 0.8 M$_\odot$  in this analysis, but note that if 
1.0 M$_\odot$ were used, this would change the physical gravities 
by $+0.10$ dex with no effect on temperature; 
this change has only a small effect on the 
chemical abundances.


\subsection{Spectroscopic Parameters \label{analysis}}

The CaT metallicity and the photometric parameters for temperature 
and gravity were adopted as the initial stellar parameters for a model 
atmospheres analysis.   Spectroscopic indicators were used 
to refine the effective temperature and metallicity, as well as to 
determine the microturbulence values.   These values are listed in
Table~\ref{spect}.  Only the physical gravity is
unchanged from the photometric parameters analysis.

\subsubsection{Model Atmospheres}

The new MARCS spherical models have been adopted in this analysis
(Gustafsson \etal  2003, 2008, further expanded by B. Plez, private
communications).
The grid covers the range of parameters (temperature, luminosity,
microturbulence, and metallicity) of RGB stars.  
The models used in this analysis 
adopt the Galactic abundance pattern, 
i.e., \afe = +0.4 for [Fe/H] $\le -1.0$, 
which is the metallicity range of the stars in Carina.
We adopt the standard version of the 
MOOG\footnote{The 2009 version of MOOG is available from http://www.as.utexas.edu/~chris/moog.html.}
(Sneden 1973) spectral synthesis code to determine the individual line
abundances.   Corrections were applied when necessary for continuum 
scattering effects that can affect the results for lines below 5000 \AA,
as described in Section~\ref{scat}.
The photometric gravities were 
retained throughout the analysis (see Section~\ref{gravity}).

\begin{center}
\begin{deluxetable*}{lrccclc} 
\footnotesize
\tablecaption{Spectroscopic Parameters (photometric gravities) \label{spect}}
\tablewidth{0pt}
\tablehead{ 
\colhead{Target} & \colhead{\teff } & \colhead{\logg$_*$} & 
\colhead{$\xi$} & \colhead{12+log(\ion{Fe}{1})} & \colhead{[\ion{Fe}{1}/H] $\pm \sigma$ (\#)} & 
\colhead{CaT} \\ 
\colhead{ } & \colhead{ (K) } & \colhead{ (cgs) } & 
\colhead{ (km/s) } &  \colhead{$\pm\delta$(Fe)\tablenotemark{a} } & \colhead{} & \colhead{(S10)\tablenotemark{b}} 
} 
\startdata
M68-6022 & 4550 $\pm$44 & 1.53 $\pm$0.04 & 1.50 $\pm$0.12 & 4.99 $\pm$0.02 & $-$2.51 $\pm$0.23 (92)  & $-$2.16 \\
M68-6023 & 4667 $\pm$55 & 1.55 $\pm$0.04 & 1.60 $\pm$0.12 & 5.09 $\pm$0.03 & $-$2.41 $\pm$0.24 (92)  & $-$2.16   \\
M68-6024 & 4650 $\pm$56 & 1.61 $\pm$0.04 & 1.80 $\pm$0.12 & 4.86 $\pm$0.03 & $-$2.64 $\pm$0.24 (81)  & $-$2.16   \\
Car-1087 & 4600 $\pm$98 & 0.95 $\pm$0.10 & 2.45 $\pm$0.25 & 4.69 $\pm$0.06 & $-$2.81 $\pm$0.34 (38)  &  $-$3.10 \\
Car-5070 & 4550 $\pm$81 & 0.83 $\pm$0.10 & 2.30 $\pm$0.20 & 5.35 $\pm$0.05 & $-$2.15 $\pm$0.38 (55)  &  $-$2.57 \\
Car-7002 & 4548 $\pm$96 & 1.16 $\pm$0.10 & 2.00 $\pm$0.24 & 4.64 $\pm$0.05 & $-$2.86 $\pm$0.33 (38)  &  $-$3.29 \\
Car-484 &  4400 $\pm$66 & 0.65 $\pm$0.10 & 2.30 $\pm$0.07 & 5.97 $\pm$0.02 & $-$1.53 $\pm$0.21 (114) &  $-$1.62 \\
Car-524 &  4500 $\pm$64 & 0.70 $\pm$0.10 & 2.40 $\pm$0.07 & 5.75 $\pm$0.02 & $-$1.75 $\pm$0.21 (121) &  $-$1.61 \\
Car-612 &  4500 $\pm$61 & 0.75 $\pm$0.10 & 2.10 $\pm$0.08 & 6.20 $\pm$0.02 & $-$1.30 $\pm$0.23 (126) &  $-$1.84 \\
Car-705 &  4500 $\pm$61 & 0.77 $\pm$0.10 & 2.10 $\pm$0.07 & 6.15 $\pm$0.02 & $-$1.35 $\pm$0.25 (127) &  $-$1.63 \\
Car-769 &  4600 $\pm$76 & 0.86 $\pm$0.10 & 2.30 $\pm$0.08 & 5.82 $\pm$0.02 & $-$1.68 $\pm$0.25 (125) &  $-$1.63 \\
Car-1013 & 4600 $\pm$105 & 0.75 $\pm$0.10 & 2.20 $\pm$0.11 & 6.20 $\pm$0.03 & $-$1.30 $\pm$0.37 (125) &  $-$1.58 \\
\enddata
\tablenotetext{a}{$\delta$(Fe) is the error in the mean of the \ion{Fe}{1} lines (see Section~\ref{abuerr}).}
\tablenotetext{b}{CaT metallicities from the calibration
by Starkenburg \etal (2010).   For the three M68 stars,
CaT metallicities are from Lee \etal (2005).  }
\end{deluxetable*}
\end{center}

\subsubsection{Effective Temperatures, Microturbulence, \& Metallicity \label{temps}}

The effective temperatures were refined from the photometric values
through examination of the \ion{Fe}{1} lines.
Only lines redwards 5000 \AA\ were considered to 
avoid uncertainties in the standard MOOG treatment of 
scattering in the continuous opacity (see Section~\ref{scat}). 
Microturbulence values were found by minimizing the slope in
the \ion{Fe}{1} line abundances with equivalent width.
Similarly, the excitation temperatures were found by minimizing 
the slope in the same \ion{Fe}{1} line abundances with 
excitation potential ($\chi$, in eV).   
While the M68 stars had lower excitation than colour temperatures 
(by an average of 67~K), the Carina stars had higher excitation 
temperatures (by an average of 200~K).   The final
metallicity (for which we adopt the [Fe I/H] values)
is also found at the end of these minimization iterations.

\begin{figure*}[t]
\plottwo{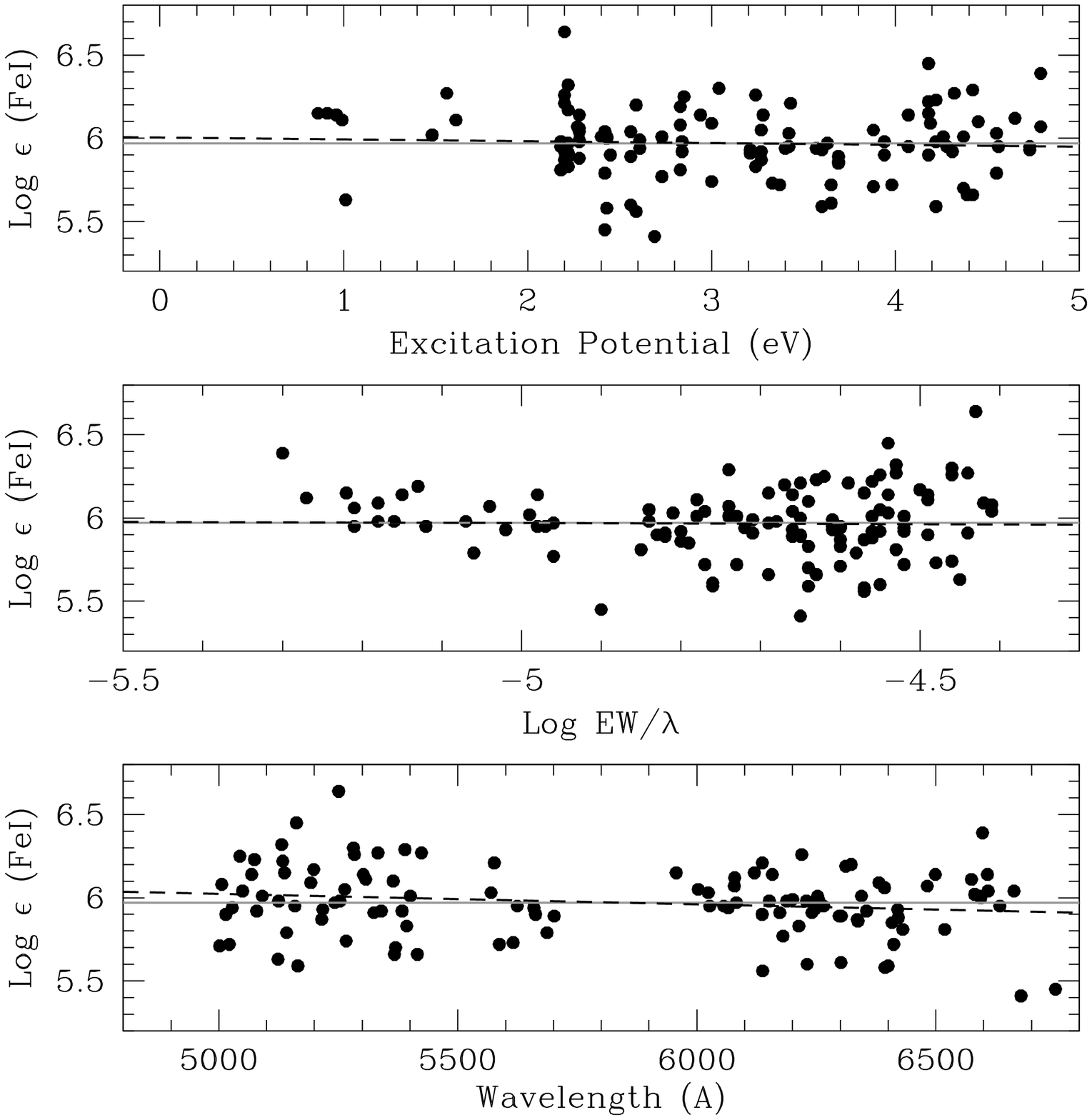}{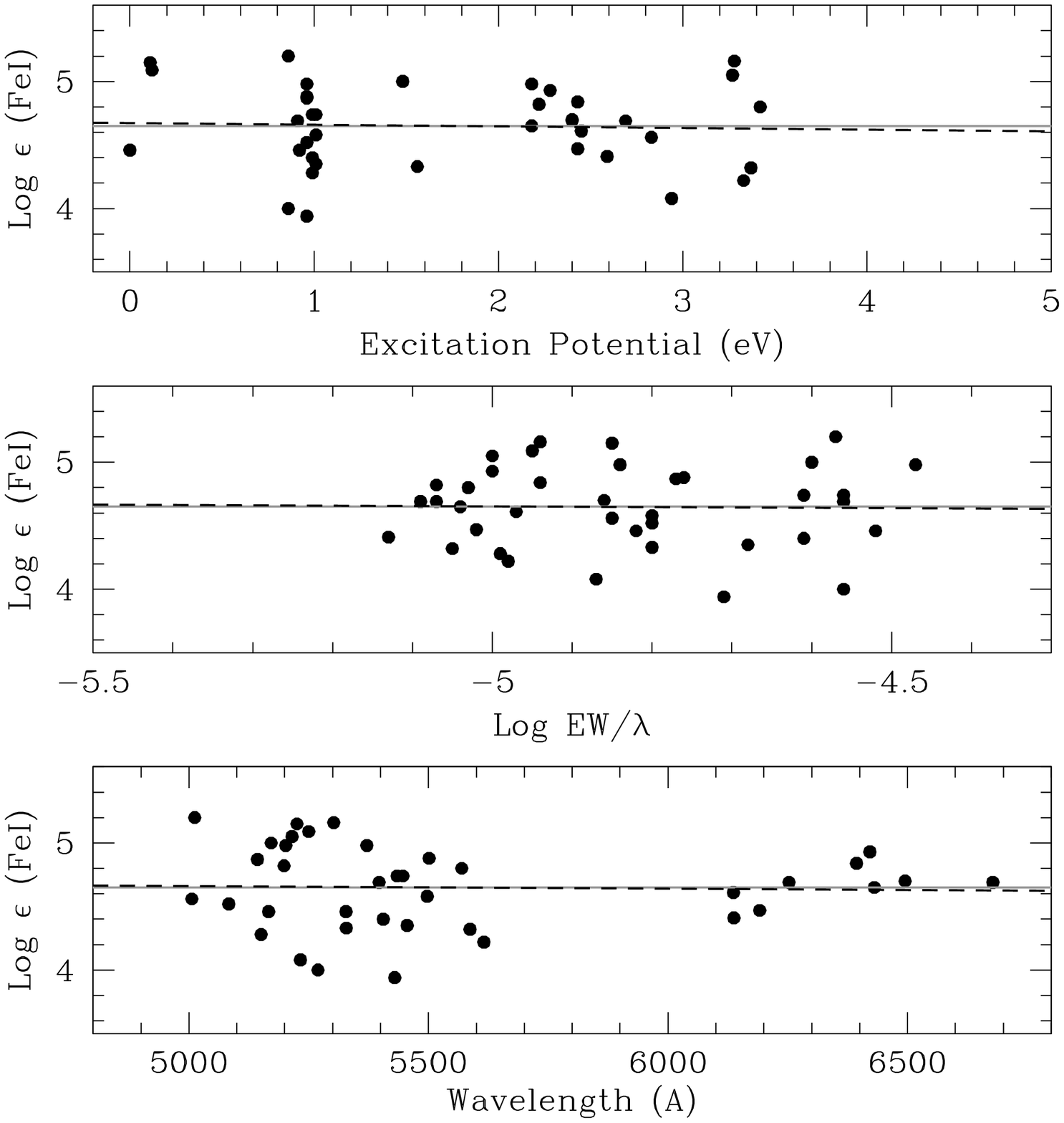}
\caption{MOOG results for Car-484 (left panel), a moderate SNR high 
resolution FLAMES/UVES spectrum, and Car-7002, a moderate SNR and 
moderate resolution Magellan/MIKE spectrum.   These plots show that 
the \ion{Fe}{1} line abundances are minimized 
with excitation potential (to determine surface temperature),
equivalent width (to define microturbulence), and wavelength 
(a check on the sky subtraction). \\
}
\label{v484}
\end{figure*}

No significant relationship was found in the \ion{Fe}{1} abundances
versus wavelength ($>$5000 \AA) for either the globular cluster or the
Carina targets;  this indicates that the sky subtraction and overall
data reduction methods were successful.
Eliminating lines with $\chi < 1.4$ eV, which can be particularly
sensitive to departures from local thermodynamic equilibrium (LTE) 
effects, also had no significant effect on the 
temperature or microturbulence values.
Samples of these relationships are shown in Fig.~\ref{v484} 
for two stars, one observed with the FLAMES/UVES spectrograph 
and the other observed with the MIKE spectrograph. 
Adjusting the metallicities from the initial CaT values 
had no significant effect on the excitation temperatures 
($\Delta$T $<$10 K). 
Adopting the spectroscopic temperatures had only a small effect on 
the photometric gravities 
($\le +0.08$ in the Carina stars, and $\le -0.02$ for the M68 stars).

Four stars in Carina have been observed and analysed independently
by Koch \etal (2008).   Their analysis with ATLAS9 models atmospheres
results in excitation temperatures that are in excellent agreement 
with ours, with differences ranging from $\Delta$~T = +116 to -70~K, 
and with an average offset of +32~K.   The differences in 
microturbulence are more significant, with Koch \etal's values 
being $\sim 0.50$ km\,s$^{-1}$ lower.    This is linked to differences 
in our gravity determinations discussed in Section~\ref{gravity}.

\subsection{Stellar Parameter Error Estimates}

Stellar parameters and their uncertainties are listed in 
Table~\ref{spect}.
The one sigma uncertainty in the effective temperature is
determined from the slope in the \ion{Fe}{1} line abundances 
vs excitation potential ($\chi$) allowing the change 
in the iron abundance to equal the standard deviation 
$\sigma$(\ion{Fe}{1}.
Similarly, the one sigma uncertainty in microturbulence
is determined by setting the slope 
of the \ion{Fe}{1} line abundances vs equivalent width 
equal to $\sigma$(\ion{Fe}{1}).

For the physical gravities, the uncertainties are determined 
from errors in the distance moduli, reddening, and stellar mass
(see Section~\ref{photometry}). 
For M68, the uncertainty in the turn off 
mass is tiny ($\pm$0.03 M$_\odot$), which has a negligible effect 
on the gravity ($\Delta$\logg = $\pm$0.04).
The uncertainties in gravity are also very 
small when the short distance modulus reported by Dinescu \etal (1999)
is adopted for M68 ($\Delta$\logg = $\pm$0.07).
For Carina, the dominant uncertainty is from the 
stellar mass.  The range in turn off masses for 
Carina is 0.8 to 1.0 M$_\odot$ due to its episodic star 
formation history;   the resulting uncertainties in \logg\ 
can be as large as 0.10 dex.

Uncertainties in metallicity are adopted from the 
error in the mean of the \ion{Fe}{1} abundances
(see Section~\ref{abuerr}).

\subsection{Spectrum Syntheses and Hyperfine Structure Corrections }

Spectrum syntheses were carried out for \ion{Na}{1} (5688, 5889, 5895),
\ion{Mg}{1} (3829, 3832), \ion{Sr}{2} (4077, 4205), 
\ion{Ba}{2} (4554, 5853, 6141, 6496), and \ion{Eu}{2} (4129, 4205, 6645).  
In each case, the instrumental resolution of the spectrum was adopted as 
the main source of broadening, and checked with the shapes and EW abundance
results from nearby spectral lines (Fe, Ca, Ti).     When an EW could 
be measured, the synthetic and EW abundance results were compared; when
excellent agreement was found then the EW result was taken.   Line
abundances taken from the synthetic results are noted in Table~\ref{ukvlines}. 

Hyperfine structure (HFS) corrections were determined from spectrum 
syntheses for  
the odd-Z elements (\ion{Sc}{2}, \ion{V}{1}, \ion{Mn}{1}, \ion{Co}{1}),
\ion{Cu}{1}, and the neutron capture elements 
\ion{Ba}{2}, \ion{La}{2}, and \ion{Eu}{2}.
HFS components were taken from Lawler \etal (2001a; \ion{La}{2}) , 
Booth \etal (1983; \ion{Mn}{1}), 
Prochaska \etal (2001; \ion{Sc}{2}, \ion{V}{1}, \ion{Mn}{1},
and \ion{Co}{1}), and the Kurucz 
database\footnote{http://kurucz.harvard.edu/LINELISTS/GFHYPERALL}
for the remaining lines.   
HFS components and isotopic ratios for \ion{Cu}{1} 
are from \citet{Beh76}, for \ion{Eu}{2} from Lawler \etal (2001b), 
and for \ion{Ba}{2} from McWilliam \etal (1998; r-process isotopic 
ratios were adopted for the most metal-poor stars, whereas the 
solar ratios were used for stars with [Fe/H]$>-2$).

\subsection{Other Considerations}

\subsubsection{Spherical vs Plane Parallel \label{sphere}}

Spherical MARCS models represent a significant improvement in the
modeling of stellar atmospheres.
%
e.g., Heiter \& Eriksson (2006) recommend
the use of spherical model atmospheres in abundance analyses for
stars with \logg $< 2$ and 4000 $\le$ \teff $\le$ 6500 K, which encompasses 
the range in stellar parameters of our target stars.   
%
Tests performed by Tafelmeyer \etal (2010) showed systematic offsets 
between the spherical and plane-parallel models are below 0.15 dex
(this includes iron lines and other elements, e.g., [C/Fe]).
 
We performed similar tests, comparing the metallicities we determined from
spherical MARCS models to those from plane parallel MARCS (pp-MARCS) 
and Kurucz models.   
The plane parallel MARCS models had the effect of reducing the excitation 
temperature results by $\sim$100 K, but the Kurucz models had a different effect, 
to increase the microturbulence values by $\sim$0.1 dex.   In both cases,
all other parameters were unchanged.   
These offsets, when applied to the abundance analysis, changed
the \ion{Fe}{1} abundances by only $\sim$0.05 dex
(+0.05 with pp-MARCS models, and -0.05 with Kurucz models).    
This is very similar to the offsets found by Tafelmeyer \etal (2010) 
and Heiter \& Eriksson (2006) for iron.

\subsubsection{Continuum Scattering Effects \label{scat}}

\begin{figure}[t]
\includegraphics[width=90mm, height=100mm]{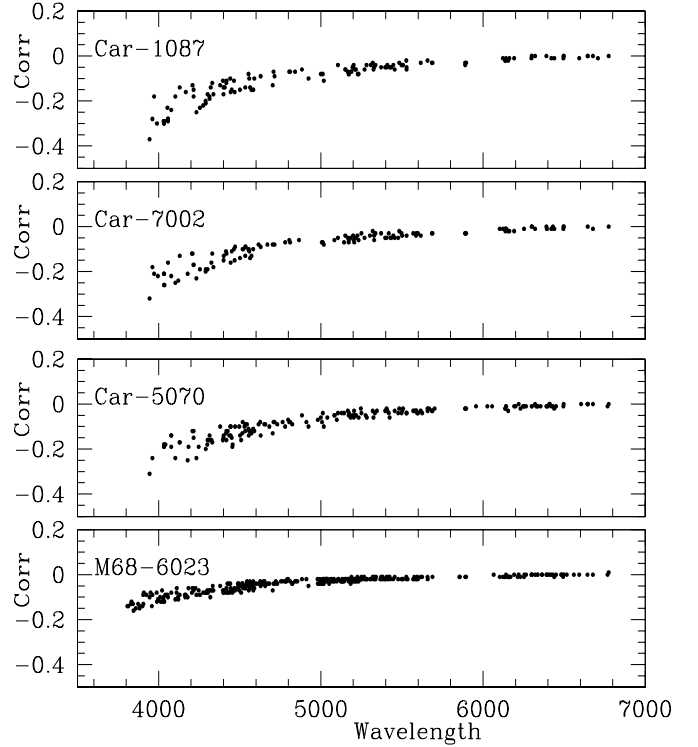}
\caption{
A comparson of the line abundance {\it corrections} from 
MOOG-SCAT for one of the M68 standard star and the three 
most metal-poor stars in Carina.
The y-label ``Corr'' = MOOG-SCAT $-$ MOOG. \\
}
\label{moogscat}
\end{figure}

The standard version of MOOG treats continuum scattering 
($\sigma_\nu$) as if it were absorption ($\kappa_\nu$)
in the source function, 
i.e., S$_\nu$ = B$_\nu$ (the Planck function), 
an approximation that is only valid at long wavelengths
in RGB stars.   At shorter wavelengths ($\le$5000 \AA), 
Cayrel \etal (2004) have shown that the scattering term must be 
taken into account such that the source function becomes 
S$_\nu$ = 
($\kappa_\nu \, B_\nu + \sigma_\nu \, J_\nu$)/($\kappa_\nu + \sigma_\nu$).
A new version of MOOG (MOOG-SCAT, Sobeck \etal  2010) has been 
used to test our results and calculate corrections 
to our abundances line by line.
As shown in Fig.~\ref{moogscat}, the scattering corrections 
are negligible for red lines, but can approach 0.4~dex in the blue.
Lines with the largest corrections are the resonance lines of
\ion{Mn}{1}.   Notice also that the corrections are sensitive to
the atmospheric parameters, in particular metallicity; the three
stars in Carina that are shown were choosen because they have the
largest corrections and they are the most metal-poor.

We compared the elemental abundances between the programs MOOG
and CALRAI (Spite 1967, Cayrel \etal  1991, Hill \etal  2012),
and also the scattering corrections calculated with MOOG-SCAT
to those determined with the program 
Turbospectrum (Alvarez \& Plez 1998). The standard abundance
results and scattering corrections were nearly identical for 
all lines between these codes, in particular, the scattering
corrections were in agreement to within $\sim$0.01 dex.  
Only the resonance lines of \ion{Mn}{1} showed slightly larger 
differences in the scattering corrections ($\sim$0.04 dex);  
since resonance lines form over more atmospheric layers, 
we expect these lines are more sensitive to the details in
the models and the line formation calculations between the codes. 
Thus, we consider the consistency in the abundance results from 
the line formation codes MOOG and CALRAI, and the scattering
corrections between MOOG-SCAT and Turbospectrum to be in 
excellent agreement.

As a final note, the scattering corrections had no effect on our 
spectroscopic temperature or microturbulence determinations, 
nor the metallicities for [\ion{Fe}{1}/H], since only \ion{Fe}{1} 
lines with $\lambda > 5000$ \AA\ were used in those steps.

\subsubsection{Gravity is not from \ion{Fe}{1}/\ion{Fe}{2} \label{gravity}}

The analysis of ten stars in Carina by Koch \etal (2008) and five more
stars by Shetrone \etal (2003) used the \ion{Fe}{2} lines to determine 
spectroscopic gravities from ionization equilibrium 
(where [\ion{Fe}{1}/H] = [\ion{Fe}{2}/H]).   
For four stars in common with Koch \etal, our gravities are smaller 
than theirs by \logg $\sim 0.5$ dex.   Since turbulent velocity and
ionization equilibrium are correlated in RGBs, then the higher 
microturbulent values found by Koch \etal (see Section~\ref{temps})
are correlated with their higher spectroscopic gravities. 


There are sufficient \ion{Fe}{2} lines in many of the stars in our analysis 
to examine the ionization equilibrium of iron and determine spectroscopic 
gravities, however we do not use this method in this analysis.  For 
consistency with the FLAMES/GIRAFFE analysis, where a shorter wavelength 
interval meant fewer \ion{Fe}{2} lines, then we adopt photometric gravities 
using the same methods as in Lemasle \etal (2012).
On the other hand, we notice that the \ion{Fe}{1} and \ion{Fe}{2} 
abundances are in good agreement for our Carina targets, 
with a mean [Fe II/Fe I] = +0.02, and range of -0.24 to +0.20, 
which is $\sim$1$\sigma$(\ion{Fe}{1}).   The stars
in M68 have higher offsets, with a mean [Fe II/Fe I] = +0.35, which
may reflect uncertainties in its distance or reddening.
When the abundance of \ion{Fe}{2} is found to be larger than
\ion{Fe}{1} in red giants, it can be due to the overionization of \ion{Fe}{1} 
by the radiation field which is neglected under the assumption of
LTE.  Corrections for this effect can be as large as 0.3 dex 
(Mashonkina \etal  2010), which is similar to the offset in the 
M68 stars. Thus, we consider our \ion{Fe}{1} and \ion{Fe}{2}
results to be in good agreement for all of our targets when 
physical gravities are adopted.    
 
Note, \ion{Fe}{2} lines at all wavelengths were examined in this analysis,
i.e., they were not restricted to $\lambda <$ 5000 \AA\ like \ion{Fe}{1}.

\subsection{Evaluating the CaT Metallicity }

The initial metallicities in our analysis are taken from a
calibration of the near-IR \ion{Ca}{2} triplet near 8500~\AA 
(= CaT). 
A direct comparison between the low resolution CaT
metallicity index and high resolution iron abundances for
large samples of RGB stars in the dwarf galaxies Fornax
and Sculptor was shown by \citet{Bat08} to
provide consistent abundances in the range 
$-2.5 <$ [Fe/H] $< -0.5$.    More recently, 
Starkenburg \etal (2010) found that the metallicities 
deviate strongly from the linear relationship 
for RGB stars with [Fe/H] $< -2.0$ and
have developed a new calibration for metal-poor stars
that also considers the [Ca/Fe] ratio.

\begin{figure}[t]
\includegraphics[width=90mm, height=90mm]{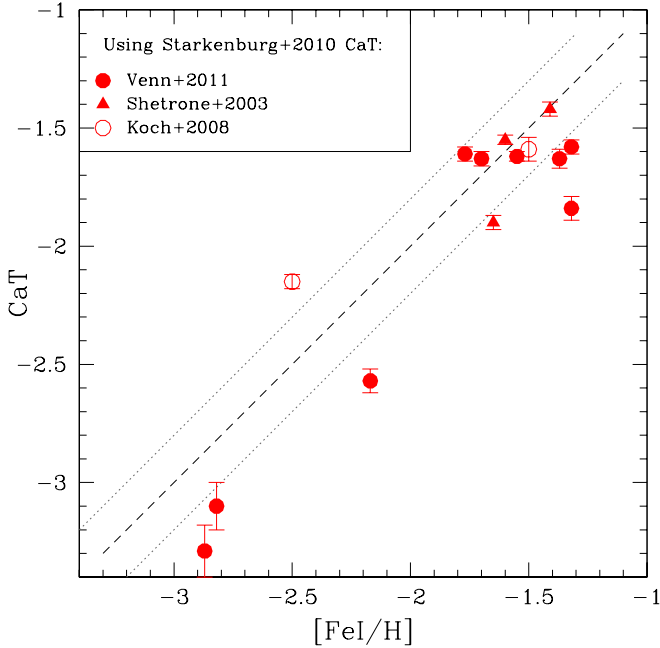}
\caption{Iron abundances ([FeI/H]) from high resolution spectral 
analyses versus those derived from the CaT calibration by   
Starkenburg \etal (2010).   A line of equal metallicities
is shown (dashed), $\pm$0.2 dex (dotted).
}
\label{cat}
\end{figure}

The CaT metallicities for our targets are listed in 
Table~\ref{spect}, and plotted against the [\ion{Fe}{1}/H] 
values in Fig.~\ref{cat}.
The high resolution [\ion{Fe}{1}/H] abundances tend to be 
slightly larger 
than the metallicities from the CaT index,
however, as Starkenburg \etal (2010) show, the
ratio of [Ca/Fe] 
becomes more important 
in the calibration of low metallicity stars.    The two 
most metal-poor stars in this analysis are found to have
lower [Ca/Fe] ratios than the Galactic trend which was
assumed in the Starkenburg \etal calibration.


\section{Abundance Results }

\subsection{Solar Abundances and Galactic Comparisons}

For comparison purposes, solar abundances from 
\citet{Asp09} are adopted.   Galactic comparisons 
are from the 
compilation by Venn \etal (2004), supplemented with 
Si, V, Cr, and Nd from Fulbright (2000).  The compilation
was updated with thick disk stars from Reddy \etal (2006),
La from the compilation by Roederer \etal (2010), 
and Cu and Zn from Mishenina \etal (2002).   All of
these were rescaled to \citet{Asp09} solar abundances
from the Grevesse \& Sauval (1998) values. 
The compilation of abundances of metal-poor stars in the
Galactic halo and Local Group dwarf galaxies by Frebel (2010b, 
after eliminating the upper limit values) was also used 
here for comparison purposes.   
Previously published abundances for Carina and M68 were 
re-scaled to the \citet{Asp09} solar abundances, 
i.e., Shetrone \etal (2003) 
and Lee \etal (2005) from Grevesse \& Sauval (1989), and
Koch \etal (2008) from \citet{Asp05} values.

\subsection{Abundance Error Estimates \label{abuerr}}

\begin{deluxetable}{lrrrrr}
\tablecolumns{5} 
\tablewidth{0pt}
\tablecaption{Abundance Uncertainties for M68-6023 \label{uncs6023}}
\tablehead{
\colhead{Elem} & \colhead{$\Delta$\teff} & \colhead{$\Delta$\logg} & \colhead{$\Delta\xi$} & \colhead{$\Delta$[Fe/H]} & \colhead{Added in}  \\
\colhead {} & \colhead{+55} & \colhead{$-$0.04 } & \colhead{+0.12} & \colhead{$-$0.02} & \colhead{Quadrature} 
}
\startdata
Fe I  & +0.09   &  0.00   & $-$0.05 &    0.01 & 0.10 \\
Fe II & $-$0.02 & $-$0.02 & $-$0.05 & $-$0.01 & 0.06 \\
CH    & +0.05 & $-$0.04   & 0.00    & 0.00    & 0.06 \\
Na I  & +0.08 & +0.01     & $-$0.03 &  0.01   & 0.09 \\
Mg I  & +0.08 & +0.02     & $-$0.03 &  0.01   & 0.09 \\
Ca I  & +0.08 & +0.01     & $-$0.04 &  0.01   & 0.09 \\
Sc II & +0.02 & $-$0.01   & $-$0.05 & $-$0.01 & 0.06 \\
Ti I  & +0.09 & +0.01     & $-$0.04 &  0.01   & 0.10 \\
Ti II & +0.02 & $-$0.01   & $-$0.05 & $-$0.01 & 0.06 \\
V I   & +0.06 &  0.00     & $-$0.03 &  0.01   & 0.07 \\
Cr I  & +0.10 & +0.01     & $-$0.05 &  0.01   & 0.11 \\
Mn I  & +0.10 &  0.00     & $-$0.07 &  0.01   & 0.12 \\
Co I  & +0.10 & +0.01     & $-$0.08 &  0.01   & 0.13 \\
Ni I  & +0.09 & +0.01     & $-$0.05 &  0.01   & 0.11 \\
Zn I  & +0.01 &  0.00     & $-$0.02 &  0.00   & 0.02 \\
Sr II & +0.04 & $-$0.01   & $-$0.08 &  0.01   & 0.09 \\ 
Y II  & +0.02 & $-$0.01   & $-$0.03 & $-$0.01 & 0.04 \\
Zr II & +0.02 & $-$0.02   & $-$0.02 & $-$0.01 & 0.03 \\
Ba II & +0.04 & $-$0.02   & $-$0.09 & $-$0.01 & 0.10 \\
Eu II & +0.03 & $-$0.02   & $-$0.09 & $-$0.01 & 0.10 \\
\enddata  
\end{deluxetable}

\begin{deluxetable}{lrrrrr}
\tablecolumns{5} 
\tablewidth{0pt}
\tablecaption{Abundance Uncertainties for CAR-1087 \label{uncs1087}}
\tablehead{
\colhead{Elem} & \colhead{$\Delta$\teff} & \colhead{$\Delta$\logg} & \colhead{$\Delta\xi$} & \colhead{$\Delta$[Fe/H]} & \colhead{Added in}  \\
\colhead {} & \colhead{+98} & \colhead{$-$0.10 } & \colhead{$-$0.25} & \colhead{$-$0.06} & \colhead{Quadrature} 
}
\startdata
Fe I  & +0.10 &  0.00 & +0.08 & +0.01 & 0.13 \\
Fe II & $-$0.03 & $-$0.02 & +0.08 & $-$0.01 & 0.09 \\
Na I  & +0.13 &  0.00 & +0.06 & +0.01 & 0.15 \\
Mg I  & +0.05 & +0.02 & +0.04 & +0.01 & 0.07 \\
Ca I  & +0.08 & +0.01 & +0.02 & +0.01 & 0.09 \\
Sc II & +0.03 & $-$0.02 & +0.06 & $-$0.01 & 0.07 \\
Ti I  & +0.17 & +0.01 & +0.08 & +0.01 & 0.19 \\
Ti II &  0.00 & $-$0.02 & +0.08 & $-$0.01 & 0.08 \\
V I   & +0.13 & +0.02 & +0.04 & +0.01 & 0.14 \\
Cr I  & +0.18 & +0.01 & +0.09 & +0.02 & 0.20 \\
Mn I  & +0.18 & +0.02 & +0.10 & +0.02 & 0.21 \\
Ni I  & +0.07 & +0.01 & +0.03 & +0.01 & 0.08 \\
Ba II & +0.08 & $-$0.04 & +0.02 & $-$0.01 & 0.09 \\
Eu II & +0.04 & $-$0.03 & +0.02 & $-$0.01 & 0.06 \\
\enddata  
\end{deluxetable}

\begin{deluxetable}{lrrrrr}
\tablecolumns{5} 
\tablewidth{0pt}
\tablecaption{Abundance Uncertainties for Car-484 \label{uncs484}}
\tablehead{
\colhead{Elem} & \colhead{$\Delta$\teff} & \colhead{$\Delta$\logg} & \colhead{$\Delta\xi$} & \colhead{$\Delta$[Fe/H]} & \colhead{Added in}  \\
\colhead {} & \colhead{+66} & \colhead{$-$0.1 } & \colhead{+0.07} & \colhead{$-$0.02} & \colhead{Quadrature} 
}
\startdata
Fe I  & +0.06   & $-$0.01 & $-$0.04 &  0.00 & 0.07 \\
Fe II & $-$0.07 & $-$0.05 & $-$0.03 &  0.00 & 0.09 \\
O I   &  0.00   & $-$0.05 & $-$0.01 & $-$0.01 & 0.05 \\
Na I  & +0.06   &   +0.01 & $-$0.01 &  0.00 & 0.06 \\
Mg I  & +0.05   &   +0.01 & $-$0.03 &  0.00 & 0.06 \\
Si I\tablenotemark{*}  & 0.00 & $-$0.01 &  0.00 & $-$0.01 & 0.01 \\
Ca I  & +0.09   &   +0.01 & $-$0.03 &  0.00 & 0.10 \\
Sc II & $-$0.02 & $-$0.03 & $-$0.02 &  0.00 & 0.04 \\
Ti I  & +0.13   & $-$0.01 & $-$0.04 &  0.00 & 0.14 \\
Ti II & $-$0.03 & $-$0.05 & $-$0.04 &  0.00 & 0.07 \\
V I   & +0.14   & $-$0.01 & $-$0.01 &  0.00 & 0.14 \\
Cr I  & +0.14   & $-$0.01 & $-$0.05 &  0.00 & 0.15 \\
Mn I  & +0.10   &    0.00 & $-$0.03 &  0.00 & 0.11 \\
Co I  & +0.08   & $-$0.02 & $-$0.03 &  0.00 & 0.08 \\
Ni I  & +0.04   & $-$0.01 & $-$0.03 &  0.00 & 0.05 \\
Cu I  & +0.08   & $-$0.02 & $-$0.04 &  0.00 & 0.08 \\
Zn I  & $-$0.04 & $-$0.02 & $-$0.03 &  0.00 & 0.05 \\
Y II  &  0.00   & $-$0.04 & $-$0.03 &  0.00 & 0.05 \\
Ba II & +0.01   & $-$0.06 & $-$0.06 &  0.00 & 0.08 \\
La II & +0.01   & $-$0.06 & $-$0.01 & $-$0.01 & 0.06 \\
Nd II &  0.00   & $-$0.05 & $-$0.03 &  0.00 & 0.06 \\
Eu II & $-$0.01 & $-$0.05 & $-$0.01 & $-$0.01 & 0.05 \\
\enddata  
\tablenotetext{*}{\ion{Si}{1} errors determined from the lines and EWs in Car-524.  } 
\end{deluxetable}

Errors in the abundances have been determined in two ways,
with the maximum of these errors adopted for the analysis 
(Tables~\ref{uncs6023} to \ref{uncs484}).
Firstly, the error in the equivalent width measurements are 
conservatively estimated as
$\Delta$EW = EW$_{\rm rms}$ +  10\% x EW, where
EW$_{\rm rms}$ is from the Cayrel formula (Section~\ref{eqws}).   
This $\Delta$EW was propagated through the abundance analysis
to provide a $\sigma(EW)$ in each line abundance.
This error is not necessarily symmetric, and we adopt the average.   
For synthetic spectra, $\sigma(EW)$ was estimated from the range 
of abundances for which a good fit of the observed line profile 
could be achieved.
The second method for estimating the error in the abundances is 
simply to calculate the standard deviation when more than five 
lines of an element are available, $\sigma(X)$;  
when fewer lines are available, 
we set $\sigma(X)$ = $\sigma$(\ion{Fe}{1}).
The final error in [X/H] is adopted as the maximum of
$\sigma(X)$ or $\sigma(EW)$.  
The corresponding error on the mean, $\delta(X)$, 
is calculated as 

\vspace{0.1in}
$\delta(X)$ = $\sigma(X)$/$\sqrt{N_X}$

\vspace{0.1in}
\noindent where N$_X$ is the number of lines available
of element X.
To get the final error in [X/Fe], i.e., the ratio with iron, then
the error on the mean ($\delta$(FeI)) was added in quadrature.

These line measurement errors do not take into account the errors 
due to the uncertainties in the stellar parameters.   These 
errors are difficult to ascertain and combine properly since they 
are not independent and their relationship(s) to one another are not 
well defined.  We report these errors 
for three representative stars, 
Car-6023, Car-1087, and Car-484, 
in Tables~\ref{uncs6023}, \ref{uncs1087}, and \ref{uncs484}.
These three stars represent the range in stellar parameters,
resolution, and SNR of all of our spectra.   It can be seen
that the stellar parameter errors are similar to or smaller 
than the line measurement errors 
(exceptions include \ion{Ca}{1}, \ion{Ti}{1}, \ion{Cr}{1}, and \ion{Co}{1}
where the stellar parameter errors can be slightly larger, but still 
$\le$0.15 dex).

The stellar parameter errors reported for Car-6023 can be 
applied to Car-6022 and Car-6024, while those for Car-1087 
can be applied to Car-5070 and Car-7002.
Those for Car-484 can be applied to the rest of the FLAMES/UVES 
sample, although the lower SNR for Car-1013 means the errors for that 
target are larger by $\sim$2x (as seen in Table~\ref{spect}).

\subsection {M68 Comparison Stars}

\begin{deluxetable}{llll}
\tablecolumns{4} 
\tablewidth{0pt}
\tablecaption{M68 Chemical Abundances \label{m68}}
\tablehead{
\colhead{Elem} & \colhead{6022} & \colhead{6023} & \colhead{6024}  \\[.4ex]
\colhead{} & \colhead{[X/Fe] $\pm \sigma$ (\#)} & 
                        \colhead{[X/Fe] $\pm \sigma$ (\#)} & 
                        \colhead{[X/Fe] $\pm \sigma$ (\#)}  
}
\startdata
Fe I   &   \phs4.99 $\pm$0.02 (92) & \phs5.09 $\pm$0.03 (92)    &  \phs4.86 $\pm$0.03 (81) \\
Fe II  &   \phs0.40 $\pm$0.05 (19) & \phs0.26 $\pm$0.05 (19)    &  \phs0.39 $\pm$0.06 (17) \\
CH     & $-$0.82  $\pm$0.20  (S)   &  $-$0.62 $\pm$0.20 (S)     &    $-$0.59  $\pm$0.20 (S) \\
O I   &  $<$ 1.36                   &   $<$ 1.33                &  $<$ 1.54        \\
Na I\tablenotemark{a}   &  \phs 0.07 $\pm$0.10 (S) & \phs0.27 $\pm$0.10 (S) & $-$0.03 $\pm$0.10 (S) \\
Mg I   &  \phs 0.01 $\pm$0.20 (S)  &  $-$0.09 $\pm$0.20 (S)     &  \phs0.04  $\pm$0.20 (S) \\
Si I   &  $<$ 1.22                  &  $<$ 1.15                 &  $<$ 1.36        \\
Ca I   &  \phs 0.48 $\pm$0.04 (17) & \phs0.39 $\pm$ 0.04 (16) & \phs0.48 $\pm$0.05 (15) \\
Sc II  &  \phs 0.57 $\pm$0.05 (10) & \phs0.28 $\pm$ 0.10 (10) & \phs0.44 $\pm$0.06 (9) \\
Ti I   &  \phs 0.11 $\pm$0.05 (23) & \phs0.12 $\pm$ 0.04 (21) & \phs0.30 $\pm$0.07 (20) \\
Ti II  &  \phs 0.77 $\pm$0.05 (52) & \phs0.54 $\pm$ 0.05 (51) & \phs0.65 $\pm$0.05 (46) \\
V I    & $-$0.06  $\pm$0.16  (2)   & \phs0.02 $\pm$ 0.17 (2)  & \phs0.35 $\pm$0.17 (2) \\
Cr I   & $-$0.17  $\pm$0.06  (12)  &   $-$0.25  $\pm$ 0.09 (12) &   $-$0.17  $\pm$0.08 (9) \\
Mn I\tablenotemark{a}   & $-$0.48  $\pm$0.07  (7)   &   $-$0.61 $\pm$0.12 (7)    &    $-$0.31 $\pm$0.10 (4) \\
Co I   & $-$0.27  $\pm$0.09  (6)   &   $-$0.26 $\pm$0.06 (7)    & \phs0.10 $\pm$0.10 (6) \\
Ni I   &  \phs 0.06 $\pm$0.03 (8)  & \phs0.20 $\pm$0.10 (8)   & \phs0.04 $\pm$0.14 (7) \\
Zn I   &  \phs 0.47 $\pm$0.16 (2)  & \phs0.33 $\pm$0.17 (2)   & \phs0.43 $\pm$0.24 (1) \\
Sr II  & $-$0.08  $\pm$0.16  (2)   &   $-$0.26 $\pm$0.17 (2)    &    $-$0.23 $\pm$0.17 (2) \\
Y II   & $-$0.07  $\pm$0.23  (1)   &   $-$0.43 $\pm$0.17 (2)    &    $-$0.21 $\pm$0.17 (2) \\
Zr II  &  \phs 0.20 $\pm$0.23 (1)  &   $-$0.05 $\pm$0.24 (1)    & \phs0.00 $\pm$0.24 (1) \\
Ba II  &  \phs 0.33 $\pm$0.12 (4)  & \phs0.14 $\pm$0.12 (4)   & \phs0.36 $\pm$0.12 (4) \\
La II  &  $<$ 1.37                 &  $<$ 1.36                  &  $<$ 1.56       \\
Nd II  &  \phs \nodata             & \phs \nodata             & \phs1.27 $\pm$0.24 (1) \\
Eu II  &  \phs \nodata             & \phs0.24 $\pm$0.17 (2)   & \phs0.47 $\pm$0.17 (2) \\
\enddata  
\tablecomments {We calculate [X/Fe] = [X/H] $-$ [\ion{Fe}{1}/H].  
For \ion{Fe}{1}, the abundance listed is 12$+$log(Fe/H). }
\tablenotetext{a}{A correction of $\Delta$[Na/Fe] = $-0.5$ has been applied to
account for NLTE effects on the Na D lines \citep{And07}.
Similarly, we apply a correction of $\Delta$log (Mn/H) = $+0.5$ to
the \ion{Mn}{1} resonance lines to account for NLTE effects 
(Bergemann \& Gehren 2008). \\ 
} 
\end{deluxetable}

The stellar parameters and chemical abundances for our three stars 
in M68 are compared to the one star analysed by Shetrone \etal (2003)
and eight stars by Lee \etal (2005).    The targets in this analysis
are fainter by $\sim$1.4 magnitudes, thus they are located further 
down the RGB and have slightly higher temperatures and gravities. 
Our microturbulence values are comparable or lower.

The metallicities from \ion{Fe}{1} for these three stars are in 
excellent agreement with one another, as expected for a single 
stellar population.  The mean \ion{Fe}{1} abundance is 
log(\ion{Fe}{1}/H) = 5.04 $\pm$0.10, where the error is the 
line to line abundance scatter in each star, added in quadrature,
and divided by $\sqrt{3}$.
This is similar to the mean [\ion{Fe}{1}/H] 
abundance determined by Lee \etal (2005), where 
[\ion{Fe}{1}/H] = 4.96 $\pm$0.06
when they adopt the photometric gravities.
Higher iron abundances are found with lower spectroscopic gravities;
this is also seen by Shetrone \etal (2003) for one star.

Most of the elemental abundances determined here are in 
excellent agreement with those from Lee \etal (2005) and 
Shetrone \etal (2003), and our \ion{Na}{1} values are 
within the ranges of the others.
To compare the [X/Fe] $ratios$ for these elements with those 
from Lee \etal  requires knowing that their [X/Fe] results use 
\ion{Fe}{1}  for ratios with neutral species and \ion{Fe}{2} 
for ratios with ionized species and 
oxygen
In our analysis, we use only the \ion{Fe}{1} lines to indicate the
metallicity, thus we adjust their [X/\ion{Fe}{2}] abundances by adding
their [\ion{Fe}{2}/\ion{Fe}{1}] ratio, = $+0.40$.

The most significant differences are that our Mg and Mn abundances
are lower than Lee \etal's (2005), and the Shetrone \etal's (2003)  
values for Zn, Y, Ba, and Eu are lower (see Figures in the next Sections).
For the heavy elements examined by Shetrone \etal, 
we track the differences to the model atmosphere parameters,
e.g., ionized species that are more sensitive to gravity, 
and Zn abundances that are $\sim$2x more sensitive to temperature 
than the other elements.  
For Mn, we note that Lee \etal  examined only a single, very 
weak line (Mn I 6021.8), whereas ours are from several lines that
are in good agreement with one another, thus we consider our result
more reliable.    
For Mg, our abundances have been determined from four spectral lines 
$-$ three lines from spectrum synthesis due to severe blending and 
difficulties in the continuum placements (two near 3830 \AA, and 
one at $\lambda$5172), and a fourth line at $\lambda$5528 which is 
sufficiently unblended to be analysed from EW measurements (note 
that a fifth line is available in the FLAMES/UVES spectra at
$\lambda$5711 but is too weak to be measured in any of the 
Magellan/MIKE spectra).  
The Mg abundances are consistent between these four lines and from
star to star in M68, however we assign a larger uncertainty to 
account for the difficulties in setting the continuum in the three 
line syntheses.   
The average [Mg/Fe] is slightly lower than found in the other 
analyses (Lee \etal 2005, Shetrone \etal 2003), than the Galactic
trends at that metallicity, and than found in the other 
$\alpha$-elements.   While the offset is within the estimated 
uncertainties, we note that our [Na/Fe] are slightly higher than
the Galactic trends $-$ it is possible these stars exhibit some
deep mixing (e.g., Gratton \etal 2010).
%

\begin{deluxetable}{llll}
\tablecolumns{4} 
\tablewidth{0pt}
\tablecaption{Carina Chemical Abundances \label{ukv1}}
\tablehead{
\colhead{Elem}  & \colhead{484} & \colhead{524} & \colhead{612} \\[.4ex]
\colhead{} &  \colhead{[X/Fe] $\pm \sigma$ (\#)} & 
                        \colhead{[X/Fe] $\pm \sigma$ (\#)} &
                        \colhead{[X/Fe] $\pm \sigma$ (\#)}
}
\startdata
Fe I  &  \phs5.97 $\pm$0.02 (114) &  \phs5.75 $\pm$0.02 (121) & \phs6.20 $\pm$0.02 (126)  \\
Fe II &  \phs0.07 $\pm$0.04 (14)  &  \phs0.09 $\pm$0.03 (15)  & \phs0.12 $\pm$0.05 (18) \\
O I   &  \phs0.40 $\pm$0.21 (1)   &  \phs0.41 $\pm$0.21 (1)   &  $<$ 0.09      \\
Na I  &  $-$0.25 $\pm$0.21 (1)    &  $<-$0.26                 &  $<-$0.72     \\
Mg I  &  \phs0.19 $\pm$0.15 (2)   &  \phs0.27 $\pm$0.15 (2)   & $-$0.50 $\pm$0.16 (2) \\
Si I  &  \phs\nodata              &  \phs0.60 $\pm$0.15 (2)   & \phs0.13 $\pm$0.23 (1) \\
Ca I  &  \phs0.16 $\pm$0.04 (20)  &  \phs0.10 $\pm$0.04 (19)   & $-$0.17 $\pm$0.04 (19) \\
Sc II & \phs0.06 $\pm$0.04 (12)   & $-$0.08 $\pm$0.04 (10)    & $-$0.74 $\pm$0.07 (8) \\
Ti I  &  \phs0.23 $\pm$0.04 (25)  & $-$0.02 $\pm$0.03 (22)    & $-$0.42 $\pm$0.04 (18) \\
Ti II & \phs0.18 $\pm$0.05 (13)  &  \phs0.03 $\pm$0.06 (13) & $-$0.29 $\pm$0.05 (12) \\
V I   & \phs0.01 $\pm$0.03 (10)  &  \phs0.21 $\pm$0.15 (2)  & $-$0.67 $\pm$0.23 (1) \\
Cr I  & \phs0.00 $\pm$0.12 (4)   & $-$0.09 $\pm$0.10 (4)    & $-$0.20 $\pm$0.12 (4) \\
Mn I  & $-$0.27 $\pm$0.05 (7)    & $-$0.34 $\pm$0.10 (4)    & $-$0.51 $\pm$0.10 (6) \\
Co I  & $-$0.13 $\pm$0.21 (1)    & $-$0.27 $\pm$0.21 (1)    & $<-$0.86        \\
Ni I  & $-$0.07 $\pm$0.04 (15)   & $-$0.08 $\pm$0.04 (14)   & $-$0.46 $\pm$0.04 (14) \\
Cu I  & $-$0.47 $\pm$0.21 (1)    &  $<-$0.85                & $<-$1.60       \\
Zn I  & \phs0.16 $\pm$0.21 (1)   & $-$0.17 $\pm$0.21 (1)    & $-$0.83 $\pm$0.23 (1) \\
Y II  & $-$0.32 $\pm$0.12 (3)    & $-$0.46 $\pm$0.12 (3)    & $-$1.37 $\pm$0.16 (2) \\
Ba II & \phs0.16 $\pm$0.26 (1)   &  \phs0.09 $\pm$0.21 (1)  & $-$0.64 $\pm$0.16 (S) \\
La II & \phs0.48 $\pm$0.12 (3)   &  \phs0.55 $\pm$0.15 (2)  & $<$ 0.11        \\
Nd II & \phs0.63 $\pm$0.15 (2)   &  \phs0.52 $\pm$0.15 (2)  & $-$0.56 $\pm$0.23 (1) \\
Eu II & \phs0.47 $\pm$0.21 (1)   &  $<$ 0.63           & $<$ 0.26        \\
\enddata  
\tablecomments {We calculate [X/Fe] = [X/H] $-$ [\ion{Fe}{1}/H].  
For \ion{Fe}{1}, the abundance listed is 12$+$log(Fe/H). }
\end{deluxetable}

\begin{deluxetable}{llll}
\tablecolumns{4} 
\tablewidth{0pt}
\tablecaption{Carina Chemical Abundances continued \label{ukv2}}
\tablehead{
\colhead{Elem}  & \colhead{705} & \colhead{769} & \colhead{1013} \\[.4ex]
\colhead{} &  \colhead{[X/Fe] $\pm \sigma$ (\#)} & 
                        \colhead{[X/Fe] $\pm \sigma$ (\#)} &
                        \colhead{[X/Fe] $\pm \sigma$ (\#)}
}
\startdata
Fe I  & \phs6.15 $\pm$0.02 (127) & \phs5.82 $\pm$0.02 (125) &  \phs6.20 $\pm$0.03 (125) \\
Fe II & \phs0.02 $\pm$0.06 (16) &  \phs0.06 $\pm$0.05 (15)  &   $-$0.24 $\pm$0.06 (17) \\
O I   & $<$0.12                 &  \phs\nodata              &  $<$0.08  \\
Na I  &  $-$0.22 $\pm$0.25 (1)  &  \phs0.04 $\pm$0.25 (1)   &   $-$0.25 $\pm$0.37 (1) \\
Mg I  & \phs0.13 $\pm$0.18 (2)  &  \phs0.42 $\pm$0.18 (2)   &  \phs0.09 $\pm$0.26 (2) \\
Si I  & \phs0.23 $\pm$0.25 (1)  &  \phs0.45 $\pm$0.18 (2)   &  \phs0.06 $\pm$0.26 (2) \\
Ca I  & \phs0.11 $\pm$0.04 (19) &  \phs0.24 $\pm$0.06 (21)  &   $-$0.06 $\pm$0.05 (20) \\
Sc II &  $-$0.03 $\pm$0.05 (11) &  \phs0.15 $\pm$0.06 (11)  &   $-$0.24 $\pm$0.07 (13) \\
Ti I  & \phs0.02 $\pm$0.05 (25) &  \phs0.27 $\pm$0.05 (23)  &  \phs0.27 $\pm$0.06 (24) \\
Ti II &  $-$0.01 $\pm$0.06 (11) &  \phs0.13 $\pm$0.06 (11)  &  \phs0.11 $\pm$0.09 (13) \\
V I   & \phs0.02 $\pm$0.06 (10) &  \phs0.41 $\pm$0.11 (5)   &  \phs0.09 $\pm$0.08 (11) \\
Cr I  &  $-$0.23 $\pm$0.12 (4)  &  \phs0.08 $\pm$0.12 (4)   &  \phs0.25 $\pm$0.21 (3) \\
Mn I  &  $-$0.30 $\pm$0.07 (7)  &  \phs0.02 $\pm$0.11 (5)   &   $-$0.51 $\pm$0.17 (5) \\
Co I  &  $-$0.26 $\pm$0.25 (1)  &  \phs0.02 $\pm$0.25 (1)   &   $-$0.01 $\pm$0.26 (2) \\
Ni I  &  $-$0.04 $\pm$0.04 (17) &  \phs0.03 $\pm$0.04 (13)  &   $-$0.10 $\pm$0.06 (17) \\
Cu I  &  $-$0.80 $\pm$0.25 (1)  & $-$0.55 $\pm$0.25 (1)     &    \phs\nodata   \\
Zn I  &  $-$0.05 $\pm$0.25 (1)  &  \phs\nodata              &   $-$0.46 $\pm$0.37 (1) \\
Y II  &  $-$0.63 $\pm$0.18 (2)  & $-$0.27 $\pm$0.14 (3)     &   $-$0.32 $\pm$0.21 (3) \\
Ba II &  $-$0.61 $\pm$0.14 (3)  & $-$0.17 $\pm$0.14 (3)     &   $-$0.15 $\pm$0.26 (S) \\
La II & $<$0.15                &  \phs\nodata              &  $<$0.14  \\
Nd II & \phs0.26 $\pm$0.18 (2)  &  \phs0.27 $\pm$0.18 (2)   &  \phs0.11 $\pm$0.26 (2) \\
Eu II & $<$0.31                 &  $<$0.63                  &  \phs0.33 $\pm$0.37 (1) \\
\enddata  
\tablecomments {We calculate [X/Fe] = [X/H] $-$ [\ion{Fe}{1}/H].  
For \ion{Fe}{1}, the abundance listed is 12$+$log(Fe/H). }
\end{deluxetable}

\begin{deluxetable}{llll}
\tablecolumns{4} 
\tablewidth{0pt}
\tablecaption{Carina Chemical Abundances \label{ngc}}
\tablehead{
\colhead{Elem} & \colhead{1087} & \colhead{5070} & \colhead{7002} \\[.4ex]
\colhead{} &  \colhead{[X/Fe] $\pm \sigma$ (\#)} & 
                        \colhead{[X/Fe] $\pm \sigma$ (\#)} &
                        \colhead{[X/Fe] $\pm \sigma$ (\#)}
}
\startdata
Fe I  & \phs4.69 $\pm$0.06 (38)  & \phs5.35 $\pm$0.05 (55)  & \phs4.64 $\pm$0.05 (38) \\
Fe II & \phs0.01 $\pm$0.07 (7)   & $-$0.14 $\pm$0.10 (11)   & \phs0.20 $\pm$0.14 (6) \\
CH    & $<-$1.02 (S)            & $<-$0.98  (S)         &  $<-$1.07 (S) \\
O I   & $<$ 1.55                 & $<$ 0.90                  &  $<$ 1.61  \\
Na I\tablenotemark{a}  & $-$0.82 $\pm$0.24 (S)    & $-$1.18 $\pm$0.38 (S)    &  $-$0.45 $\pm$0.23 (S) \\
Mg I  & \phs0.48 $\pm$0.34 (1)   & $-$0.36 $\pm$0.38 (1)    & \phs0.22 $\pm$0.23 (2) \\
Si I  & $<$ 1.70                 & $<$ 1.01                 &  $<$ 1.76  \\
Ca I  & $-$0.03 $\pm$0.20 (3)    & $-$0.01 $\pm$0.10 (9)    & \phs0.18 $\pm$0.10 (6) \\
Sc II & $-$0.01 $\pm$0.15 (5)    & $-$0.20 $\pm$0.24 (8)    & \phs0.10 $\pm$0.15 (4) \\
Ti I  & \phs0.65 $\pm$0.17 (4)   & $-$0.08 $\pm$0.15 (8)    & \phs0.11 $\pm$0.18 (3) \\
Ti II & \phs0.29 $\pm$0.07 (24)  & \phs0.04 $\pm$0.07 (30)  & \phs0.39 $\pm$0.08 (20) \\
V I   & \phs0.72 $\pm$0.34 (1)   & $-$0.26 $\pm$0.38 (1)    & \phs\nodata \\
Cr I  & $-$0.82 $\pm$0.17 (4)    & $-$0.47 $\pm$0.08 (7)   &  $-$0.34 $\pm$0.16 (4) \\
Mn I  & $-$0.72 $\pm$0.24 (2)    & $-$0.81 $\pm$0.18 (3)   &  $-$0.33 $\pm$0.23 (2) \\
Co I  & $<$ 1.11                 &  $<$ 0.31                &  $-$0.26 $\pm$0.23 (2) \\
Ni I  & \phs0.36 $\pm$0.34 (1)   & \phs0.18 $\pm$0.13 (7)   & \phs0.35 $\pm$0.18 (3) \\
Cu I  &  $<$ 0.37                & $<-$0.43                &  $<$ 0.39  \\
Zn I  &  $<$ 0.66                & $<$ 0.01                 &  $<$ 0.77  \\
Sr II & $-$1.56 $\pm$0.24 (S)    & \phs\nodata              &  $-$1.21 $\pm$0.32 (S) \\
Y II  &  $<-$0.05               &  $<-$0.69               &  $<-$0.12  \\
Zr II &  $<$ 0.24                &  $<-$0.41               &  $<$ 0.36  \\
Ba II & $-$1.07 $\pm$0.34 (S)    &  $-$1.12 $\pm$0.27 (S)   &  $-$0.98 $\pm$0.18 (S) \\
La II &  $<$ 1.77                &  $<$ 1.15                &   $<$ 1.75  \\
Nd II &  $<$ 0.99                &  $<$ 0.34                &   $<$ 1.09  \\
Eu II &  $<$ 1.85                &  \phs0.13 0.38 (1)       &  $<$ 1.90  \\
\enddata  
\tablecomments {We calculate [X/Fe] = [X/H] $-$ [\ion{Fe}{1}/H].  
For \ion{Fe}{1}, the abundance listed is 12$+$log(Fe/H). }
\tablenotetext{a}{A correction of $\Delta$[Na/Fe] = $-0.6$ has been applied to
account for NLTE effects on the Na D lines \citep{And07}.
Similarly, we apply a correction of $\Delta$log (Mn/H) = $+0.5$ to
the \ion{Mn}{1} resonance lines to account for NLTE effects 
(Bergemann \& Gehren 2008). \\ 
}
\end{deluxetable}

New elemental abundances not previously determined in M68 
(\ion{Sr}{2} and \ion{Zr}{2})
are in good agreement with the Galactic distributions.   
Finally, our upper limits for \ion{Si}{1}, \ion{O}{1}, and \ion{La}{2} 
are in agreement with the abundances determined by Lee \etal (2005)  
and Shetrone \etal (2003).   
  
Thus, our M68 results appear to be in good agreement with other 
abundance determinations for this cluster, and Galactic cluster
and field stars in general.  We conclude that our stellar parameter 
determinations and model atmospheres analyses can therefore be used
reliably for elemental abundances in the Carina targets.

\subsection{Carina Stars}

The chemical abundances of the stars in Carina are presented in Tables~\ref{ukv1}
to \ref{ngc}, and discussed element by element in the following Sections.

\subsubsection{Carbon}

Carbon forms during the helium burning phases, whether the hydrostatic
helium core burning phases in massive stars, or helium shell burning
phases in AGB stars (Woosley \& Weaver 1995).  
The chemical evolution of carbon is further complicated 
by CNO cycling, where carbon is reduced when exposed to hot protons 
and the CN-cycle runs to equilibrium values.    This evolution in carbon 
can be seen in RGB stars in globular clusters, where the [C/H] abundance 
and $^{12}$C/$^{13}$C ratio are both reduced as stars ascend the RGB due 
to internal (self) mixing (Gratton \etal 2004). 
 
\begin{figure}[t]
\includegraphics[width=90mm, height=50mm]{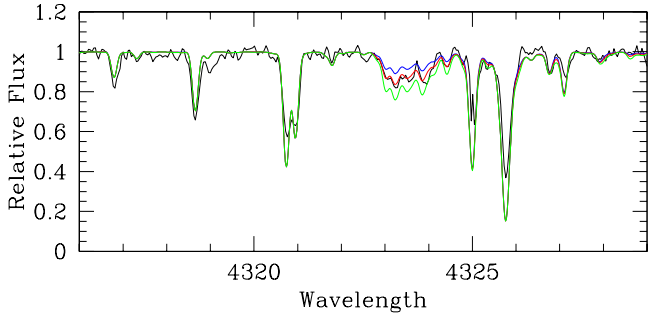}
\caption{Spectrum synthesis of the CH 4320 \AA\ feature in
M68-6023.   The chemical composition determined in this paper
was used for the syntheses, along with three carbon abundances, 
[C/Fe] = $-0.4, -0.6, -0.8$.   Broadening parameters were
initially taken from the resolution of the Magellan spectra
(Gaussian FWHM $\sim$ 0.15)
and checked against the best fit well defined iron-features. \\
}
\label{ch}
\end{figure}

Carbon abundances in this analysis were determined from spectrum 
syntheses of portions of the $A^2\Pi$ - $X^2\Delta$ CH G-band 
near $\lambda$4320 in the Magellan spectra.    The best fit
was taken as the carbon abundance from the CH line list by 
Brown \etal (1987) and Carbon \etal (1982), with acceptable fits 
as the abundance uncertainty.   A C$^{\rm 12}$/C$^{\rm 13}$ ratio 
of six was adopted (the typical value for bright RGB stars that have 
undergone the first dredge up); this ratio made only a small 
difference such that increasing the value to 50 caused 
$\Delta$log(C/H) = 0.06.   Similarly, reducing the oxygen
abundance had only a small effect (since C can be locked into the CO 
molecule); we adopted [O/Fe] = +0.4, but reducing this 
by 3x lowered the carbon abundance by only $\le 0.05$ dex.
The spectrum of M68-6023 around the CH 4320 \AA\ feature is
shown in Fig.~\ref{ch}, with spectrum syntheses for three 
carbon abundances.

\begin{figure}[t]
\includegraphics[width=90mm, height=90mm]{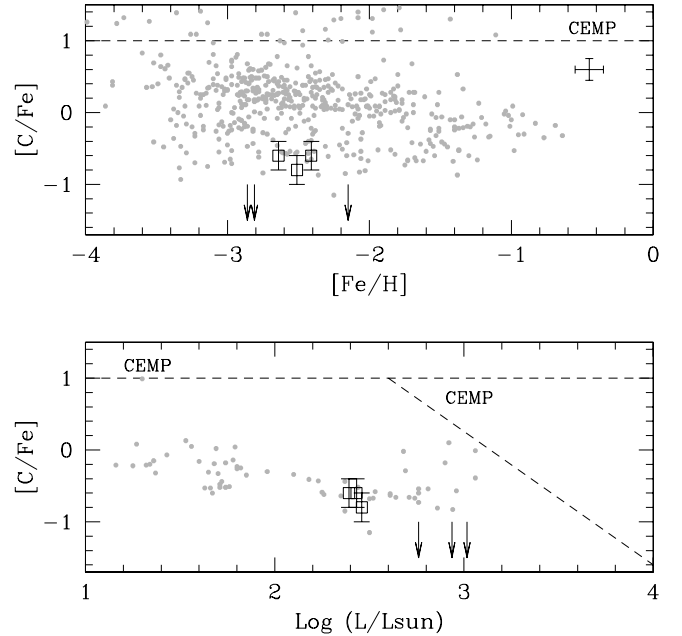}
\caption{Carbon abundances in Galactic stars (small grey 
dots; Frebel 2010b, Gratton \etal  2000), the three stars in
M68 (empty black squares), and upper limits on the three metal 
poor Carina stars ({\it upper panel}). 
CEMP stars are usually defined as having [C/Fe] $> 1.0$, although
extra mixing at the RGB tip can lead to a luminosity dependence
as noted in the {\it lower panel}.   The low carbon abundances in
the Carina stars are consistent with those of luminous RGB stars 
in globular clusters (data from Gratton \etal 2000), which are not
similar to CEMP stars.   
}
\label{carbon}
\end{figure}

The carbon abundances for the M68 stars are in good agreement
with typical red giants in globular clusters above the RGB bump
(e.g., Smith \etal  2005, Gratton \etal 2000), 
where [C/Fe] $\sim -0.4$ at the RGB bump and slowly decreases 
to $\sim -1.0$ dex at the RGB tip due to mixing of CNO-cycled 
gas.  This trend can be seen in Fig.~\ref{carbon} (lower panel,
data from Gratton \etal 2000). 
The carbon abundances compiled by Frebel (2010b) suggests 
a much larger scatter in carbon in metal-poor stars than
seen in clusters (Fig.~\ref{carbon}, upper panel), however 
Frebel's compilation includes field stars of all evolutionary
stages, including carbon enhanced metal-poor (CEMP) stars. 
CEMPs have [C/Fe] $> 1.0$ and are thought to be both stars 
that have undergone mass transfer in a binary system with 
an AGB companion (CEMP-s, e.g., Beers \& Christlieb 2005) 
and stars with high primordial carbon 
(e.g., from high mass rotating stars, Meynet \etal  2006).
\citet{Aoki07} suggested that mixing on the RGB 
and extra mixing at the tip of the RGB will lower the surface 
carbon abundance, such that the definition of a carbon 
enhanced metal-poor (CEMP) star should depend on luminosity.
This luminosity dependent range is shown in Fig.~\ref{carbon}
(lower panel), and clearly shows that stars in M68 and the 
Carina dSph are not CEMP stars by any definition.
 
The mean carbon abundance in M68 is $<$[C/Fe]$>$ = $-0.7 \pm 0.1$, 
in agreement with typical upper RGB stars.  Upper limits on 
the carbon abundances in the metal-poor Carina stars 
are even lower ([C/Fe] $\le -1.0$); this value is typical 
of upper RGB stars in globular clusters, but on the 
low side of the values found for the bright, metal-poor 
{\it field} stars in the Galaxy, as seen in Fig.~\ref{carbon}.

\subsubsection {Alpha Elements}

The alpha-elements (O, Mg, Si, Ca, Ti) are built from multiple 
captures of He nuclei ($\alpha$ particle) during various stages
in the evolution of massive stars ($>$ 8 M$_\odot$, 
e.g., carbon burning, neon burning, complete and incomplete Si burning),
and dispersed during SN~II events.    Thus, the [$\alpha$/Fe] ratio
in a star
is a way to trace the relative contributions from SN~II to SN~Ia 
products in the ISM when it formed (the nucleosynthesis of iron is
discussed in the next Section). 
Although Ti is not a true alpha-element, the dominant isotope 
$^{48}$Ti forms through explosive Si-burning and the 
$\alpha$-rich freeze out\footnote{The $\alpha$-rich freeze out 
occurs when a strong shock travels through infalling matter at the 
inner most shell of a core-collapse supernova.   After being shocked,
material is heated to $>5$ billion degrees, so that nuclei decompose
into neutrons and tightly bound $\alpha$ particles.}, 
thus it behaves like an alpha-element (Woosley \& Weaver 1995). 

\begin{figure}[t]
\includegraphics[width=90mm, height=75mm]{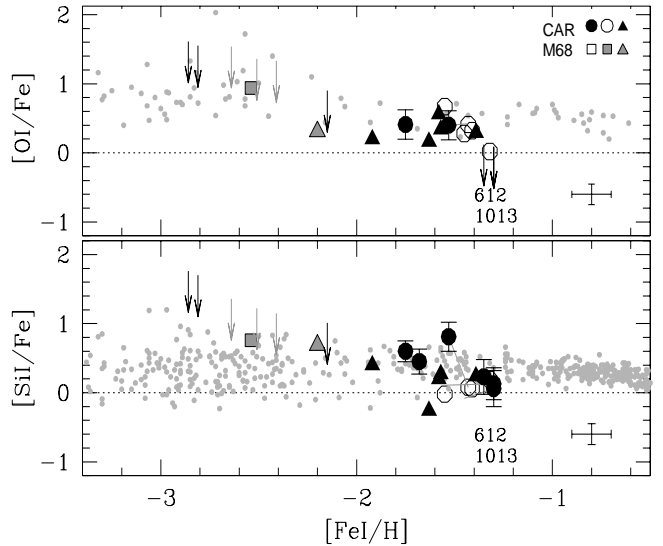}
\caption{The LTE oxygen and silicon abundances of stars
in Carina and M68, compared to the Galactic distribution.   
Results for Carina are from this paper (filled circles), 
Koch \etal (2008, empty circles), and 
Shetrone \etal (2003, filled triangles); 
for four stars in common between this analysis 
and Koch \etal, the abundance results are connected by a 
grey line. 
M68 results are shown for one star from 
Shetrone \etal (2003, grey triangle), 
the mean of 7 RGBs from Lee \etal (2005, grey square).  
Usually our M68 results are shown by empty black squares,
but for O and Si only upper limits are available, shown
as grey arrows.   A representative error bar 
is shown ($\Delta$[Fe/H] = $\pm0.1$, $\Delta$[X/Fe] = $\pm$0.15), 
although the actual errors per star from 
this analysis are also plotted per point. 
Small grey points show the Galactic distributions summarized
by Venn \etal (2004) and Frebel (2010b). \\
}
\label{osi}
\end{figure}

\begin{figure}[t]
\includegraphics[width=90mm, height=95mm]{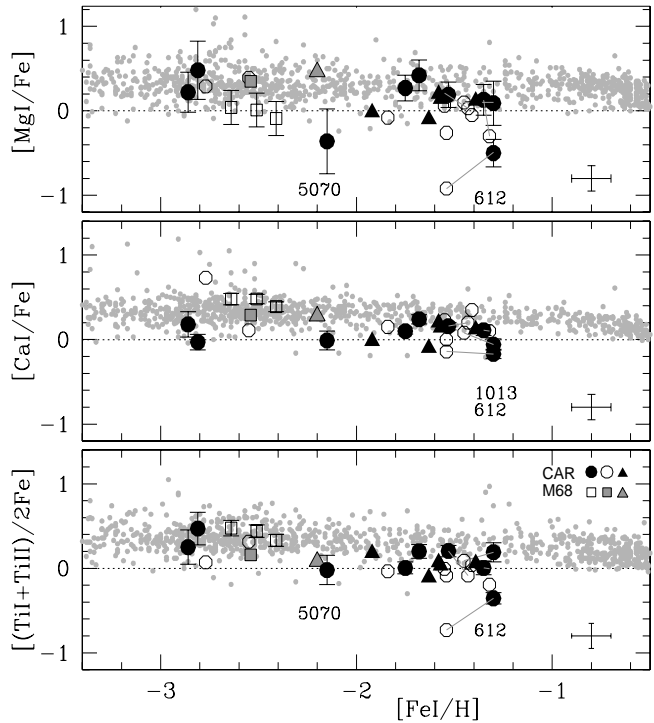}
\caption{The LTE magnesium, calcium and titanium abundances
of stars in Carina and M68, compared to the Galactic distribution. 
Symbols are the same as in Fig.~\ref{osi}. \\
}
\label{mgcati}
\end{figure}

Our [$\alpha$/Fe] results are plotted in Figs.~\ref{osi} and \ref{mgcati}.
Koch \etal (2008) determined $\alpha$ element abundances for 4
stars in common with ours, and a solid line connects those results
in the Figures.
In general, the $\alpha$-element abundances in Carina tend to be lower 
than the Galactic distributions at all intermediate metallicities, 
i.e., $-2.2 <$ [Fe/H] $< -1.3$.  
In the most metal-poor stars at 
[Fe/H] = $-2.9$, most of the $\alpha$ elements are in good 
agreement with the Galactic distributions.   
Two stars stand out in their [$\alpha$/Fe] distributions,
Car-612 and Car-5070, which have very low ratios of
[Mg,Ca,Ti/Fe] for their metallicities.   These two stars
will be discussed further in the following Sections. 

{\it Oxygen:} Oxygen abundances in this analysis are solely from 
the [\ion{O}{1}] 6300 line.    Oxygen could only be measured in the 
higher metallicity stars in the sample where the line is stronger.    
Excellent agreement was found for one star in common with Koch \etal (2008;
Car-484, where they found [O/Fe]=+0.39 and we find =+0.48).

{\it Silicon:} Silicon abundances are only available for the 
higher metallicity stars.    Line measurements of the
\ion{Si}{1} lines at 5684 \AA\ and 6155 \AA\ are similar
to those in Koch \etal (2008), however our oscillator strength for one line
is lower by 0.4 dex.   
Only upper limits were available from the Magellan spectra
taken of the metal-poor stars.

{\it Magnesium:}
\ion{Mg}{1} abundances were calculated from the $\lambda$5528 line
in all of the Carina stars, combined with $\lambda$5172 and $\lambda$5183
when $<$200 m\AA, and $\lambda$5711 in the FLAMES/UVES spectra
(it was too weak to measure in the Magellan spectra).  
The abundances between these lines are in good agreement.
We note that the Carina Mg abundances are in good
agreement with those from Koch \etal (2008) and 
Shetrone \etal (2003), see Fig.~\ref{mgcati}.
%

{\it Calcium:}  Many lines of \ion{Ca}{1} were available in all of
the program stars, over a wide range of wavelengths, and of appropriate line 
strengths.    Results are in good agreement with Koch \etal (2008),
see Fig~\ref{mgcati}.

{\it Titanium:}
Titanium abundances were determined from a range of 
\ion{Ti}{1}  and \ion{Ti}{2} lines across the entire 
spectral region.   The abundances between
the two species are in good agreement, 
typically within 1~$\sigma$ of each other, despite the
possibility that titanium can be overionized by the 
radiation field
resulting in lower \ion{Ti}{1} abundances when
LTE is assumed.   
Letarte \etal (2010) suggested that better agreement is found
when [\ion{Ti}{1}/\ion{Fe}{1}] is compared to
 [\ion{Ti}{2}/\ion{Fe}{2}] in their sample of Fornax RGB
stars, due to similarities in the NLTE effects and possibly 
temperature scale variations.  Examination of our M68 comparison stars 
are in agreement with their finding, such that the agreement improves
when \ion{Ti}{2} is compared with \ion{Fe}{2}, however this had 
no effect on our Carina abundances. 
Thus, in this paper, the abundances from both species are 
averaged together.
We note that our Carina Ti abundances are in good
agreement with those from Koch \etal (2008) and Shetrone \etal (2003); 
see Fig.~\ref{mgcati}. 

\subsubsection {Sodium \label{sodium}} 
Sodium is produced primarily during the carbon burning stages, but 
as metallicity increases then a sufficient amount of neutron rich material
also allows sodium to be produced through the Ne-Na cycle during H-burning
(Woosley \& Weaver 1995).   Thus, the chemical evolution of Na initially
follows the $\alpha$ elements, but then deviates from this (rising)
when AGB stars contribute.
Furthermore, Herwig \etal (2004) have shown that {\it metal-poor AGB stars} 
from 2 to 5 M$_\odot$ will {\it overproduce} sodium by factors of 10 to 100.    
This Na overproduction has multiple sites dependent on the AGB mass.

In the FLAMES/UVES spectra, sodium abundances were determined from a 
single \ion{Na}{1} line at 5688 \AA; this was because the resonance 
Na D lines at 5895 and 5889 \AA\ were too strong ($\sim$ 300 m\AA), 
and the line at 5682 \AA\ (used by Koch \etal 2008) was too weak.  
Only the most metal-poor stars in Carina have Na D lines that are 
weak enough to be used in this analysis, but since those stars were 
observed with Magellan/MIKE spectra, their lower SNR spectra required
spectrum syntheses for the line abundance determinations.
The \ion{Fe}{1} 5615 and 5634 lines were 
examined to check the broadening parameters 
(Gaussian broadening with FWHM = 0.25~\AA\ was adopted, which is
the effective resolution of the Magellan MIKE spectra).
When the Na D lines could not be fit simultaneously, then an average of
the best fit abundances for each line was adopted, 
and the range used to estimate 
the uncertainty.

\begin{figure}[t]
\includegraphics[width=90mm, height=50mm]{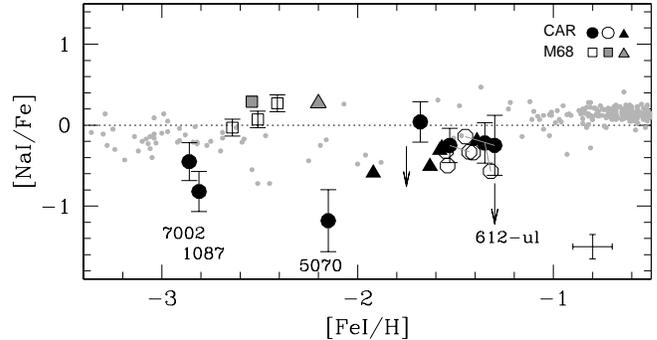}
\caption{Sodium abundance ratios of stars in Carina and M68,
compared to the Galactic distribution.   For the metal-poor
stars in Carina and M68, NLTE corrections have been applied 
to the Na D line abundances; the symbols are the same as in 
Fig.~\ref{osi}.     For the Galactic data, only a limited
amount of available data are shown; the results from 
\citep{And07} and Gehren \etal (2006) are included
since these two studies considered NLTE effects on the NaD
lines, and results from Reddy \etal (2003, 2006) are shown
since they only used the subordinate \ion{Na}{1} lines.
}
\label{na}
\end{figure}

\citep{And07} have examined the NLTE corrections for the 
Na D resonance lines in metal-poor RGB stars.   Using their Table~2,
combined with their Fig.~6 showing the effect of the correction on
the line equivalent widths, then we estimated a NLTE correction for 
our abundances.    For the M68 stars, the higher gravities and strong
line strengths imply corrections of $\sim -0.5$ dex.   For the three
metal-poor Carina stars (Car-1087, Car-5070, and Car-7002) then the
atmospheric parameters and line strengths suggest corrections of
$\sim -0.6$ dex.    These corrections
are included in Fig.~\ref{na}.    
Of course, no corrections are applied to the other Carina objects 
since we did not use the Na D lines for those.    

For Car-612, we only determined an upper limit, EW $<$ 30 m\AA\ for
\ion{Na}{1} $\lambda$5688, which corresponds to a very low upper limit 
of [Na/Fe] = $-0.8$.  This result is comparable to that from 
Koch \etal (2008) who measured an EW (= 24 m\AA) for the same weak 
line and determined [Na/Fe] = $-0.5$.  Our lower upper limit is due
to small differences in the atmospheric parameters and adopted solar 
Na abundance.

\subsubsection {Iron peak Elements }

In the early Universe, the iron-peak elements (Sc to Zn:
23 $\le$ Z $\le$ 30) are exclusively synthesized during 
Type II supernovae explosions by explosive oxygen and
neon burning, and complete and incomplete explosive Si
burning.     Abundances of these elements show a strong
odd-even effect (odd nuclei have lower abundances than
the even nuclei).  Their yields depend on the mass of
the progenitors (e.g., Woosley \& Weaver 1995), the 
mass cut adopted (mass expelled relative to the mass that 
falls back onto the remnant, e.g., Nakamura \etal  1999) 
and SNe~II explosive energies (e.g., Umeda \& Nomoto 2005). 
%
Only at later times ($\sim$ 1 Gyr, e.g., Maoz \etal  2010), 
when lower mass stars reach the end of their life time, 
do SNe~Ia become a significant, possibly
the dominant, contributor to the total iron-group inventory.
The onset of SNe~Ia in the chemical evolution of the Galaxy
is observed as a {\it knee} in the [$\alpha$/Fe] vs [Fe/H]
near [Fe/H] $\sim -1.0$ (e.g., McWilliam \etal  1995). At
lower metallicities, SNe~Ia [X/Fe] yields could vary
(e.g., Kobayashi \& Nomoto 2009). 
We have measured several of the Fe-peak elements 
in the Carina stars to compare with the Galactic trends.  

\begin{figure}[t]
\includegraphics[width=90mm, height=95mm]{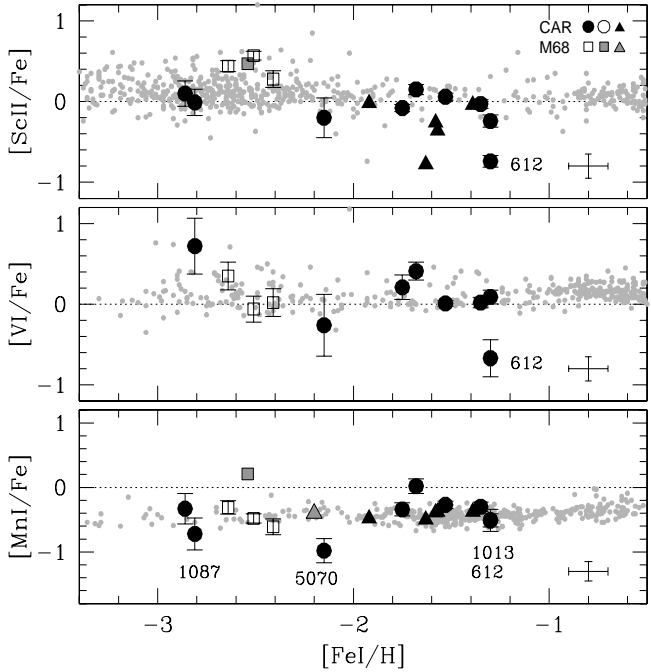}
\caption{
The odd-Z LTE abundance ratios for scandium, vanadium,
and manganese in Carina.  HFS corrections have been applied,
as well as a NLTE correction to the resonance lines of 
\ion{Mn}{1}.  Symbols are the same as in Fig.~\ref{osi}
with the exception of the Galactic standards for [Mn/Fe]
which are taken from Sobeck \etal (2006) and Cayrel \etal (2004).  \\
}
\label{scvmn}
\end{figure}

\begin{figure}[t]
\includegraphics[width=90mm, height=95mm]{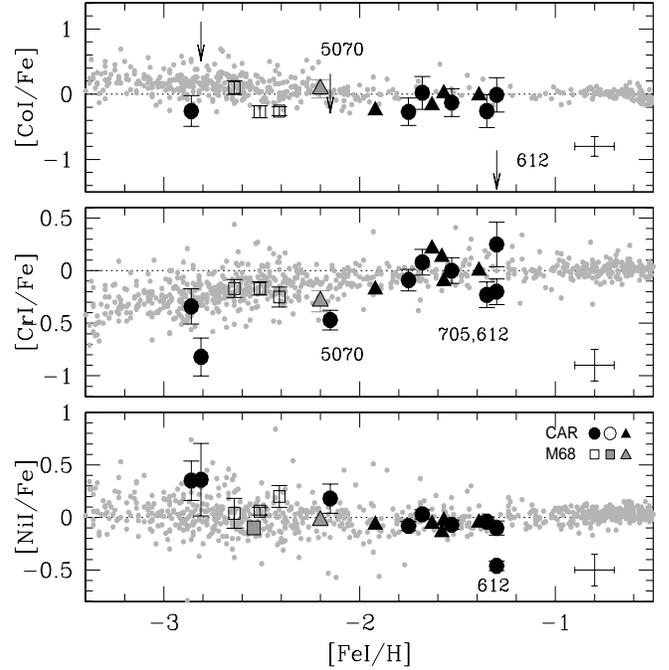}
\caption{
LTE abundance ratios for chromium, cobolt, and nickel in Carina.
The low abundances for Car-612, and low Cr in Car-5070, are
exceptional, like in Fig.~\ref{scvmn}.
Symbols are the same as in Fig.~\ref{osi}.  \\
 }
\label{crconi}
\end{figure}

{\it Scandium:} Several (4-13) lines of \ion{Sc}{2} were 
measured and HFS corrections applied.   The distribution 
in the Sc abundances at intermediate metallicities is larger 
than most of the other iron peak elements 
(see Fig.~\ref{scvmn} and \ref{crconi}), 
although this includes Car-612 which has several chemical 
peculiarities and stars analysed by Shetrone \etal (2003).   
Other than those objects, the Sc abundances are similar
to the Galactic distribution.

{\it Vanadium:} For most of the intermediate metallicity stars,
the V abundance is well determined from $\sim$10 \ion{V}{1} lines
and HFS corrections were applied.  However, in the metal-poor stars 
and Car-612, it is determined from only 1-2 lines.  

{\it Manganese:} The \ion{Mn}{1} abundances were determined from 
several lines (4-7) in the intermediate metallicity stars, and 
these line abundances showed good agreement with one another.   
Generally, [Mn/Fe] in Carina follows the Galactic distribution
above [Fe/H] = $-2$.   In the metal-poor Carina stars and M68 
stars, a different set of spectral lines was used (because
of the wavelength coverage and resolution of the Magellan/MIKE 
spectra).  The blue resonance lines at $\lambda$4030, 4033, and 
4034, as well as several additional subordinate lines were measured, 
with only   $\lambda$4823 in common between the MIKE/Magellan and 
FLAMES/UVES datasets.   The Mn results from the resonance lines 
were lower than from $\lambda$4823 and the other subordinate lines
in the M68 standard stars, most likely due to non-LTE effects.
Bergemann \& Gehren (2008) have calculated non-LTE corrections 
for \ion{Mn}{1} lines and do find large corrections for the blue 
resonance lines that are metallicity dependent.   They do not
give corrections for red giants, but for red main sequence stars
the corrections can be as large as +0.5 dex at [Fe/H] = $-3$.   
These corrections are similar to the offsets that we find between 
resonance and subordinate lines, therefore we apply a correction
of $\Delta$log(Mn) = +0.5 dex to the \ion{Mn}{1} resonance line
abundances $-$ this correction should be checked with proper modelling
of these metal-poor red giant atmospheres.   Correcting the resonance 
line abundances improves the mean [Mn/Fe] results, but the mean
values in Car-5070 and Car-1087 remain lower than for similar 
stars in the Galactic halo.



{\it Chromium:} Cr abundances were determined from 4-7
lines of \ion{Cr}{1}, which showed good agreement from line to line.
A recent calculation of the NLTE 
effects on the \ion{Cr}{1} lines by Bergemann \& Cescutti 
(2010) showed that the corrections are small ($\le 0.1$ dex), 
but that \ion{Cr}{1}/\ion{Cr}{2} ionization 
equilibrium and the solar [Cr/Fe] ratio is regained for metal 
poor stars, rather than the downward trend seen in Fig.~\ref{crconi}.  
Our Carina data generally follow the Galactic trend, with
the exception of very low abundances in 
Car-5070, Car-1087, Car-705, and Car-612.

{\it Cobalt:}  
Co was determined from only 1-2 lines in this analysis.
The [Co/Fe] ratios lie slightly below the Galactic 
abundances, although they are not well constrained from so
few lines.  The only exception is the very low upper limit 
to [Co/Fe] for Car-612 (see Fig.~\ref{crconi}).
\citet{Ber10} calculated NLTE corrections for 
\ion{Co}{1} and \ion{Co}{2}; they suggest the NLTE 
corrections depend on metallicity and can become as large as 
+0.6 to +0.8 dex at [Fe/H] $\sim -3.0$.   This would  not
affect the very low Co upper limit found for Car-612.

{\it Nickel:} 
Several \ion{Ni}{1} lines (13-17) were measured in the intermediate
metallicity stars over a range of wavelengths, showing that [Ni/Fe]
 is in agreement with the Galactic distribution (Fig.~\ref{crconi}).   
Only 1-3 \ion{Ni}{1} lines were available in the most metal-poor 
stars though, thus the uncertainties are large because $\sigma$(\ion{Fe}{1}) 
was adopted for the error estimates. 
Only the Ni result for Car-612 stands out, but this result is
robust, determined from 14 \ion{Ni}{1} lines with a small error
in the mean.   Combined with the low Na upper limit, then Car-612
fits the Na-Ni relationship observed for some $\alpha$-poor stars
in the Galaxy by Nissen \& Schuster (1997, 2010).

\subsubsection{Cu \& Zn}

While contiguous on the Periodic Table, the main nucleosynthetic
sites of copper and zinc are difficult to ascertain.
In massive stars, Cu and Zn form during complete Si-burning, 
the $\alpha$-rich freeze out, and even the weak s-process
(e.g., Timmes \etal 1995).   
In SN~Ia models, the yields of Cu and Zn are sensitive to the 
neutron excess and thus metallicity (Matteucci \etal 1993, 
Travaglio \etal  2005, Kobayashi \& Nomoto 2009).
The contributions
to the production of Cu and Zn in AGB stars are uncertain, though
recent calculations suggest small amounts of Cu and even smaller 
amounts of Zn can be produced in more massive (5 M$_\odot$) AGB stars
(Karakas \etal 2008, Karakas 2010).   Precise estimates of the AGB 
yields can also depend on uncertain parameters such as the mass loss 
law and number of dredge up episodes (Travaglio \etal 2004).


\begin{figure}[t]
\includegraphics[width=90mm, height=75mm]{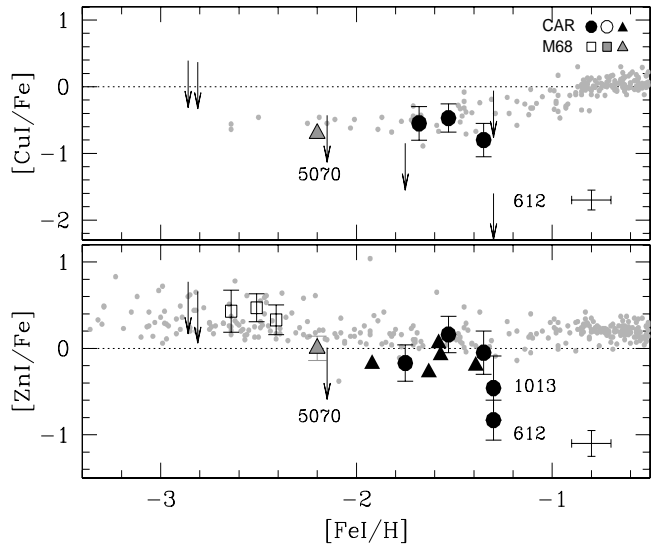}
\caption{The LTE copper and zinc abundances in the Carina stars.
The Cu and Zn abundances in Car-612 are remarkably low.
Symbols are the same as in Fig.~\ref{osi}.  \\
}
\label{cuzn}
\end{figure}

{\it Copper:}
The \ion{Cu}{1} line at 5105 \AA\ was observed in some of our target stars.
Upper limits were calculated adopting an EW of 30 m\AA\ for
the FLAMES/UVES spectra and 50 m\AA\ for the most metal-poor Carina stars.
Hyperfine structure corrections were applied.
Generally, copper follows the Galactic trend, and though we have very
few actual measurements, the upper limit for Car-612 is very low and
provides a very strong constraint.

{\it Zinc:}
The \ion{Zn}{1} line at $\lambda$4810 was observed in most of our 
sample, or used to determine the upper limits to the Zn abundance 
(adopting $<$40 m\AA); see Fig~\ref{cuzn}.    
The zinc abundances at intermediate metallicities were slightly lower 
than the Galactic distribution, especially Car-612.

\subsubsection {Neutron Capture Elements \label{heavy}}

Neutron-capture elements (Sr to U, 38 $\le$ Z $\le$ 92) originate
in the rapid neutron capture process (r-process) that occurs during 
explosive nucleosynthesis in SNe~II. 
The r-process has a {\it main} source in 
8-10 M$_\sun$ stars that form elements with Z$>$50, and 
a {\it weak} source in more massive 
progenitors ($>$20 M$_\sun$) that contribute primarily 
to the lighter neutron-capture elements (e.g., Sr, Y, Zr,
Travaglio \etal  2004).  A more detailed examination of
the r-process shows that a more accurate representation 
is in terms of a high entropy wind model 
(Farouqi \etal 2009, Roederer \etal 2010); 
nucleosynthesis in low entropy winds proceeds primarily
through a chared-particle ($\alpha$-) process, whereas
a neutron-capture component (the classical weak and main 
r-processes) occur in the high entropy winds.
Thermal pulsing in AGB stars also contributes to these elements
through the slow neutron capture process (s-process).
With intermediate masses (2-4 M$_\odot$), AGB stars have 
longer lifetimes than the sites of the r-process, and 
therefore in simple chemical evolution models AGB stars do 
not produce any of the heavy elements in the most 
metal-poor stars, but dominate the formation of some
of these elements at later times and higher metallicities.
In the Galaxy, this is seen as the {\it rise in the s-process} 
which begins near
[Fe/H] = $-2.5$ (McWilliam \etal  1998, \citealt{Bur00}, 
Francois \etal  2007);  this is in contrast to the knee in
[$\alpha$/Fe] due to contributions from SNe~Ia 
near [Fe/H] $= -1.0$. 
%
%
%
In the Sun, the r-process contributes (11\%, 15\%, 25\%, 28\%, 53\%, 
and 97\%) of (Sr, Ba, La, Y, Nd, and Eu) respectively \citep{Bur00}.
Thus Eu is critically important as a nearly pure r-process indicator.

\begin{figure}[t]
\includegraphics[width=90mm, height=95mm]{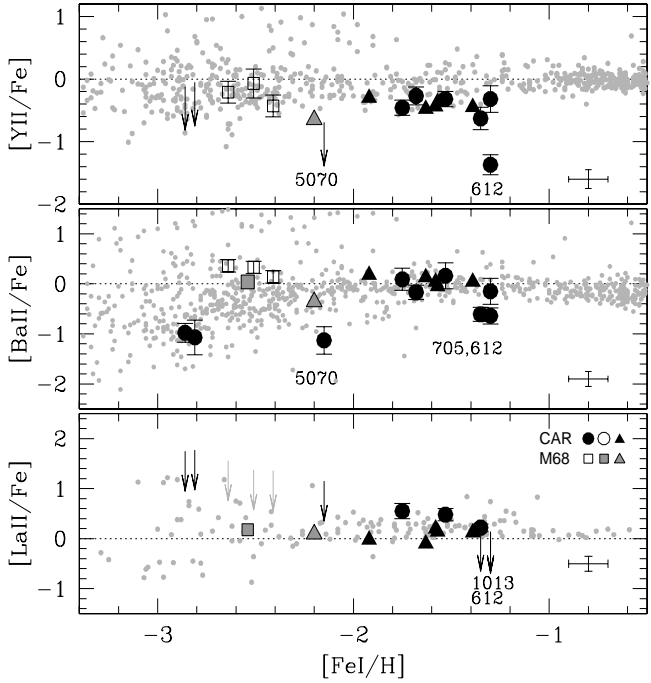}
\caption{The LTE abundances for the heavy 
elements Y, Ba, and La in Carina and M68, compared to the 
Galactic distribution. 
The Y abundance in Car-612 is remarkably low, but 
verified by a comparison of the spectra of Car-612 and
Car-705, e.g., in Fig.~\ref{y2}.
The Galactic star abundances for La are from the critically examined
compilation by Roederer \etal (2010).  
These ratios are again quite low for Car-612, and Ba in Car-5070.
Symbols are the same as in Fig.~\ref{osi}.
}
\label{ybala}
\end{figure}

\begin{figure}[h]
\includegraphics[width=90mm, height=50mm]{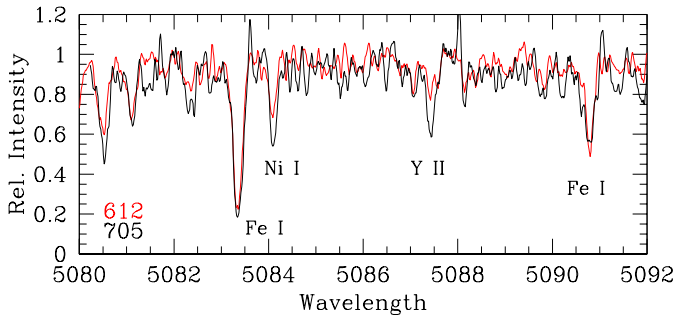}
\caption{
Spectra around the \ion{Y}{2} 5087 \AA\ lines in Car-612 and Car-705.  
Since the atmospheric parameters and metallicities of these two
stars are very similar, then this figure shows yttrium is truly weaker 
in Car-612.  Spectrum synthesis in this region confirms the different
\ion{Y}{2} abundances derived from the EW analyses.  The lower [Ni/Fe]
abundances in Car-612 can also be seen directly by 
comparing the 5084 \AA\ lines. \\
}
\label{y2}
\end{figure}

{\it Yttrium:}  
[Y/Fe] in the intermediate metallicity stars was determined from
\ion{Y}{2} 4883, 5087, and 5200, which gave very consistent abundances
from line to line and star to star.  
The \ion{Y}{2} 4900 line could not be used because it is severely 
blended in all of the spectra.  
The [Y/Fe] ratios are similar to those found by Shetrone \etal (2003), 
i.e., slightly below the Galactic values at intermediate metallicities,
see Fig.~\ref{ybala}.
One exception is Car-612 which has an extremely low [Y/Fe] result, and
another is Car-5070 which has a low upper limit value.
The low [Y/Fe] in Car-612 can be seen directly in 
comparison with Car-705, where Fig.~\ref{y2} which shows that the 
\ion{Fe}{1} line strengths are similar but the \ion{Y}{2} 5087 line 
is much weaker in Car-612.
Upper limits were determined in the most metal-poor stars
using 40 m\AA\ as the EW upper limit for the 
\ion{Y}{2} $\lambda$4883 line.

{\it Barium:}
Five lines of \ion{Ba}{2} were analysed 
(at 4554, 4934, 5853, 6141, and 6496 \AA), 
which yielded consistent abundances from line to line and star to star, 
although they were not all observed in one star.
Spectrum syntheses were used to confirm the abundances when 
the SNR was low.  Hyperfine structure corrections for three lines 
(5853, 6141, and 6496 \AA) are negligible ($\le0.02$).
NLTE corrections are also negligible ($\le0.03$, Short \& Hauschlidt 2006), 
with the exception of the $\lambda$4554 resonance line, however we have 
chosen to {\it not correct} that line since the estimated correction 
is not large ($\le0.15$ dex, Short \& Hauschlidt 2006, \citealt{And09}) 
and the LTE abundance from that line is consistent with the other \ion{Ba}{2} 
line results.
Most stars in Carina have the same [Ba/Fe] distribution as
stars in the Galaxy, with the exception of three stars,
Car-612, Car-5070, and Car-705; see Fig.~\ref{ybala}.

{\it Lanthanum:} La was determined from three lines of \ion{La}{2} at 
$\lambda$6320, 6390, 6774 (additional lines at $\lambda$4333, 5301, 5303 
could not be used due to the SNR of the spectra).  Negligible 
HFS ($\le 0.02$) were calculated for 
two La lines ($\lambda$6320 and 6390).  [La/Fe] follows the
Galactic distribution for the few stars where we could 
measure it (see Fig.~\ref{ybala}).

\begin{figure}[t]
\includegraphics[width=90mm, height=95mm]{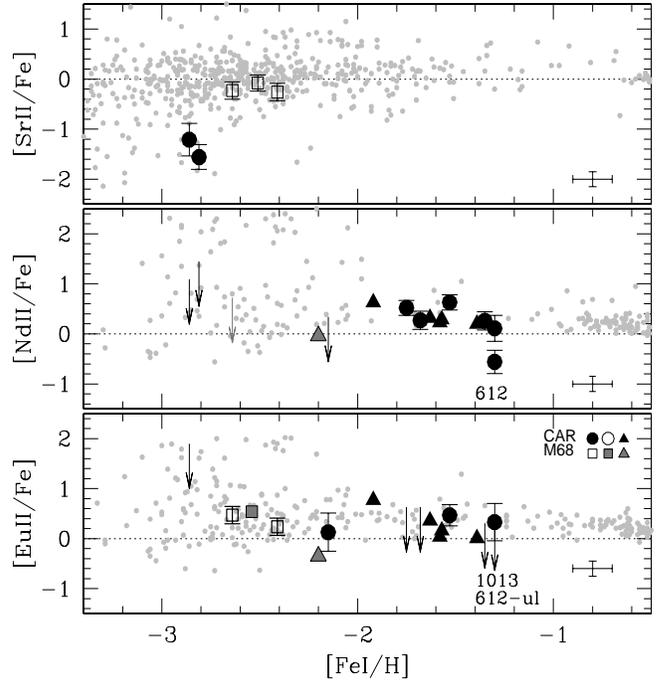}
\caption{LTE abundances for the heavy elements 
Sr, Nd, and Eu in Carina and M68, compared to the Galactic 
distribution.  
Symbols the same as in Fig.~\ref{osi}. \\
}
\label{srndeu}
\end{figure}

\begin{figure}[h]
\includegraphics[width=90mm, height=65mm]{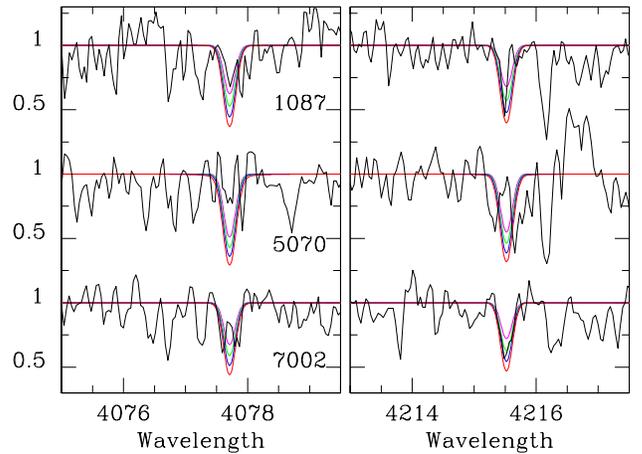}
\caption{
Spectrum syntheses of the \ion{Sr}{2} 4077 and 4215 lines in the 
three most metal-poor Carina stars, 
for [Sr/Fe] = $-0.5, -1.0, -1.5, and -2.0$.
The low SNR makes accurate Sr abundances very difficult.    
}
\label{srsyn}
\end{figure}

{\it Strontium:} 
[Sr/Fe] was determined from two \ion{Sr}{2} lines at very blue wavelengths
(4077 and 4215 \AA) reached only by our Magellan/MIKE spectra.   
The SNR is poor in that region, thus we adopted the results from 
spectrum syntheses for those lines in the three Carina stellar spectra, 
see Fig.~\ref{srsyn}.   Nevertheless, the interpretation of the synthetic
results remains difficult; the $\lambda$4077 line is the stronger of the
\ion{Sr}{2} resonance lines, and yet the detection of the $\lambda$4215 line 
is more clear in Car-7002, and possibly Car-1087. 
The best estimates for the Sr abundances
in these two stars are listed in Table~\ref{ngc}.
A Sr upper limit for Car-5070 could not be determined due to
noise near both the $\lambda$4215 and 4077 lines.
NLTE corrections were considered, but not applied to our results.
Short \& Hauschildt (2006) estimate abundance corrections of $-0.07$ dex, 
whereas \citep{And11} 
estimate corrections of $\sim 0.0$ and +0.1 dex for $\lambda$4077 and
4215, respectively.   We chose to not apply either correction 
since these estimates are small and in opposite directions.

{\it Zirconium:}
Upper limits for Zr were determined from the \ion{Zr}{2} $\lambda$4208
line in the Magellan spectra.
These upper limits are quite high, and do not add the discussion on
the heavy element abundances.

{\it Neodymium:}
Nd was determined from two \ion{Nd}{2} lines ($\lambda$5319 and 5249) for 
all stars with FLAMES/UVES spectra, and these lines were used to calculate
the upper limits from the Magellan spectra.  
Nd is consistent with the Galactic distribution, though it is very
low in Car-612 (Fig.~\ref{srndeu}).

{\it Europium:}
The europium abundance in the Sun is nearly entirely due to the
r-process, therefore Eu is identified as an important indicator 
in any stellar analysis to establish the ratio of r-process to 
s-process contributions of the neutron capture elements.
In this analysis, three \ion{Eu}{2} lines were examined
($\lambda$4129, 4205, \& 6645).      Unfortunately, all three 
lines were not observed in any one star;  the Magellan spectra 
have too low resolution to detect the weak 6645 \AA\ line, and
the FLAMES/UVES data do not cover the bluer wavelengths.     
Nevertheless, the Eu results are consistent 
with the Galactic distribution (see Fig~\ref{srndeu}).
Hyperfine structure and isotopic splitting
corrections are negligible ($<$0.05 dex).

\begin{figure}[t]
\includegraphics[width=90mm, height=90mm]{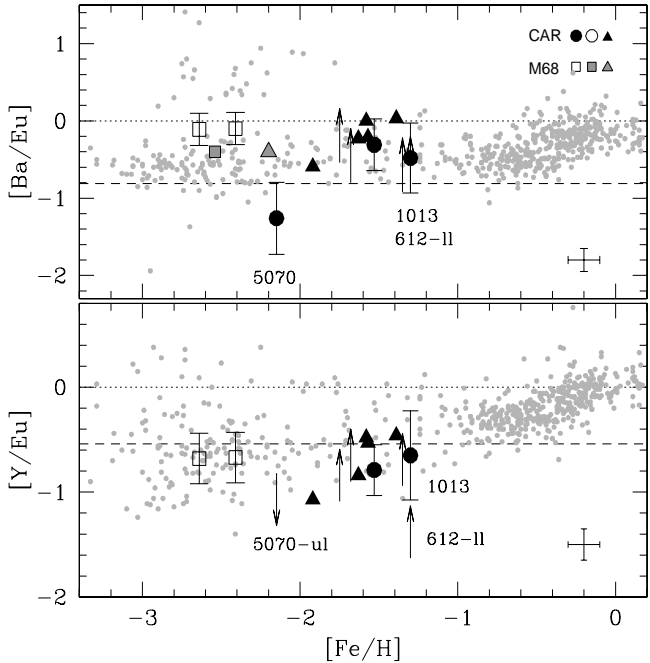}
\caption{Neutron capture ratios with Eu of stars in Carina and M68,
compared to the Galactic distribution.  The ratio with Eu permits
an assessment of the s-process contributions relative to the r-process
contents in the Sun (Burris \etal  2000, dashed line).   
The rise in the s-process can be seen in the Carina stars and the 
Galactic distributions.   
Upper and lower limits are shown for some stars. 
Symbols the same as in Fig.~\ref{osi}. \\
}
\label{xeu}
\end{figure}


\subsubsection {s-process to r-process Ratios: [X/Eu] \label{s2r}}

To assess the relative amounts of s-process to r-process 
contributions amongst the neutron capture elements, then abundance 
ratios relative to Eu are examined.    In Fig.~\ref{xeu},
[Y/Eu] and [Ba/Eu] of stars in Carina and the Galaxy are shown
and compared to the solar r-process fractions from 
Burris \etal (2000; also see \citealt{Arl99}.
%
In both [Y/Eu] and [Ba/Eu] (as well as [La/Eu] and [Nd/Eu],
not shown), the gradual rise in these ratios can be seen
with increasing metallicity.  McWilliam \etal (1998) showed
that the rise in the s-process starts at [Fe/H] = $-2.5$ in 
the Galaxy.  In Carina, this rise appears to begin at a
higher metallicity, [Fe/H] $\ge -2.0$.   This is most likely
due to the metallicity dependence of the AGB yields, as 
modelled by Travaglio \etal (2004, also Pignatari \etal  2010),
where fewer iron seed nuclei in a high neutron density wind can
collect more neutrons, thus underproducing the first s-process 
peak elements\footnote{The s-process peaks are defined by 
the neutron magic numbers for filling nuclear shells (e.g.,
N=50, 82, 126 are full nuclear shells).  A full nuclear shell
lowers  the cross section for further neutron captures, thus
elements collect at these neutron numbers defining the first,
second, and third s-process peaks.},
and overproducing the second and/or third s-process peak elements.
The slightly lower [Y/Eu] and slightly higher [Ba/Eu] ratios in 
Carina are consistent with metal-poor AGB stars contributions.

As a final note, the high Ba abundance in M68 is somewhat surprising.
This was also found by Lee \etal (2005).  Since [Ba/Eu] 
is high, but [Eu/Fe] is normal, 
this suggests M68 has had an unusual chemical evolution,
possibly having been enriched (or self-enriched) by metal-poor 
AGB stars.


\section{Discussion}

Detailed chemical abundances for up to 23 elements in nine stars 
in the Carina dSph have been presented in this paper.   Previously,
only five stars in this dSph had detailed abundance determinations
for so many elements (Shetrone \etal 2003), while another ten had
[$\alpha$/Fe] determinations from high resolution spectroscopy
(Koch \etal 2008).
In a companion paper by Lemasle \etal (2012), the VLT FLAMES/GIRAFFE 
spectra for 36 additional RGB stars are presented, but due to the 
limited wavelength coverage of those spectra, only four elements 
are analysed in detail (Mg, Ca, Ba, and Fe).
For the first time, Lemasle \etal (2012) calculated the ages of all 
stars with detailed chemical abundances in the Carina dSph, including
those in this paper, from isochrone fitting.   While the uncertainties 
in the ages can be quite large for any one star, the differential ages
should be better, and any discussion regarding ages is therefore 
limited to two
age bins, an {\it old} population and an {\it intermediate-aged} 
population.  These ages (two age bins) are adopted in this discussion.

In the previous Section, our results were compared to stars in the 
Galaxy, where some chemical peculiarities were noted.   In this 
Section, we will compare our abundance results to those of stars 
in other dwarf galaxies. 
Three dSphs have been chosen for comparison: Sculptor, Fornax, and Sextans.
The datasets for the three dSphs are taken from Shetrone \etal (2001, 2003), 
Geisler \etal (2005), \citet{Aoki09}, Letarte \etal (2006, 2010), 
Tafelmeyer \etal (2010), and Frebel \etal (2010c),
all scaled to the \citet{Asp09} solar abundances.
These three dSphs were chosen since large datasets of detailed abundances 
from high resolution spectroscopy are available, and the analyses of
those spectra used similar methods and model atmospheres to our analysis.
We do not include results from Kirby \etal (2009, 2010, 2011) in 
these comparisons to avoid systematic differences due to the lower 
resolution of their spectra and differences in the analysis methods.
Also, we do not include higher mass dwarfs (e.g., the LMC, SMC, or Sgr) 
because the metallicity range of the stars in those galaxies do not 
overlap well with the stars in the Carina dSph.  

We also compare several elemental abundances of the most metal-poor stars
in Carina to stars in the ultra faint dwarf galaxies (UFDs) and the
metal-poor stars in the Draco dSph.   
Datasets for the UFDs are from the compilation by Frebel (2010b), and
for Draco are from Cohen \etal (2009), Fulbright (2004), and Shetrone \etal (2001).
These comparisons are valuable because Carina has a tiny dynamical mass within its 
half light radius (M$_{1/2}$), e.g.,  
M$_{1/2}$ (Carina) = 6.1 $\pm$2.3 $\times$ 10$^6$ M$_\odot$ 
(Walker \etal 2009),
which is a factor of 2, 4, and 9 times smaller than
Sculptor, Sextans, and Fornax, respectively, 
but is comparable to those of the more massive UFDs.
%
As an example, 
B\"ootes~I has M$_{1/2}$ = 5.9 $\pm$3.7 $\times$ 10$^6$ M$_\odot$,
which is essentially the same as Carina, 
although the masses of the UFDs continue to be revised downwards, 
e.g., the mass of Bo\"otes I has been revised downwards by a factor of 
14 due to the lower velocity dispersions found by Koposov \etal (2011).  
Of course, Carina is not considered an UFD because of its luminosity 
- Carina is 8x brighter than Bo\"otes I (Walker \etal 2009).

\subsection {The Metallicity Distribution of the High Resolution Spectroscopic Sample}

As shown in Fig.~4 of Lemasle \etal (2012), the stars that have been
analysed with high resolution spectroscopy are biased towards metallicities 
near [Fe/H] = $-1.5 \pm0.3$.   This is the same as the mean metallicity 
and metallicity range predicted for the dominant intermediate-aged population 
by \citet{Bon10} from their CMD analysis; however this is not 
scientifically significant for two reasons.   
Firstly, more than half of the high resolution sample comes from the 
FLAMES/GIRAFFE spectral analysis presented by Lemasle \etal (2012), and as
discussed in that paper, the low SNR of those spectra meant that 
the weaker lines of the more metal-poor stars were harder to
detect.   This effectively removed those stars from the analysis.  
Secondly, although the V magnitude at the tip of the RGB may be slightly 
brighter with higher metallicities, this is only a small effect 
and maximizing the fiber placement was a more significant concern.  
Also, while it is true that the three stars targeted for Magellan/MIKE observations
are located in the outer fields of Carina and are the most metal-poor stars
sampled, this does not imply a population gradient in Carina as those stars 
were purposely selected for their low metallicities.
Koch \etal (2006) found no significant difference in the mean metallicities
of RGB stars in the inner and outer fields of Carina.
Thus, the results presented in this paper are not used to constrain the 
distributions in location or metallicity of the stars in Carina.

\subsection{ Dispersions in [$\alpha$/Fe] \label{dmgca}}

\begin{figure}[t]
\includegraphics[width=90mm, height=90mm]{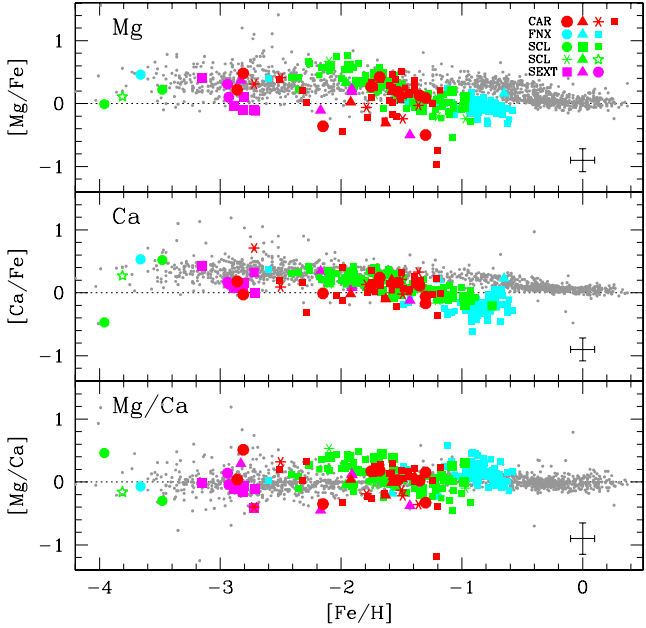}
\caption{
Mg and Ca abundances of stars in Carina (red),
Sculptor (green), Fornax (cyan), and Sextans (magenta).   
Symbols for Carina: 
{\it solid circles} are from this paper, 
{\it squares} are from Lemasle \etal (2012),
{\it triangles} are from Shetrone \etal (2003), 
and {\it asterisks} are from Koch \etal (2008). 
Symbols for Fornax: 
{\it circles} are from Tafelmeyer \etal (2010), 
{\it triangles} are from Shetrone \etal (2003), 
{\it squares} are from Letarte \etal (2010).
Symbols for Sculptor: 
{\it large/small squares} are UVES/GIRAFFE FLAMES data from Hill \etal (2012), 
{\it solid circles} are from Tafelmeyer \etal (2010),
{\it asterisk} are from Giesler \etal (2005), 
{\it triangles} are from Shetrone \etal (2003), 
and the {\it star} is from Frebel \etal (2010c). 
Symbols for Sextans:
{\it squares} are from \citet{Aoki09},
 {\it triangles } are from Shetrone \etal (2001),
and {\it circles} are from Tafelmeyer \etal (2010).
Tafelmeyer \etal (2010) note that the very low [Ca/Fe]
ratio reported for the most metal-poor dSph star 
(at [Fe/H] $\sim -4$) is likely due to NLTE effects 
in the formation of the strong resonance \ion{Ca}{1} 4227\AA\
line, a line that was not used throughout the rest of 
their analysis, and therefore should be regarded with
caution. 
Representative error bars of $\Delta$[Fe/H] = $\pm$0.1 and
$\Delta$[X/Fe] = $\pm$0.15 are shown.  \\
}
\label{dmgca2012}
\end{figure}

In Fig.~\ref{dmgca2012}, we show [Mg/Fe] and [Ca/Fe] for Carina, compared to 
the three other dSphs and the Galaxy, over the full metallicity range examined.   
Looking at [Ca/Fe] alone suggests that the chemical evolution of Carina has been 
similar to Sculptor.  In Sculptor, a noticeable downward trend 
in these abundances begins at [Fe/H] $\sim -1.8$, whereas in Galactic
stars this occurs at a higher metallicity.   This knee in the [$\alpha$/Fe] 
ratios is usually interpreted as the onset of contributions to [Fe/H] from SN~Ia, 
with the shift in the knee to lower metallicities in dwarf galaxies 
attributed to their slower chemical evolution (e.g., Lanfranchi \etal 2008,
Kirby \etal 2011).  
In Carina, it is not clear where or if there is a knee.

The more remarkable result seen in Fig.~\ref{dmgca2012} 
is the dispersion in [Mg/Fe] ($\Delta$[Mg/Fe]) in Carina, which
is observed in both the FLAMES/UVES and FLAMES/GIRAFFE data, and is much larger 
than the dispersion in [Ca/Fe] ($\Delta$[Ca/Fe]).  
A calculation of the intrinsic spreads\footnote{We have estimated the intrinsic
spread (N$_i$) from the formula N$_i^2$ = N$^2$ - $\langle\sigma\rangle^2$, where
N is the range in [X/Fe] and $\langle\sigma\rangle$ the average error in [X/Fe].} 
suggests that this difference is real and signficant when considering the whole 
dataset, the data from Lemasle \etal (2012) only, 
our nine UVES targets alone,
or the data at a specific intermediate metallicity (such as [Fe/H]$=-1.2\pm0.1$),
i.e., 0.4 $\le$ [N$_i$([Mg/Fe])$-$N$_i$([Ca/Fe])] $\le$ 0.7. 
Only for [Fe/H] $< -2.0$ are the intrinsic spreads between [Mg/Fe] and [Ca/Fe] similar.

Differences in the dispersions of [Mg/Fe] and [Ca/Fe] may be partially due to 
differences in their nucleosynthetic sites, but also may be the result of 
inhomogeneous mixing of the interstellar gas and therefore poor statistical 
sampling of the SN contributions when forming stars.   In terms of nucleosynthesis, 
SN yields of Ca and especially Mg depend on the progenitor mass
(e.g., Woosley \& Weaver 1995, Iwamoto \etal 1999); for example,  
Mg forms in hydrostatic core C and O burning, 
whereas Ca has contributions from the $\alpha$-rich 
freeze out and explosive Si burning during the SN~II explosion
Differences in the SN~Ia models can also lead to differences 
in the [Si-Ca/Fe] ratios, e.g., the central density, 
metallicity, ignition source, flame speed, and even type of SN~Ia
model can play a role
(e.g., Maeda \etal 2010, R\"opke \etal  2006, Iwamoto \etal  1999).  
In terms of inhomogeneous mixing, models by Revaz \& Jablonka 
(2011, also Revaz \etal 2009) predict a large spread in the [Mg/Fe] 
ratios in low mass dwarf galaxies due to longer gas cooling times 
and subsequently longer mixing timescales for the interstellar medium.
As an example, the hot gas from SNe~II can be subject to buoyant forces 
requiring up to 2 Gyr to cool and mix through a low mass galaxy.   
This hot gas can also quench star formation.   Therefore, it is possible
to find models of low mass dwarf galaxies that predict both a high dispersion 
in element ratios and an episodic star formation history, like Carina.
 
Surprisingly, the [Mg/Ca] ratios in Carina
do not show a larger dispersion then the other dSphs in Fig.~\ref{dmgca2012}. 
This may imply that inhomogeneous mixing plays the dominant role 
(over differences in nucleosynthetic sites). 
The one star with the extremely low [Mg/Ca] ratio is Car-743, 
analysed by Lemasle \etal (2012) from their lowest SNR spectrum; 
the formal uncertainty in that one result is $\sim2$x larger
than the representative errorbar shown, and should be considered with caution.
The one star with the highest [Mg/Ca] ratio is Car-1087, 
analysed in this paper; with a metallicity of [Fe/H] $=-2.9$ and 
high [Mg/Ca] ratio then this star is very similar to the unusual 
star Draco-119 (discussed below).

\begin{figure}[t]
\includegraphics[width=90mm, height=90mm]{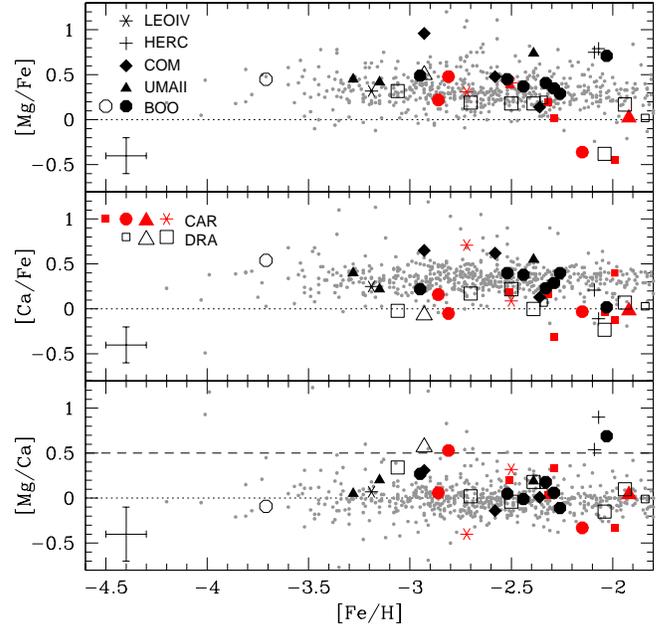}
\caption{[Mg/Fe] and [Ca/Fe] in the most metal-poor stars in
the Galaxy, Carina, Draco, and UFD galaxies.
The data for Carina are the same as in Fig.~\ref{dmgca2012}, 
while for Draco, the small/large squares are from 
Shetrone \etal (2001)/Cohen \etal (2009). 
The large empty triangle is Draco-119, 
the metal-poor star found by Shetrone \etal (2001) 
and reanalysed with higher SNR spectra by Fulbright (2004). 
The UFDs are presented as follows:
{\it solid/empty circles} are for B\"ootes~I from 
Feltzing \etal (2009)/Norris \etal (2010), 
{\it solid triangles} are for Ursa Major II from Frebel \etal (2010a),
{\it solid diamonds} are for Com Ber from Frebel \etal (2010a),
{\it plus signs} are for Hercules from Koch \etal (2008b),
and the {\it asterisk} for one star in Leo IV by Simon \etal (2010).
The dashed line in the [Mg/Ca] panel represents stars with the
more extreme enhancements (3 in UFDs, 2 in dSphs).  
\\
}
\label{dufdmgca2012}
\end{figure}

\subsubsection{ Metal-poor Stars and the UFDs }

In Fig.~\ref{dufdmgca2012}, we show only the metal-poor tail
in the Carina abundances and compare the [Mg/Fe] and [Ca/Fe] ratios to those
of metal-poor stars in five UFD galaxies and the Draco dSph.   
Most of the UFD stars have high values of [Mg/Fe] and [Ca/Fe], 
like the Galactic halo, whereas Carina and Draco show a range 
extending to very low values.    

Some stars in all of these systems show very high [Mg/Ca] values ($>0.5$) 
e.g., two stars in the Hercules UFD (Koch \etal 2008), one in B\"ootes I 
(Feltzing \etal 2009), one in Draco (Fulbright 2004, Shetrone \etal 2001),
and one in Carina (this paper).
In the UFDs, it has been proposed that high [Mg/Ca] may be due to the chemical
enrichment of its interstellar gas by as few as one SN~II 
explosion from a massive progenitor (e.g., a $>$35 M$_\sun$ star).
In this scenario, a unique chemical signature can be imprinted onto the gas
that is used to form stars, while the rest of the gas is expelled to quench further 
star formation (e.g., Koch et al 2008b, Frebel \etal  2010a, Simon \etal  2010).
This scenario is aided by peculiar neutron capture ratios, e.g., in the Hercules
stars the \ion{Ba}{2} lines are not detected (Koch \etal 2008).  Similarly, 
the \ion{Ba}{2} (and \ion{Sr}{2}) lines in the high [Mg/Ca] star in Draco 
(Draco-119) are also not detected (Fulbright \etal 2004 suggested that this star 
was enriched by a SN~II in the mass range of  $20 - 25$ M$_\odot$, 
and is lacking in contributions from higher mass progenitors).
However, \ion{Ba}{2} (and \ion{Sr}{2}) lines are detected in the high [Mg/Ca]
star in Carina.   Those lines are weak, leading to very low abundances 
of [Ba/Fe] and [Sr/Fe], but if the source of these elements is the same as
in the other dSphs, then Carina appears to have been able to retain a bit more 
of those early (high mass progenitor) SN~II enrichments.   


\subsubsection{ Measurement Errors? }

One question worth considering is whether the large dispersion in [Mg/Fe] is
simply due to measurement errors.    
For example, there are only 1-2 \ion{Mg}{1} lines available in each star, 
whereas there are 3-21 lines of \ion{Ca}{1} (depending on the metallicity) in this analysis. 
Similarly, due to the limited wavelength coverage of the FLAMES/GIRAFFE spectra, 
there are $\le$13 \ion{Ca}{1} lines measured by Lemasle \etal (2012), and only 
one \ion{Mg}{1} line at $\lambda$5528.    A larger number of spectral lines reduces 
the random measurement errors in an elemental abundance (by $\sqrt{N}$).
This is reflected in the smaller average error in [Ca/Fe] than [Mg/Fe] in
Table~10 of Lemasle \etal (2012).
In this paper, even if the measurement error from fitting a 
line is small, as in a high SNR spectrum, the analysis of a single line 
means that the adopted error is the standard deviation in the \ion{Fe}{1} 
line abundance ($\sigma$(Fe); see Section~\ref{abuerr}).
These (conservative) errors are about as large as the [Mg/Ca] dispersion,
but smaller than the [Mg/Fe] dispersion, therefore the large dispersion 
in [Mg/Fe] does not appear to be due to measurement errors alone.
 
Another consideration is that variations in the [Mg/Fe] and [Mg/Ca] ratios 
have been found in a number of metal-poor stars in the ultra faint dwarf galaxies, 
and attributed to incomplete mixing and poor sampling of the full mass function,
as shown in Fig.~\ref{dufdmgca2012}.  Only the largest outliers are examined in
terms of pollution by a single SN~II, and similarly large outliers are found in
our [Mg/Ca] and [Mg/Fe] data as well.
The consistency between these analyses suggests that the signatures in the
largest outliers are not due only to measurement errors.

\subsection {Age Considerations}

The Carina dSph is an interesting galaxy partially because of its low mass and
partially because of its episodic star formation history.    Lemasle \etal (2012)
were the first to attempt to interpret the chemical evolution of this dwarf galaxy
after separating the stars into two populations, old ($>$ 10 Gyr) and 
intermediate-aged ($<$ 6 Gyr); 
stars with ages between 6 and 10 Gyr were examined
by Lemasle \etal 2012, but we do not include those in this discussion.   
The old population shows a large range in 
[Fe/H] and [Mg/Fe], whereas the intermediate-aged (IA) population appears to 
have a very small range in [Fe/H] and [Mg/Fe].     Lemasle \etal (2012) were 
also struck by the significant overlap in [Fe/H] between these populations,
which would imply that the IA stars formed from gas that was more metal-poor 
than the end-point metallicity of the old population.     Combined with
the higher mean [Mg/Fe] abundance of the IA population, Lemasle \etal (2012)
suggested that the second epoch of star formation in Carina may have 
occurred after the late accretion of metal-poor, $\alpha$-element rich gas.

\begin{figure}[t]
\begin{center}
\includegraphics[width=90mm, height=90mm]{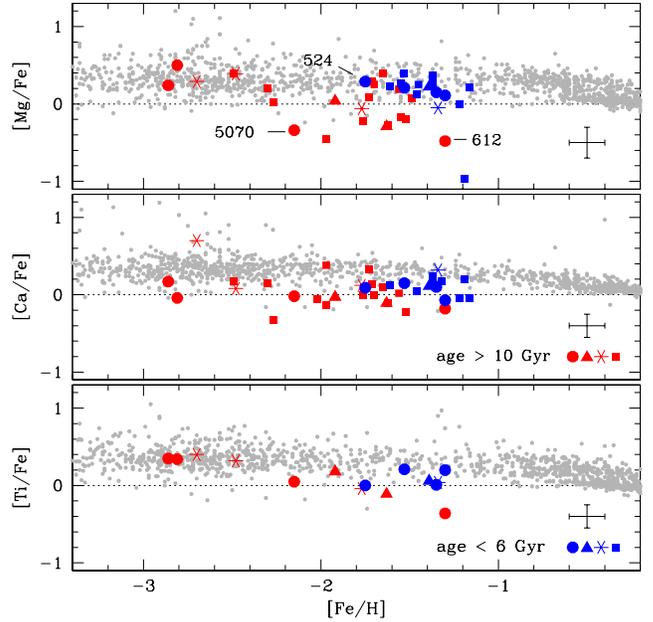}
\caption{[$\alpha$/Fe] ratios of stars in the Carina dSph
separated into age groups.  
 Symbols the same as in Fig.~\ref{dmgca2012}, 
though now red is used for the old population 
and blue for the IA population; 
ages are from Lemasle \etal (2012).
Galactic comparison stars are shown as small grey dots (including
data from Venn \etal 2004, Frebel \etal 2010b, and
Reddy \etal 2003, 2006).   Representative errorbars are shown
based on the mean error in the Carina dataset.
}
\label{ages1}
\end{center}
\end{figure}

The overlap in the metallicities of the old and IA population can be seen
in Figs.~\ref{ages1} to \ref{ages2}.    We note that the overlap 
is heavily weighted by the age assigned to Car-612, which is the chemically 
peculiar star discussed in Section~4.  Differences in the specific chemistry 
of stars are known to affect their isochrone ages (Dotter \etal 2007), 
and thus the age assignment to this star should be considered with caution.
The overlap is also heavily weighted by Car-524, which has a large 
uncertainty in its age assignment.  If the ages for these two stars are neglected, 
then there is a sharp transition at $-1.4 <$ [Fe/H] $< -1.6$ between
the age groups, i.e., equivalent to the measurement errors in metallicity.  
Thus, the overlap in metallicity between the two age groups is not sufficiently 
clear to indicate an infall of metal-poor gas to form the second generation of stars.

\subsubsection{[$\alpha$/Fe] Ratios Between the Age Groups}

An offset in the mean [Mg/Fe] abundance in the IA population relative to the
old population can clearly be seen in Fig.~\ref{ages1}. 
A Kolmogorov-Smirnov (K-S) test shows that the distribution of the [Mg/Fe] 
values for the old population is broader and with lower mean values, i.e.,
<[Mg/Fe]>$_{\rm old}$ = 0.05 $\pm$0.32 
and <[Mg/Fe]>$_{\rm IA}$ = 0.22 $\pm$0.15, both with moderately high 
probabilities of having ``normal'' distributions once the outlier in the 
IA dataset (at [Mg/Fe] = -0.95) is removed. 
This was also seen by Lemasle \etal (2012) for both [Mg/Fe] and [Ca/Fe],
although the offset in [Ca/Fe] is not as large, which we confirm with a
K-S test. 
We also note that these offsets in the [Mg/Fe] and [Ca/Fe] ratios with 
age are still present when only the high resolution data are examined 
(this paper, Koch \etal 2008, and Shetrone \etal 2003), although they are
not as clear.
The same signature is also hinted at in the [Ti/Fe] ratios 
(from the high resolution data), but there is insufficient data in 
the IA population for a meaningful K-S test. 
Since the large disperions in the [$\alpha$/Fe] ratios discussed above
(larger than seen in most dSph galaxies) are an indication of inhomogeneous
mixing, the simplest explanation for the offset in [$\alpha$/Fe]
between the old and IA population is that the second epoch of star formation
occurred in $\alpha$-enriched gas.   The small range in the [$\alpha$/Fe] 
and [Fe/H] ratios suggests this gas was well mixed.

\subsubsection {The Rise in the s-Process}

\begin{figure}[t]
\begin{center}
\includegraphics[width=90mm, height=90mm]{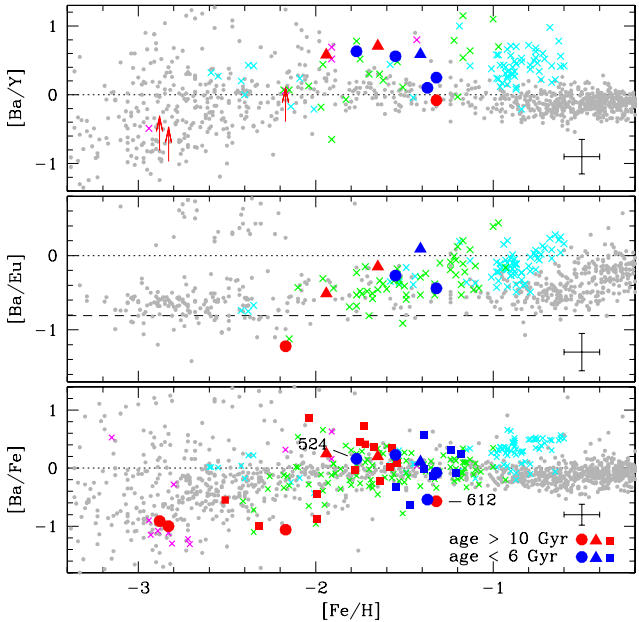}
\caption{The neutron capture abundance ratios of stars in the Carina 
dSph separated into age groups.  Data references and symbols for Carina 
and the Galaxy are the same as in Figs.~\ref{ages1}. 
Data for the other dwarf galaxies are noted only by colour
(cyan for Fornax, green for Sculptor, magenta for Sextans) 
with references as in Figs.~\ref{dmgca2012}. \\
}
\label{ages3}
\end{center}
\end{figure}

\begin{figure}[t]
\begin{center}
\includegraphics[width=90mm, height=90mm]{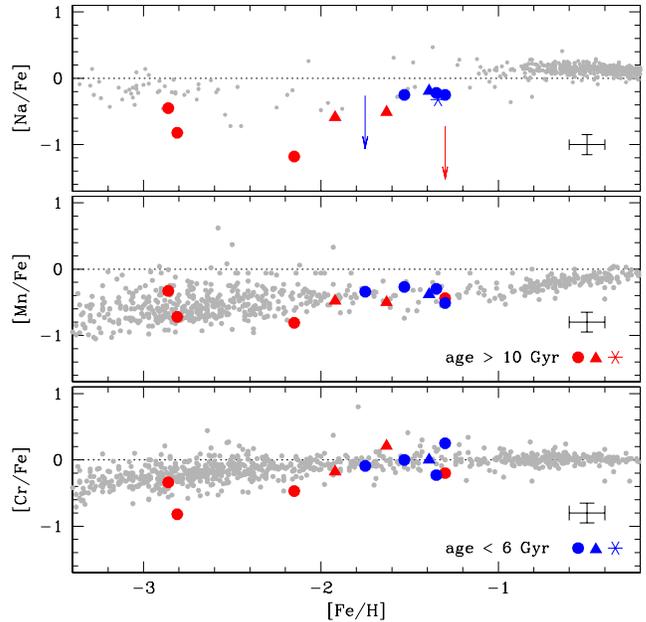}
\caption{Abundance ratios of [Na/Fe], [Mn/Fe], and [Cr/Fe]
for stars in the Carina dSph separated into age groups. 
Data references and symbols are the same as in 
Fig.~\ref{ages1}, with the exception of Na where 
Galactic data are from \citep{And07},
Gehren \etal (2006), Edvardsson \etal (1993), and
Reddy \etal (2003, 2006), and also Mn where the
Galactic data are from Sobeck \etal (2006) and
Cayrel \etal (2004). \\
}
\label{ages2}
\end{center}
\end{figure}

Clearly the old and IA stellar populations in Carina show different chemical
signatures, and it is interesting to investigate how the two populations
might be connected.  While we have discussed the $\alpha$-elements, 
possibly the most valuable abundance ratios for studying the chemical
evolution of a dwarf galaxy are of the neutron capture elements.

In Fig.~\ref{ages3}, it is possible to see that AGB stars have contributed 
to both age groups in Carina, but not until [Fe/H] $> -2$.    
This can be seen by the [Ba/Eu] ratios, where the lower dashed line represents 
the r-process contributions (in the Sun), such that stars with this ratio
have been enriched by SNe~II products only.  
Above [Fe/H] $= -2$, the [Ba/Eu] ratios slowly increase from the low
r-process value through contributions to Ba from the s-process 
(the {\it rise} in the s-process).   This is seen in the Galaxy,
Fornax, and Sculptor as well, although the rise begins at higher 
metallicities in those systems.
From the [Ba/Fe] ratios, it is also possible to see how the 
AGB contributions begin to dominate the nucleosynthesis of Ba, 
e.g., the dispersion in [Ba/Fe] at low metallicities in the Galaxy
is interpreted as the stochastic sampling of the (small amounts of)
Ba from SNe~II products early on, 
but this scatter lessens when the dominant contributions 
from AGB stars come at later times.   The flatness of the relationship 
in [Ba/Fe] at higher metallicities in the Galaxy implies that 
the timescale and yields of Ba from the AGB stars and Fe from 
SNe~Ia are similar, and the small scatter implies that the 
gas is well mixed.
In Carina, [Ba/Fe] in the IA population may have a large scatter.
If the large scatter is real, then this would suggest that the 
AGB contributions were not well mixed in the ISM at any age,
which would be an interesting contrast to the uniformity in
the [$\alpha$/Fe] ratios in the IA population.   However, several
of these data points are from Lemasle etal (2012) where the 
measurements of a single \ion{Ba}{2} line were very difficult 
due to the low SNR in that spectral region.    The data from our
high resolution analysis is far more uniform above [Fe/H] = $-2$, 
as shown in Fig.~\ref{ybala}.
 
Below [Fe/H] = $-2$ it is difficult to ascertain if there are any 
stars with AGB contributions.   There is one star (Car-5070) 
with [Ba/Eu] below the r-process ratio, implying that this star 
shows no signs of AGB products; however Carina was poorly mixed 
at early times, and metal-poor AGB stars could produce very 
little Ba.   The high [Ba/Y] ratios in stars over 
[Fe/H] = $-2$ show that most of the s-process elements came from 
metal-poor AGB stars, where the first s-process peak (Y) was 
bypassed in favour of the second s-process peak (Ba; as 
described further in Section~\ref{s2r}) $-$ at even lower 
metallicities, both can be bypassed for the third s-process peak.  
Therefore we cannot ascertain the precise metallicity when AGB stars
began to contribute in Carina, but certainly there are contributions
from metal-poor AGB stars in the stars with [Fe/H] $> -2$, and a hint
of the rise in the s-process between -2.0 $<$ [Fe/H] $< -1.6$. 

Further evidence for the rise in the s-process can be seen in
the evolution in the [Na/Fe]; see Fig.~\ref{ages2}.   
Cayrel \etal (2004) and \citep{And07}
showed that the [Na/Mg] and [Na/Fe] ratios are flat and low in metal-poor
stars in the Galaxy, suggesting that Na is produced with the $\alpha$-elements
in SNe~II at a (low) fixed ratio.  The Na abundance rises when AGB stars 
begin to contribute (see Section~\ref{sodium}).   
We see a very similar trend in the [Na/Fe] ratios
in Carina; in fact the rise in [Na/Fe] is very similar to the rise in
the s-process elements.    One significant difference though is that
the initial [Na/Fe] value appears to be much lower in Carina than in 
the Galaxy.    The final [Na/Fe] ratios may be similar in Carina 
as in the Galaxy (and Sculptor), and we see no offset in the Na
abundances between the old and IA populations.    A dispersion in
Na is not clear in this small sample.
 
It is interesting that the evolution of [Mn/Fe] and [Cr/Fe] are also 
similar to Na in Carina.   The abundances of these elements are much 
lower than in the Galaxy at low metallicities, but then rise above 
[Fe/H] = $-2.0$ to abundances that are similar to stars in the Galaxy 
with the same metallicity.  There is also no difference in these 
elements between the old and IA populations.    Like Na, once AGB stars
and SNe~Ia begin to produce iron-group elements, those
contributions will dominate over the initial (low) SNe~II yields. 
The timescale for AGB contributions is thought to be equal to or 
shorter than that of SNe~Ia ($\sim$ 1~Gyr, Maoz \etal 2010), thus it is
likely that both AGB stars and SNe~Ia begin to contribute to the ISM in
Carina near [Fe/H] $= -2.0$.   The SNe~Ia contributions appear to
be well mixed in the ISM, and there is no offset between the old
and IA stars.


\subsubsection {Early Chemical Evolution by SNe~II}

The earliest stages of chemical evolution in Carina can be examined
from the element ratios of the stars below [Fe/H] = $-2$.  As discussed above,
these stars appear to have been enriched only in SNe~II products.   While
there is a large range in the [$\alpha$/Fe] and [Mg/Ca] ratios 
 (when the whole dataset is examined, including previously published 
results) , 
the ratios of Mn, Cr, and Na all start out very low but then rise to the 
Galactic values at intermediate metallicities.   

\begin{figure}[t]
\includegraphics[width=90mm, height=90mm]{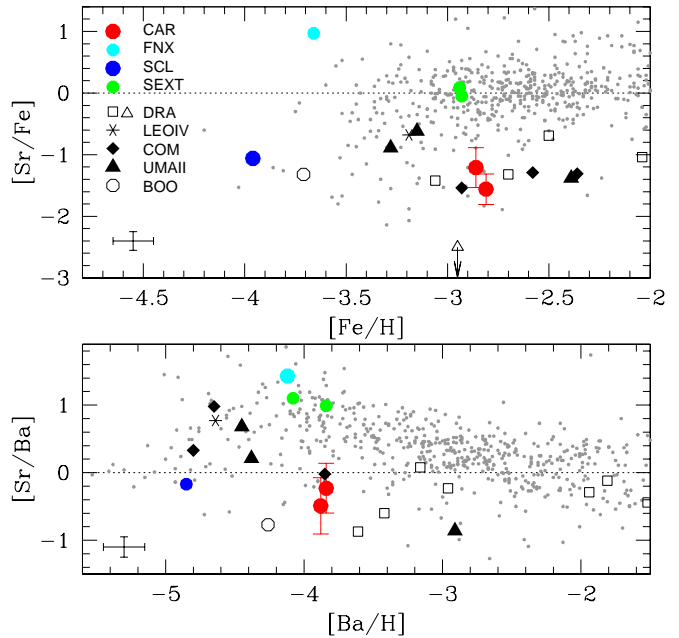}
\caption{[Sr/Fe] and [Sr/Ba] in low metallicity stars in the Galaxy,
classical dSphs, and UFDs.  The [Sr/Ba] ratio is enhanced in metal-poor 
stars in the Galaxy, which is not always seen in the dwarf galaxies. 
The UFD symbols are the same as in Fig.~\ref{dufdmgca2012}, while
the dSph data is from this paper and Tafelmeyer \etal (2010).
}
\label{dufdsrba}
\end{figure}

In addition, the [Sr/Fe] values are amongst the lowest of all stars 
analysed in the Galaxy, other dSphs, and the UFDs (see Fig.~\ref{dufdsrba}).
The [Sr/Fe] values for Car-1087 and Car-7002 are similar to the lowest
values found for metal-poor stars in Draco, Bo\"otes I, and Com Ber.
Tafelmeyer \etal (2010) suggested that dwarf galaxies may have a lower floor
in the [Sr/Fe] ratio ($\sim -1.2$) when their mass is equal to or lower than Draco.
The low [Sr/Fe] ratios in Carina lead to very low [Sr/Ba] ratios as well.
[Sr/Ba] ratios are quite interesting because they tend to be high in metal-poor 
stars in the Galactic halo, which has been interpreted as evidence for an extra 
nucleosynthetic source in the formation of these elements in massive stars
(Travaglio \etal  2004, Ishimaru \etal  2005, 
Qian \& Wasserburg 2008, Farouqi \etal 2009, Pignatari \etal  2010).

The absence of a [Sr/Ba] enhancement at low metallicities in Carina 
suggests a {\it lack} of the excesses seen in the Galactic stars.
It is not clear from which stellar mass range these excesses arise:
Travaglio \etal (2004) suggest the main r-process occurs in 8-10 M$_\odot$
stars, and that the excess must arise from the more massive SNe~II,
whereas Farouqi \etal (2009) suggest the excess occurs in SN~II with
lower entropy winds where a neutron-rich, $\alpha$-freezeout can occur.
The specific mass or energy range in Farouqi's models is not clear.
If we follow Travaglio's suggestions, then the excess Sr may form in
more massive, or higher energy SNe~II (i.e., hypernovae),
and therefore those SNe~II seem to be missing in Carina $-$ 
either those stars did not form or their ejecta was driven out of 
Carina\footnote{Low upper limits on the [Sr/Fe] and 
[Ba/Fe] are also reported for Draco-119, however these cannot be used together
to constrain [Sr/Ba].  As an exercise, if the upper limits are taken as the
actual values, then [Ba/Fe] $=-2.6$ and [Sr/Ba] $=+0.1$, which places this
star squarely amongst the Galactic distribution.  Thus, even the lack of a
detection of the \ion{Ba}{2} and \ion{Sr}{2} spectra lines in Draco-119 does 
{\it not} provide a strong constraint on the one shot hypernova model.   The
low values {\it could} be consistent with inhomogeneous mixing and/or gas
driven out by supernova winds.}.

Mn and Cr also form in hypernovae as the decay products of complete 
and incomplete Si-burning (Umeda \& Nomoto 2002, Nomoto \etal 2011), thus 
a lack of hypernovae might also explain the very low initial values of these 
elements.  A neutron-rich, $\alpha$-freezeout in hypernovae would also 
contribute to Na production, and thus a lack of hypernovae (i.e., if this is
the source of the Sr excess, as in Farouqi \etal's 2009 models) 
could explain the very low [Na/Fe] ratios 
in the most metal-poor stars.
Evolution models (e.g., Salvadori \etal 2008,
Brooks \etal 2009, 2007, Ferrara \& Tolstoy 2000) 
suggest that very low mass galaxies can lose gas after the first star 
forming epoch through SN~II driven winds (possibly reaccreting cold gas 
for later star formation events). 
Koch \etal (2006) also suggested that the loss of metal rich winds would help 
to explain the MDF of Carina, which suffers from the well known G dwarf 
problem.    Detailed models of the star formation and chemical evolution of
Carina by Lanfranchi \& Matteucci (2004) invoked two major epochs of outflows through
winds.  Their best model included a wind efficiency that was 7x the star formation
rate and associated with the initial and IA star formation episodes.
Thus, the detailed chemical composition of the most metal-poor stars 
in Carina suggest a lack of hypernovae contributions, possibly 
implying that these stars did not form, but more likely indicating 
that their gas was removed from Carina by SN driven winds.

One alternative to this scenario is that the metal-poor stars are 
displaying the imprint of the hypernovae, rather than the lack of hypernovae,
which would provide a very strong constraint on the nucleosynthetic models.
An excellent test of this alternative would be the [Zn/Fe] abundances, which
are predicted to be enriched by hypernovae because of an increase in the 
mass ratio between the complete and incomplete Si-burning regions
(Nomoto \etal 2011).  
We are only able to determine upper limits to the [Zn/Fe] ratios in three
stars in Carina below [Fe/H] $= -2$ from our moderate SNR Magellan/MIKE spectra; 
however these limits are tantalizingly close to providing an interesting 
constraint (see Fig.~\ref{cuzn}).  
In particular, Car-5070 has a low [Zn/Fe] upper limit, and suggests that 
Car-5070 {\it lacks} hypernova enrichments.

\subsubsection{Summary}

Carina and Draco (and possibly Bo\"otes I) appear to be quite different in their
early chemical evolution from the other dSphs and UFDs.    These two (three) 
galaxies may be at the critical mass where SN driven winds remove the gas 
from the most massive or energetic SNe~II progenitors, but the products of 
the remaining SNe~II are retained, and contribute to the 
(inhomogeneous) chemical evolution of the host.    
This is unlike the dSphs, which appear to retain the gas from the earliest 
epochs and undergo a smooth chemical evolution  that is not too different
from stars in the Galactic halo (other than contributions occurring at lower
metallicities, e.g., the AGB yields are from metal-poor stars, and the SNe~Ia
contribute at lower metallicities).  Both of these galactic systems are 
unlike the UFDs, where a single massive SN~II may remove all of the gas, 
quenching the star formation event, and imprinting their unique chemical 
signatures on the few stars that will complete the star formation process 
- in this case, any abundance variations within an UFD are due to stochastic 
sampling of that hypernova and not chemical evolution.

\subsection {Car-612: A pocket of SN~Ia Enriched Gas}

\begin{figure}[b]
\begin{center}
\includegraphics[width=90mm, height=40mm]{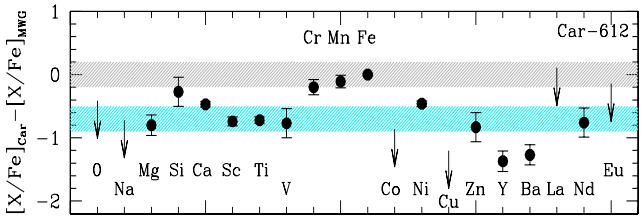}
\caption{Abundance distribution for Car-612 relative to Galactic stars at 
its metallicity of [Fe/H] = $-1.3$.  The Galactic abundances are estimated 
from the distributions in Figs.~\ref{osi} to \ref{srndeu}.  
The grey shaded area represents a (mean) error in 
the Galactic abundance of $\pm$0.2 dex.  The cyan shaded area  represents an
offset by $-0.7$ dex, i.e., if Car-612 were to have a metallicity of $-2.0$.
Note that at [Fe/H] = $-2.0$, the stars in classical dSphs 
have similar [X/Fe] values as stars in the Galaxy, thus the
apparent iron enrichment does not depend on differences in 
the chemical evolution of the comparison dataset.
}
\label{a612}
\end{center}
\end{figure}

As discussed in Section~4, Car-612 is underabundant in nearly 
every element when examined relative to iron, i.e., [X/Fe].    
Therefore, we propose that this star is iron enhanced, most 
likely due to an excess of SN~Ia 
contributions in the gas cloud from which it formed.  

In Fig.~\ref{a612}, we show the abundance distribution relative to 
the Galactic averages at [Fe/H] = -1.3 (estimated from the Figures in 
Section 4).    Notice that very few element ratios lay in the grey
band that would describe the [X/Fe] ratios of a typical Galactic 
star at that metallicity.   The only exceptions are Cr, Mn, and Fe, 
and possibly Si (though the error is large for that element).    
If this star has an excess of iron by 0.7 dex (a factor of 5), 
then removing this in the [X/Fe] ratios would produce the values
in the blue band (other than Fe itself which we do not adjust).   
Now Cr, Mn, and Fe (itself) appear enhanced, however nearly
all other [X/Fe] ratios are in agreement with the Galactic stars.  
A direct comparison of spectral line strengths shows
that Car-612 has similar \ion{Fe}{1} line strengths to stars with
similar atmospheric parameters (e.g., Car-705 in Fig.~\ref{y2}; 
also to Car-1013 not shown), while the other lines are weaker.

This peculiar chemistry makes Car-612 similar to three iron-rich 
stars in the outer halo, studied by Ivans \etal (2003).  
These stars are all near [Fe/H] $= -2$ and have 
low [$\alpha$/Fe] ratios, low ratios of Y, Sr, and Ba, 
and two show enhancements in Cr, Mn, Ni, and Zn.   
Of those two stars, one is enhanced in Si and Eu.    
Ivans \etal concluded that these stars have larger SN~Ia/II 
contributions, by factors of 3 to 7 relative to the average 
halo star.  This is similar to Car-612 where we suggest 
the enhancement is by a factor of 5. 
  
A deep and thorough examination of the predicted yields from 
existing SN models showed that no combination could reproduce 
the detailed abundance patterns of the three outer halo stars, 
i.e., problems remained in the abundances of Ti, Cr, Mn, Ni, and Zn.   
Since the SN yields are able to reproduce the chemistry of the
majority of stars in the Galaxy, Ivans \etal suggested that 
perhaps the SN yields should not be integrated over a Salpeter IMF 
for these three stars and/or the degree of mixing in the ISM may 
vary from region to region.   If so, they considered that
it is also possible for these stars to have been deposited in 
the outer halo during a dwarf galaxy merger.   Variations in 
their detailed abundances are therefore related to differences 
in the chemical evolution of their hosts.   
Car-612 fits with this hypothesis $-$ it appears to
be iron-enhanced, and yet its detailed chemistry is again
different from that of the three outer halo stars, 
e.g., Ni, Zn, and Eu are not enhanced.   Identifying this
star {\it within} the Carina dwarf galaxy provides additional
evidence for inhomogeneous mixing in this low mass galaxy, and
provides a clear connection between the formation of these
chemically peculiar stars in dwarf galaxies and their
existence in the outer Galactic halo.


\section{Conclusions}

Carina is an interesting galaxy for chemical evolution studies
because of its low mass and its episodic star formation history.
In this paper, we have determined the abundances of 23 elements 
in the spectra of nine RGB stars in the Carina dSph galaxy taken 
with both the VLT/FLAMES-UVES and Magellan/MIKE spectrographs.    
This is a significant increase from the previous number of stars with 
detailed analyses, e.g., by Shetrone \etal (2003, where all 
stars had intermediate metallicities) and Koch \etal (2008, 
where only the iron and $\alpha$-element abundances were determined).  
Our analysis uses both photometric and spectroscopic techniques to
determine the stellar parameters, and we use new spherical models, an 
expanded line list, continuum scattering corrections, and 
hyperfine structure and NLTE corrections (when available) to  
improve the precision in our abundances.   Adopting the ages determined
by Lemasle \etal (2012), we are able to examine the chemical evolution
of Carina, separating chemical signatures in the old and intermediate-aged
populations. 
A summary of the most important results in this paper are as follows:

\begin{itemize}

\item  {\it Inhomogeneous mixing:}   A large dispersion in [Mg/Fe] 
       indicates poor mixing in the old population.   
       An offset in the [$\alpha$/Fe] ratios between the old 
       and intermediate-aged populations 
       (when previously published data
       are included)
       also suggests that the second 
       star formation event occurred in $\alpha$-enriched gas.  
       In addition, one star, Car-612, seems to have
       formed in a pocket enhanced in SN Ia/II products.

\item  {\it SN driven winds:} Stars with [Fe/H] $< -2$ do not appear to have
       been enriched in AGB or SNe Ia products.    Their peculiar chemistry
       includes very low ratios of [Sr/Ba], an element ratio that usually
       shows an excess in metal-poor stars in the Galaxy and dSphs.   
       Adopting a scenario where the excess Sr forms in the more massive or 
       energetic SN II, then the lack of this excess in Carina (also Draco
       and Bo\"otes I) is consistent with the loss of those products by 
       SN II driven winds.   Low ratios of [Na/Fe], [Mn/Fe], and [Cr/Fe] 
       support this scenario, with additional evidence from the low [Zn/Fe]
       upper limit for one star, Car-5070.    The $\alpha$-elements ratios
       in Car-5070 are also lower than the Galactic distribution.

\end{itemize}

It is interesting that the chemistry of the metal-poor stars in Carina are
not similar to those in the Galaxy, most of the other dSphs, or the UFDs.   
The [Sr/Fe] and [Sr/Ba] ratios are clear indicators of the differences 
in the early chemical evolution of these systems:   the Galaxy and dSphs 
appear to retain all of their SN II products, and show excesses in the [Sr/Ba]
ratios, whereas Carina (with Draco and Bo\"otes I) may be at the critical mass 
where {\it some} gas is lost through SN II driven winds, by showing
very low [Sr/Fe] and [Sr/Ba] ratios.  In the UFDs, all of the gas may be lost 
with the first SN II, quenching star formation, and imprinting 
the unique chemical signature of that SN II on the remaining stars as they 
complete the formation process, thus [Sr/Fe] is low, but [Sr/Ba] can vary.    

It is also interesting to find a star with an enhancement in the SN~Ia/II
products in Carina, similar to the three outer halo stars examined by
Ivans \etal (2003).   This provides the first direct link between the formation 
of these stars in low mass galaxies and their presence in the outer Galactic halo

\acknowledgements
We are grateful to the ESO VLT and Magellan Observatory support staff for outstanding
help and hospitality during our visitor runs associated with this work.    
Special thanks to the Stars Group at UVic for our frequent and lively discussions, and
help with Magellan/MIKE spectroscopy from Mario Mateo and Andy McWilliam.
KAV and MD thank NSERC for the Discovery Grant that funded the majority this work.  
KAV and CB thank the NSF for early support through award AST 99-84073.  

{Facilities:} \facility{VLT:Kueyen(FLAMES)}, \facility{Magellan:I(MIKE)},
              \facility{ESO:2.2m(WFI)}


\begin{thebibliography}{}
\bibitem[Andrievsky \etal (2007)]{And07} Andrievsky, S.~M., Spite, M., Korotin, S.~A., Spite, F., 
     P., Bonifacio, P., Cayrel, R., \& Hill, V.  Fran{\c c}ois, P., 2007, \aap, 464, 1081   
\bibitem[Andrievsky \etal (2009)]{And09} Andrievsky, S.~M., Spite, M., Korotin, S.~A., Spite, F., 
     Fran{\c c}ois, P., Bonifacio, P., Cayrel, R., \& Hill, V.\ 2009, \aap, 494, 1083 
\bibitem[Andrievsky \etal (2011)]{And11} Andrievsky, S.~M., Spite, F., Korotin, S.~A., 
     Fran{\c c}ois, P., Spite, M., Bonifacio, P., Cayrel, R., \& Hill, V.\ 2011, \aap, 530, 105 
\bibitem[Aoki \etal (2003)]{Aoki03} Aoki, W., et al.\ 2003, \apjl, 592, L67 
\bibitem[Aoki \etal (2009)]{Aoki09} Aoki W. et al.\ 2009, \aap, 502, 569 
\bibitem[Aoki \etal (2007)]{Aoki07} Aoki W., Beers T.C., Christlieb N., Norris J.E., Ryan S.G.,
  Tsangarides S., 2007, \apj, 655, 492 
\bibitem[Alonso \etal (1999)]{Alo99} Alonso, A., Arribas, S., \& Mart{\'{\i}}nez-Roger, C.\ 1999, \aaps, 139, 335  
\bibitem[Alvarez \& Plez (1998)]{Alv98} Alvarez R., Plez B., 1998, A\&A, 330, 1109 
\bibitem[Arlandini \etal (1999)]{Arl99} Arlandini, C., K{\"a}ppeler, F., Wisshak, K., 
       Gallino, R., Lugaro, M., Busso, M., \& Straniero, O.\ 1999, \apj, 525, 886 
\bibitem[Asplund \etal (2009)]{Asp09} Asplund, M., Grevesse, N., Sauval, A.~J., \& Scott, P.\ 2009, \araa, 47, 481  
\bibitem[Asplund \etal (2005)]{Asp05} Asplund, M., Grevesse, N., \& Sauval, A.~J.\ 2005, 
        Cosmic Abundances as Records of Stellar Evolution and Nucleosynthesis, 336, 25 
\bibitem[Battaglia \etal (2006)]{Bat06} Battaglia G., et al.\ 2006, \aap, 459, 423  
\bibitem[Battaglia \etal (2008)]{Bat08} Battaglia G., Irwin M., Tolstoy E., Hill V., Helmi A., Letarte B.,
  Jablonka P., 2008, MNRAS, 383, 183  
\bibitem[Beers \& Christlieb (2005)]{Beers05} Beers, T.~C., \& Christlieb, N.\ 2005, \araa, 43, 531
\bibitem[Bergemann \& Gehren (2008)]{Ber08} Bergemann, M., \& Gehren, T.\ 2008, \aap, 492, 823 
\bibitem[Bergemann \& Cescutti (2010)]{Ber10b} Bergemann, M., \& Cescutti, G.\ 2010, \aap, 522, A9 
\bibitem[Bergemann \etal (2010)]{Ber10} Bergemann, M., Pickering, J.~C., \& Gehren, T.\ 2010, \mnras, 401, 1334 
\bibitem[Biehl (1976)]{Beh76} Biehl, D.\ 1976, Ph.D.~Thesis    
\bibitem[Bono \etal (2010)]{Bon10} Bono, G., et al.\ 2010, \pasp, 122, 651 
\bibitem[Booth \etal (1983)]{Boo83} Booth, A.~J., Shallis, 
	M.~J., \& Wells, M.\ 1983, \mnras, 205, 191  
\bibitem[Brocato \etal (1997)]{Broc97} Brocato E., Castellani V., Piersimoni A., 1997,
  ApJ, 491, 789
\bibitem[Brown \etal (1987)]{Bro87} Brown, J.~A., Johnson, 
	H.~R., Alexander, D.~R., \& Wehrse, R.\ 1987, \baas, 19, 705
\bibitem[Brown (1997)]{Brown97} Brown, J.~A., Wallerstein, G., \& Zucker, 
	D.\ 1997, \aj, 114, 180  
\bibitem[Burris (2000)]{Bur00} Burris, D.~L., 
	Pilachowski, C.~A., Armandroff, T.~E., Sneden, C., Cowan, J.~J., 
	\& Roe, H.\ 2000, \apj, 544, 302
\bibitem[Carbon \etal (1982)]{Car82} Carbon, D.~F., 
	Romanishin, W., Langer, G.~E., Butler, D., Kemper, E., Trefzger, C.~F., 
	Kraft, R.~P., \& Suntzeff, N.~B.\ 1982, \apjs, 49, 207 
\bibitem[Cassisi \etal (1998)]{Cas98} Cassisi S., Castellani V., degl'Innocenti S.,
  Weiss A., 1998, A\&AS, 129, 267
\bibitem[Cayrel \etal (1988)]{Cay88} Cayrel, R.\ 1988, The Impact of 
	Very High S/N Spectroscopy on Stellar Physics, 132, 345  
\bibitem[Cayrel \etal (1991)]{Cay91} Cayrel, R., 
	Perrin, M.-N., Barbuy, B., \& Buser, R.\ 1991, \aap, 247, 108 
\bibitem[Cayrel \etal (2004)]{Cay04} Cayrel R., Depagne E., Spite M., Hill V., Spite F.,
  Francois P., Plez B., Beers T., Primas F., Andersen J., Barbuy B., 
  Bonifacio P., Molaro P., Nordstrom B., 2004, A\&A, 416, 1117 
\bibitem[Cescutti \etal (2008)]{Ces08} Cescutti, G., Matteucci, F., Lanfranchi, G.~A., \& McWilliam, A.\ 2008, \aap, 491, 401
\bibitem[Cohen \& Huang (2009)]{Coh09} Cohen J., Huang W., 2009, \apj, 701, 1053 
\bibitem[Cohen \etal (2008)]{Coh08} Cohen J., Christlieb N., McWilliam A., Shectman S.,
  Thompson I., Melendez J., Wisotzki L., Reimers D., 2008, \apj, 672, 320 
\bibitem[Dall'Ora \etal (2003)]{Dal03} Dall'Ora, M., et al.\ 2003, \aj, 126, 197 
\bibitem[de~Laverny \etal (2006)]{deL06} de Laverny, P., Abia, C., Dom{\'{\i}}nguez, I., 
	Plez, B., Straniero, O., Wahlin, R., Eriksson, K., \& J{\o}rgensen, U.~G.\ 2006, \aap, 446, 1107 
\bibitem[Dinescu \etal (1999)]{Din99} Dinescu D.I., van Altena W.F., Girard T.M.,
  Lopez C.E., 1999, AJ, 117, 277 
\bibitem[Dolphin \etal (2002)]{Dol02} Dolphin A.E., 2002, MNRAS, 332, 91 
\bibitem[Dufton \etal (2008)]{Duf08} Dufton, P.~L., Ryans, 
	R.~S.~I., Thompson, H.~M.~A., \& Street, R.~A.\ 2008, \mnras, 385, 2261
\bibitem[Fagotto \etal (1994)]{Fag94} Fagotto F., Bressan A., Bertelli G., Chiosi C.,
  1994, A\&AS, 104, 365
\bibitem[Feltzing \etal (2009)]{Fel09} Feltzing, S., Eriksson, K., 
	Kleyna, J., \& Wilkinson, M.~I.\ 2009, \aap, 508, L1  
\bibitem[Feltzing \etal (2009b)]{Fel09b} Feltzing, S., Primas, F., \& Johnson, 
	R.~A.\ 2009, \aap, 493, 913 
\bibitem[Ferrara \& Tolstoy (2000)]{Fer00} Ferrara A., Tolstoy E., 2000, MNRAS, 313, 291
\bibitem[Fran{\c c}ois \etal (2007)]{Fra07} Fran{\c c}ois, P., et al.\ 2007, \aap, 476, 935 
\bibitem[Frebel \etal (2010c)]{Fre10c} Frebel A., Kirby E., Simon J.D., 2010c, Nature, 464, 72 
\bibitem[Frebel \etal (2010b)]{Ftabl} Frebel A., 2010b, Astronomische Nachrichten, 331, 474
\bibitem[Frebel \etal (2010a)]{Fre10a} Frebel A., Simon J.D., Geha M., Willman B., 2010a,
   \apj, 708, 560 
\bibitem[Fulbright (2000)]{Ful00} Fulbright, J.~P.\ 2000, \aj, 120, 1841 
\bibitem[Fulbright \etal (2004)]{Ful04} Fulbright, J.~P., 
	Rich, R.~M., \& Castro, S.\ 2004, \apj, 612, 447  
\bibitem[Geisler \etal (2005)]{Gei05} Geisler D., Smith V.V., Wallerstein G., Gonzalez G., 
  Charbonnel C., 2005, \aj, 129, 1428 
\bibitem[Gilmore (2009)]{Gil09} Gilmore, G.\ 2009, IAU 
	Symposium, 254, 487 
\bibitem[Gratton \etal (2010)]{Gra10} Gratton, R.~G., Carretta, E.,  Bragaglia, A., 
        Lucatello S., D'Orazi V., \ 2010, \aap, 517, 81   
\bibitem[Gratton \etal (2004)]{Gra04} Gratton, R.~G., 
	Sneden, C., Carretta, E., 2004, \araa, 42, 385
\bibitem[Gratton \etal (2000)]{Gra00} Gratton, R.~G., 
	Sneden, C., Carretta, E., \& Bragaglia, A.\ 2000, \aap, 354, 169  
\bibitem[Gratton \etal (1997)]{Gra97}  Gratton R.G., Fusi Pecci F., Carretta E., Clementini G.,
  Corsi E.E., Lattanzi M.G., 1997, ApJ, 491, 749 
\bibitem[Grevesse \& Sauval (1998)]{Gre98} 
	Grevesse, N., \& Sauval, A.~J.\ 1998, \ssr, 85, 161  
\bibitem[Guo \etal (2011)]{Guo11} Guo, Q., et al.\ 2011, \mnras, 413, 101 
\bibitem[Gustafsson \etal (2003)]{Gus03} Gustafsson, B., 
	Edvardsson, B., Eriksson, K., Mizuno-Wiedner, M., J{\o}rgensen, U.~G., 
	\& Plez, B.\ 2003, Stellar Atmosphere Modeling, 288, 331  
\bibitem[Gustafsson \etal (2008)]{Gus08} Gustafsson, B., Edvardsson, B., 
	Eriksson, K., J{\o}rgensen, U.~G., Nordlund, {\AA}., \& Plez, B.\ 2008, \aap, 486, 951
\bibitem[Hambly \etal (1994)]{Ham94} Hambly, N.~C., Dufton, P.~L., Keenan, F.~P., Rolleston, 
	W.~R.~J., Howarth, I.~D., \& Irwin, M.~J.\ 1994, \aap, 285, 716 
\bibitem[Harris (1996)]{Ha96} Harris, W.E., 1996, AJ, 112, 1487 
\bibitem[Heger \& Woosley (2002)]{Heg02}
  Heger, A., \& Woosley, S.~E.\ 2002, \apj, 567, 532 
\bibitem[Heiter \etal (2006)]{Hei06} Heiter, U., 
	\& Eriksson, K.\ 2006, \aap, 452, 1039  
\bibitem[Helmi \etal (2006)]{Hel06} Helmi, A., et al.\ 2006, \apjl, 651, L121
\bibitem[Hill \etal (2012)]{Hill12} Hill V., \etal 2012, in preparation
\bibitem[Hernandez \etal (2000)]{Her00} Hernandez X., Gilmore G., Valls-Gabaud D., 2000,
   MNRAS, 317, 831 
\bibitem[Hurley-Keller \etal (1998)]{HK98} Hurley-Keller D., Mateo M., Nemec J., 1998,
 	AJ, 115, 1840 
\bibitem[Irwin \& Hatzidimitriou (1995)]{IH95} Irwin, M., \& Hatzidimitriou, D.\ 1995, 		
	\mnras, 277, 1354 
\bibitem[Ishimaru \etal (2004)]{Ish04} Ishimaru, Y., Wanajo, 
	S., Aoki, W., \& Ryan, S.~G.\ 2004, \apjl, 600, L47 
\bibitem[Ivans \etal (2003)]{Iva03} Ivans, I.~I., Sneden, C., James, C.~R., Preston, G.~W., Fulbright, J.~P., H{\"o}flich, P.~A., 			Carney, B.~W., \& Wheeler, J.~C.\ 2003, \apj, 592, 906 
\bibitem[Iwamoto \etal (1999)]{Iwa99} Iwamoto, K., Brachwitz, F., Nomoto, K., Kishimoto, N., Umeda, H., Hix, W.~R., 
	\& Thielemann, F.-K.\ 1999, \apjs, 125, 439 
\bibitem[Johnston \etal (2008)]{Joh08} Johnston, K.~V., Bullock, J.~S., Sharma, S., 
        Font, A., Robertson, B.~E., \& Leitner, S.~N.\ 2008, \apj, 689, 936 
\bibitem[Kappeler \etal (1989)]{Kap89} Kappeler, F., Beer, 
	H., \& Wisshak, K.\ 1989, Reports on Progress in Physics, 52, 945
\bibitem[Karakas \etal (2010)]{Kar10} Karakas A., 2010, MNRAS, 403, 1413 
\bibitem[Karakas \etal (2008)]{Kar08} Karakas A., Lee H.Y., Lugaro M., G\"orres J., Wiescher M., 2008, \apj, 676, 1254 
\bibitem[Kaufer \etal (2004)]{Kau04} Kaufer, A., Venn, K.~A., Tolstoy, E., Pinte, C., \& 
	Kudritzki, R.-P.\ 2004, \aj, 127, 2723 
\bibitem[Kirby \etal (2008)]{Kir08} Kirby E.N., Simon J.D., Geha M., Guhathakurta P., 
  Frebel A., 2008, \apj, 685, 43
\bibitem[Kirby \etal (2011a)]{Kir11a} Kirby, E.~N., Cohen, J.~G., Smith, G.~H., Majewski, 
	S.~R., Sohn, S.~T., \& Guhathakurta, P.\ 2011a, \apj, 727, 79 
\bibitem[Kirby \etal (2011b)]{Kir11b} Kirby, E.~N., Lanfranchi, G.~A., Simon, J.~D., Cohen, 
	J.~G., \& Guhathakurta, P.\ 2011b, \apj, 727, 78 
\bibitem[Kobayashi \& Nomoto (2009)]{Kob09} Kobayashi, C., \& Nomoto, K.\ 2009, \apj, 707, 1466 
\bibitem[Kobayashi \etal (2006)]{Kob06} Kobayashi, C., Umeda, 
	H., Nomoto, K., Tominaga, N., \& Ohkubo, T.\ 2006, \apj, 653, 1145
\bibitem[Koch \etal (2006)]{Koc06} Koch A., Grebel E.K., Wyse R.F.G., Kleyna J.T., 
  Wilkinson, M.I., Harbeck D.R., Gilmore G.F., Evans N.W., 2006,
  AJ, 131, 895
\bibitem[Koch \etal (2008a)]{Koc08a} Koch A., Grebel E.K., Gilmore G.F., Wyse R.F.G., 
  Kleyna J.T., Harbeck D.R., Wilkinson M.I., Wyn Evans, N., 2008a,
  \aj, 135, 1580 
\bibitem[Koch \etal (2008b)]{Koc08b} Koch A., McWilliam A., Grebel E.K., Zucker D.B.,
  Belokurov V., 2008b, \apjl, 688, 13
\bibitem[Koch \etal (2009)]{Koc09} Koch A., Wilkinson M.I., Kleyna J.T., Irwin M.J., 
  Zucker D.B., Belokurov V., Gilmore G.F., Fellhauer M., Evans N.W., 
  2009, \apj, 690, 453
\bibitem[Kraft \& Ivans (2003)]{Kra03} Kraft, R.~P., \& 
	Ivans, I.~I.\ 2003, \pasp, 115, 143  
\bibitem[Kuhn \etal (1996)]{Kuh96} Kuhn, J.~R., Smith, H.~A., \& Hawley, S.~L.\ 1996, \apjl, 469, L93 
\bibitem[Lai \etal (2008)]{Lai08} Lai, D.~K., Bolte, M., Johnson, J.~A., 
	Lucatello, S., Heger, A., \& Woosley, S.~E.\ 2008, \apj, 681, 1524  
\bibitem[Lane \etal (2009)]{Lan09} Lane, R.~R., Kiss, L.~L., 
	Lewis, G.~F., Ibata, R.~A., Siebert, A., Bedding, T.~R., 
	\& Sz{\'e}kely, P.\ 2009, \mnras, 400, 917 
\bibitem[Lanfranchi \& Matteucci (2004)]{Lan04} Lanfranchi, G.~A., \& Matteucci, F.\ 2004, \mnras, 351, 1338
\bibitem[Lanfranchi \& Matteucci (2003)]{Lan03} Lanfranchi, G.~A., \& Matteucci, F.\ 2003, \mnras, 345, 71 
\bibitem[Lanfranchi \etal (2006)]{Lan06} Lanfranchi, G.~A., Matteucci, F., \& Cescutti, G.\ 2006, \aap, 453, 67 
\bibitem[Lawler \etal (2001)]{Law01} Lawler, J.~E., 
	Bonvallet, G., \& Sneden, C.\ 2001a, \apj, 556, 452 
\bibitem[Lawler \etal (2001b)]{Law01b} Lawler, J.~E., 
	Wickliffe, M.~E., den Hartog, E.~A., \& Sneden, C.\ 2001b, \apj, 563, 1075 
\bibitem[Lee \etal (2005)]{Lee05} Lee J.W., Carney B.W., Habgood M.J., 2005, AJ, 129, 251 
\bibitem[LeMasle \etal (2012)]{LeM12} Lemasle B., \etal 2012, \aap, 538, 100
\bibitem[Letarte \etal (2010)]{LeT10} Letarte B., \etal 2010, \aap, 523, 17 
\bibitem[Letarte \etal (2009)]{LeT09} Letarte, B., et al.\ 
	2009, \mnras, 400, 1472  
\bibitem[Letarte \etal (2006)]{LeT06} Letarte, B., Hill, V., Jablonka, P., Tolstoy, E.,
	Fran{\c c}ois, P., \& Meylan, G.\ 2006, \aap, 453, 547 
\bibitem[Lokas \etal (2009)]{Lok09} {\L}okas, E.~L.\ 2009, \mnras, 394, L102 
\bibitem[Mac~Low \& Ferrara (1999)]{Mac99} Mac Low, M.-M., \& Ferrara, A.\ 1999, \apj, 513, 142 
\bibitem[Maoz \etal (2010)]{Mao10} Maoz, D., Sharon, K., \& Gal-Yam, A.\ 2010, \apj, 722, 1879 
\bibitem[Mashonkina \etal (2010)]{Mash10} Mashonkina, L., 
	Gehren, T., Shi, J., Korn, A., \& Grupp, F.\ 2010, IAU Symposium, 265, 197 
\bibitem[Mateo \etal (1993)]{Mat93} Mateo, M., Olszewski, E.~W., Pryor, C., Welch, D.~L., \& Fischer, P.\ 1993, \aj, 105, 510 
\bibitem[Mateo (1998)]{Mat98}Mateo, M. 1998, \araa, 36, 435
\bibitem[Majweski \etal (2000)]{Maj00} Majewski, S.~R., Ostheimer, J.~C., Patterson, R.~J., 
    Kunkel, W.~E., Johnston, K.~V., \& Geisler, D.\ 2000, \aj, 119, 760 
\bibitem[Majewski \etal (2002)]{Maj02} Majewski, S.~R., et al.\ 2002, Modes of Star Formation and the Origin of Field Populations, 285, 199 
\bibitem[Majewski \etal (2005)]{Maj05} Majewski, S.~R., et al.\ 2005, \aj, 130, 2677 
\bibitem[McClure \etal (1987)]{McC87} McClure R., Vandenberg D.A., Bell A.R., Hesser J.E., 
  Stetson P.B., 1987, AJ, 101, 515 
\bibitem[McWilliam \etal (1995)]{McW95} McWilliam A., Preston G.W., Sneden C., Searle L., 
  1995, \aj, 109, 2736 
\bibitem[McWilliam (1998)]{McW98} McWilliam A., 1998, \aj, 115, 1640 
\bibitem[Meynet \etal (2006)]{Mey06} Meynet, G., 
	Ekstr{\"o}m, S., \& Maeder, A.\ 2006, \aap, 447, 623 
\bibitem[Mighell (1997)]{Mig97}  Mighell K.J., 1997 AJ, 114, 1458 
\bibitem[Mishenina (2002)]{Mish02} Mishenina, T.~V., Kovtyukh, V.~V., 
	Soubiran, C., Travaglio, C., \& Busso, M.\ 2002, \aap, 396, 189 
\bibitem[Monelli \etal (2003)]{Mon03} Monelli M., Pulone L., Corsi C.E., Castellani M.,
  etal 2003 AJ, 126, 218 
\bibitem[Mu{\~n}oz \etal (2006)]{Mun06} Mu{\~n}oz, R.~R., et al.\ 2006, \apj, 649, 201 
\bibitem[Nakamura \etal (1999)]{Nak99} Nakamura, T., Umeda, 
	H., Nomoto, K., Thielemann, F.-K., \& Burrows, A.\ 1999, \apj, 517, 193 
\bibitem[Nakamura \etal (2001)]{Nak01} Nakamura, T., Mazzali, 
	P.~A., Nomoto, K., \& Iwamoto, K.\ 2001a, \apj, 550, 991 
\bibitem[Nakamura \etal (2001b)]{Nak01b} Nakamura, T., Umeda, 
	H., Iwamoto, K., Nomoto, K., Hashimoto, M.-a., Hix, W.~R., 
	\& Thielemann, F.-K.\ 2001b, \apj, 555, 880 
\bibitem[Nomoto \etal (2001)]{Nom01} Nomoto, K., Maeda, K., Umeda, H., \& Nakamura, 
	T.\ 2001, The Influence of Binaries on Stellar Population Studies, 264, 507 
\bibitem[Nissen \& Schuster (1997)]{NS97} Nissen P., Schuster W.J., 1997, \aap, 326, 751 
\bibitem[Nissen \& Schuster (2010)]{NS10} Nissen, P.~E., \& 
	Schuster, W.~J.\ 2010, \aap, 511, L10 
\bibitem[Nomoto \etal (1997a)]{Nom97a} Nomoto, K., Hashimoto, 
	M., Tsujimoto, T., Thielemann, F.-K., Kishimoto, N., Kubo, Y., 
	\& Nakasato, N.\ 1997, Nuclear Physics A, 616, 79  
\bibitem[Nomoto \etal (1997b)]{Nom97b} Nomoto, K., Iwamoto, K., 
	Nakasato, N., Thielemann, F.-K., Brachwitz, F., Tsujimoto, T., Kubo, Y., 
	\& Kishimoto, N.\ 1997, Nuclear Physics A, 621, 467  
\bibitem[Norris \etal (2008)]{Nor08} Norris, J.~E., Gilmore, G., Wyse, R.~F.~G., Wilkinson, 
	M.~I., Belokurov, V., Evans, N.~W., \& Zucker, D.~B.\ 2008, \apjl, 689, L113 
\bibitem[Norris \etal (2010a)]{Nor10a} Norris J.E., Wyse R.F.G., Gilmore G., Yong D., Frebel A., 
  Wilkinson M.I., Belokurov V., Zucker D.B., 2010 
\bibitem[Norris \etal (2010b)]{Nor10b} Norris J.E., Gilmore G., Wyse R.F.G., Yong D., Frebel A., 
   2010
\bibitem[Norris \etal (2010c)]{Nor10c} Norris J.E., Yong D., Gilmore G., Wyse R.F.G., 2010,
   \apj, 711, 350 
\bibitem[O'Brian \etal (1991)]{OB91} O'Brian T.R., Wickliffe M.E., Lawler J.E., Whaling W., 
   Brault J.W., 1991, J. Opt. Soc. Am. B, 8, 1185
\bibitem[Pasetto \etal (2010)]{Pas10} Pasetto, S., Grebel, E.~K., Berczik, P., Spurzem, R., \& Dehnen, W.\ 2010, \aap, 514, A47 
\bibitem[Pasquini \etal (2002)]{Pas02} Pasquini, L., et al. 2002, The Messenger, 110, 1 
\bibitem[Piatek \etal (2003)]{Pai03} Piatek, S., Pryor, C., Olszewski, E.~W., Harris, H.~C., Mateo, M., Minniti, D., 
	\& Tinney, C.~G.\ 2003, \aj, 126, 2346 
\bibitem[Pietrzy{\'n}ski \etal (2009)]{Pie09} 
	 Pietrzy{\'n}ski, G., G{\'o}rski, M., Gieren, W., Ivanov, V.~D., Bresolin, 
	F., \& Kudritzki, R.-P.\ 2009, \aj, 138, 459 
\bibitem[Pignatari \etal (2010)]{Pig10} Pignatari, M., Gallino, R., Heil, M., Wiescher, M., K{\"a}ppeler, F., Herwig, F., 
	\& Bisterzo, S.\ 2010, \apj, 710, 1557 
\bibitem[Pomp{\'e}ia \etal (2008)]{Pomp08} Pomp{\'e}ia, L., et al.\ 2008, \aap, 480, 379 
\bibitem[Prochaska \etal (2001)]{Pro01} Prochaska, J.~X., et al.\ 2001, \apjs, 137, 21  
\bibitem[Qian \& Wasserburg (2007)]{QW07} Qian, Y.-Z., \& Wasserburg, G.~J.\ 2007, \physrep, 442, 237 
\bibitem[Ram{\'i}rez \etal (2005)]{Ram05} Ram{\'{\i}}rez, I., \& Mel{\'e}ndez, J.\ 2005, \apj, 626, 465 
\bibitem[Reddy \etal (2006)]{RL06} Reddy, B.~E., Lambert, 
	D.~L., \& Allende Prieto, C.\ 2006, \mnras, 367, 1329  
\bibitem[Revaz \& Jablonka (2011)]{Rev11} Revaz Y., Jablonka P., 2011, in preparation
\bibitem[Revaz \etal (2009)]{Rev09} Revaz, Y., et al.  2009, \aap, 501, 189  
\bibitem[Rizzi \etal (2003)]{Riz03} Rizzi, L., Held, E.~V., Bertelli, G., \& Saviane, I.\ 2003, \apjl, 589, L85 
\bibitem[Roederer (2009)]{Roe09} Roederer, I.~U.\ 2009, \aj, 137, 272  
\bibitem[Roederer \etal (2010)]{Roe10} Roederer, I.~U., 
	Sneden, C., Thompson, I.~B., Preston, G.~W., \& Shectman, S.~A.\ 2010, \apj, 711, 573 
\bibitem[R{\"o}pke \etal (2006)]{Rop06} R{\"o}pke, F.~K., 
	Hillebrandt, W., \& Blinnikov, S.~I.\ 2006, ESA Special Publication, 637,  
\bibitem[Rutlegde \etal (1997a)]{Rut97a} Rutledge, G.~A., 
	Hesser, J.~E., Stetson, P.~B., Mateo, M., Simard, L., Bolte, M., Friel, 
	E.~D., \& Copin, Y.\ 1997, \pasp, 109, 883 
\bibitem[Rutledge \etal (1997b)]{Rub97b} Rutledge, G.~A., 
	Hesser, J.~E., \& Stetson, P.~B.\ 1997, \pasp, 109, 907 
\bibitem[Salvadori \etal (2007)]{SSF07} Salvadori S., Schneider R., Ferrara A., 2007, MNRAS, 381, 647
\bibitem[Salvadori \etal (2008)]{SFS08} Salvadori S., Ferrara A., Schneider R., 2008, MNRAS, 386, 348
\bibitem[Saha \etal (2010)]{Sah10} Saha, A., et al.\ 2010, \aj, 140, 1719 
\bibitem[Schlegel \etal (1998)]{Sch98} Schlegel, D.~J., 
	Finkbeiner, D.~P., \& Davis, M.\ 1998, \apj, 500, 525  
\bibitem[Schoerck \etal (2009)]{Sch09} Schoerck, T., \etal 2009, \aap, 507, 817
\bibitem[Shetrone \etal (1998)]{sbs98} Shetrone M.D., Bolte M., Stetson P.B., 1998, 
   \aj, 115, 1888 
\bibitem[Shetrone \etal (2001)]{scs01} Shetrone M.D., C\^ot\'e P., Sargent W.L.W., 2001,
  \apj, 548, 592 
\bibitem[Shetrone \etal (2003)]{svt03} Shetrone M.D., Venn K.A., Tolstoy E., Primas F.,
  Hill V., Kaufer A., 2003, \aj, 125, 684
\bibitem[Short \& Hauschildt (2006)]{Sho06} Short, C.~I., \& Hauschildt, 
	P.~H.\ 2006, \apj, 641, 494 
\bibitem[Simmerer \etal (2004)]{Sim04} Simmerer, J., Sneden, C., Cowan, J.~J., 
	Collier, J., Woolf, V.~M., \& Lawler, J.~E.\ 2004, \apj, 617, 1091 
\bibitem[Simon \etal (2010)]{sfm10} Simon J.D., Frebel A., McWilliam A., Kirby E.N.,
  Thompson I.B., 2010, ApJ, 716, 446 
\bibitem[Smith \& Briley (2005)]{Smi05} Smith, G.~H., Briley, 
	M.~M., \& Harbeck, D.\ 2005, \aj, 129, 1589  
\bibitem[Sneden (1973)]{1973PhDT.......180S} Sneden, C.~A.\ 1973, 
	Ph.D.~Thesis,   
\bibitem[Sneden \etal (2008)]{Sne08} Sneden, C., Cowan, J.~J., \& 
	Gallino, R.\ 2008, \araa, 46, 241 
\bibitem[Sobeck \etal (2010)]{Sob10} Sobeck, J., Frohlich, C., Truran, J., \& Kim, Y.\ 2010, 
	American Institute of Physics Conference Series, 1294, 287 
\bibitem[Spite (1967)]{Spi67} Spite, M.\ 1967, Annales 
	d'Astrophysique, 30, 685 
\bibitem[Starkenburg \etal (2010)]{Star10} Starkenburg E., et al., 2010 \aap, 513, 34 
\bibitem[Stetson (2000)]{S2000} Stetson P.B., 2000, PASP, 112, 925 
\bibitem[Stetson (2005)]{S2005} Stetson P.B., 2005, PASP, 117, 563
\bibitem[Stetson \& Pancino (2008)]{SP08} Stetson P.B., Pancino E., 2008, PASP, 120, 1332 
\bibitem[Tafelmeyer \etal (2010)]{Taf10}
	Tafelmeyer, M., et al.\ 2010, \aap, 524, A58  
\bibitem[Tautvai{\v s}iene{\. e} (2004)]{Tau04} Tautvai{\v s}ien{\.e}, G., Wallerstein, 
	G., Geisler, D., Gonzalez, G., \& Charbonnel, C.\ 2004, \aj, 127, 373 
\bibitem[Tolstoy \etal (2009)]{Tol09} Tolstoy E., Hill V., Tosi M., 2009, ARAA 47, 371 
\bibitem[Tolstoy \etal (2004)]{Tol04} Tolstoy, E., et al.\ 2004, \apjl, 617, L119 
\bibitem[Tolstoy \etal (2003)]{Tol03} Tolstoy E., Venn K.A., Shetrone M.D., Primas F.,
  Hill V., Kaufer A., Szeifert T., 2003, \aj, 125, 707 
\bibitem[Travaglio \etal (2004)]{Tra04} Travaglio, C., 
	Gallino, R., Arnone, E., Cowan, J., Jordan, F., \& Sneden, C.\ 2004, \apj, 601, 864 
\bibitem[Travaglio \etal (2005)]{Tra05} Travaglio, C., Hillebrandt, W., \& Reinecke, M.\ 2005, \aap, 443, 1007 
\bibitem[Umeda \etal (2005)]{UN05} Umeda, H., 
	\& Nomoto, K.\ 2005, \apj, 619, 427 
\bibitem[Vandenberg \& Bell (1985)]{vdb85} Vandenberg, D.~A., \& Bell, R.~A.\ 1985, \apjs, 58, 561 
\bibitem[Venn \etal (2004)]{Ven04} Venn K.A., Irwin M.J., Shetrone M.D., Tout C.A., Hill V., Tolstoy E.,
   2004 \aj, 128, 1177 
\bibitem[Venn \etal (2003)]{VX03} Venn K.A., Tolstoy E., Kaufer A., Skillman E.D., et al. 2003,
   \aj, 126, 1326
\bibitem[Venn \etal (2001)]{VX01} Venn K.A., Lennon D.J., Kaufer A., McCarthy J.K., et al. 2001,
   \apj, 547, 765
\bibitem[Walker \etal (2009)]{Wal09} Walker M., \etal 2009 
\bibitem[Walker \etal (1994)]{Wal94} Walker A.R., 1994, AJ, 108, 555 
\bibitem[Woosley \& Weaver (1995)]{Woo95} Woosley, S.~E., \& Weaver, T.~A.\ 1995, \apjs, 101, 181  

\end{thebibliography}
\end{document}